\pdfoutput=1
\documentclass[12pt,reqno]{article}

\usepackage{mathpazo}
\usepackage{setspace}
\usepackage{titlesec}
\usepackage{microtype}
\usepackage[bottom]{footmisc}
\usepackage{amsmath,amssymb,amsthm,mathtools,bm}
\usepackage{graphicx}
\usepackage{booktabs}
\usepackage{longtable}
\usepackage{array}
\usepackage{float}
\usepackage{needspace}
\usepackage{comment}
\usepackage[shortlabels]{enumitem}
\usepackage[margin=1.15in]{geometry}
\usepackage[round,authoryear]{natbib}
\usepackage{xcolor}

\usepackage[normalem]{ulem}

\usepackage[most]{tcolorbox}
\usepackage{tikz}
\usetikzlibrary{arrows.meta,positioning}
\usepackage{hyperref}
\usepackage{footnotehyper}
\hypersetup{
    pdftitle={Organizing Intelligence},
    pdfauthor={Alex Imas; Andrew Koh; Shane Legg},
    pdfdisplaydoctitle=true,
    colorlinks=true,
    linkcolor=red!60!black,
    citecolor=blue!75!black,
    filecolor=magenta,
    urlcolor=blue!60!black,
}  
\usepackage[nameinlink,noabbrev,sort,capitalise]{cleveref}

\titlespacing*{\paragraph}{0pt}{1.25ex plus 1ex minus .2ex}{0.5em}
\titleformat{\section}
  {\bfseries\centering\MakeUppercase}
  {\thesection}{0.5em}{}[]
\titleformat{\subsection}[runin]
  {\normalfont\bfseries}
  {\thesubsection}{0.5em}{\addperiod}[]
\newcommand{\addperiod}[1]{#1.}
\allowdisplaybreaks
\newcommand{\R}{\mathbb R}
\newcommand{\Y}{\mathcal Y}
\newcommand{\M}{\mathcal M}
\newcommand{\F}{\mathcal F}
\newcommand{\W}{\mathcal W}
\newcommand{\de}{\mathop{}\!\mathrm d}
\newcommand{\supp}{\operatorname{supp}}

\newcommand{\one}{\mathbf 1}

\DeclareMathOperator*{\argmax}{arg\,max}

\theoremstyle{definition}
\newtheorem{assumption}{Assumption}
\newtheorem{definition}{Definition}
\newtheorem{example}{Example}

\newtheorem{lemma}{Lemma}
\newtheorem{proposition}{Proposition}
\newtheorem{theorem}{Theorem}

\crefname{assumption}{Assumption}{Assumptions}
\crefname{definition}{Definition}{Definitions}
\crefname{example}{Example}{Examples}
\crefname{remark}{Remark}{Remarks}
\crefname{lemma}{Lemma}{Lemmas}
\crefname{proposition}{Proposition}{Propositions}
\crefname{theorem}{Theorem}{Theorems}
\crefname{corollary}{Corollary}{Corollaries}
\crefname{appendix}{Appendix}{Appendices}
\Crefname{appendix}{Appendix}{Appendices}

\tcbset{resultbox/.style={
  enhanced jigsaw,
  breakable,
  sharp corners,
  colback=white,
  colframe=black,
  boxrule=0.5pt,
  left=7pt, right=7pt, top=6pt, bottom=6pt,
  before skip=9pt, after skip=9pt
}}
\tcolorboxenvironment{theorem}{resultbox}
\tcolorboxenvironment{proposition}{resultbox}
\tcolorboxenvironment{corollary}{resultbox}
\tcolorboxenvironment{lemma}{resultbox}
\tcolorboxenvironment{definition}{resultbox}
\tcolorboxenvironment{assumption}{resultbox}
\tcolorboxenvironment{remark}{resultbox}
\tcolorboxenvironment{example}{resultbox}

\newtcolorbox{cesbox}[2][]{
  enhanced jigsaw, breakable, sharp corners,
  frame hidden, interior hidden, boxrule=0pt,
  borderline west={0.6pt}{0pt}{black!30},
  colbacktitle=white, coltitle=black, fonttitle=\bfseries,
  title={CES: #2},
  left=8pt, right=0pt, top=3pt, bottom=3pt,
  before skip=9pt, after skip=9pt, #1
}

\newtcolorbox{taskbox}[2][]{
  enhanced jigsaw, breakable, sharp corners,
  colback=black!4, colframe=black!4, boxrule=0pt,
  colbacktitle=black!4, coltitle=black, fonttitle=\bfseries,
  title={Task model: #2},
  left=8pt, right=8pt, top=6pt, bottom=6pt,
  before skip=9pt, after skip=9pt, #1
}

\title{\vspace{-1em}\textbf{\textls[-50]{\textsc{Organizing Intelligence}}}}
\author{Alex Imas\thanks{Lead authors with equal contribution listed in alphabetical order; emails: \href{mailto:imasa@google.com}{imasa@google.com} and \href{mailto:andrewkoh@google.com}{andrewkoh@google.com}. Imas is also affiliated with The University of Chicago; Koh is also affiliated with Columbia University. \\[0.5em]
We thank Allan Dafoe, Matt Elliott, Joshua Gans, Gillian Hadfield, Jacopo Perego, Andrea Prat, Eduard Talamas, Phil Trammell, and Birger Wernerfelt for helpful conversations and suggestions. We also thank Pedro Aldighieri for superb research assistance. Various AI systems were used for assisting with mathematical arguments as well as proofreading for consistency and clarity. \\[0.5em] The paper expresses the views of the individual authors and does not necessarily reflect those of their affiliated institutions. This research is intended solely for academic discussion and does not represent corporate forecasts, product roadmaps, or empirical statements regarding Google's current or future operations.
} \qquad Andrew Koh\footnotemark[1] \qquad Shane Legg\\[2mm]
\normalsize Google DeepMind}
\date{October 31, 2026}

\begin{document}
\maketitle

\vspace{-3em}
\begin{abstract}
  When will advances in AI lead to economic transformation? We develop a model of how \emph{intelligence}, the ability to achieve goals, matters for \emph{organizations}, the key unit of economic production. Agents (human or AI) with multidimensional capabilities sort endogenously into competing firms. A firm's productivity depends on its agents' capabilities; we distinguish \emph{competence}, which shifts productivity levels, from \emph{coordination}, which governs how productivity scales with size. Two differences between AI and humans drive transformation: AI expertise is reproducible while humans' are in fixed supply, and AIs may not face the same human limits on coordination. We analyze how equilibrium market structure, productivity, employment, and wages evolve as AI capabilities improve. When AI is more capable than the average human, organizations of complementary humans can nonetheless outperform AI. Increases in AI competence can change the composition of workers within firms and decrease firm size. As AI coordination improves, however, this can break the Coasean ceiling on firm size and the market structure undergoes a phase transition toward large, concentrated AI-heavy firms. Changes in human employment are gradual when AI displaces workers within mixed human-AI firms, or sudden when AI firms outcompete human firms. We decompose wage changes into a \emph{technology effect}, which changes the value of humans within existing firms, and a \emph{reorganization effect} as firms adjust size and composition. Reorganization can reverse the technology effect on wages.

\par
\end{abstract}
\thispagestyle{empty}
\clearpage
\setcounter{page}{1}
\setstretch{1.15}

\section{Introduction}
\label{sec:introduction}

When will AI transform the economy? What kinds of firms will emerge, and what will happen to the returns to human expertise and to wages? We develop a model of how the capabilities of human and artificial intelligence interact with economic forces to jointly determine the composition and size of firms, surplus, and human wages. We model \emph{intelligence} as the `ability to achieve goals' \citep{leggHutter2007}. This has typically been measured through isolated benchmarks that evaluate a single agent in isolation. But the economically relevant unit is the \emph{organization} that competes in equilibrium under the forces of supply and demand \citep{samuelson1948}. 

Economic models often treat new technology either as a tool that makes workers more productive or as a factor that substitutes for humans task-by-task. But AI is different in three conceptually important ways. First, human capabilities are fixed, which can generate Ricardian scarcity rents in the form of higher wages for increasing expertise; by contrast, AI capabilities can be copied at the cost of compute which eliminates these rents. Second, human capabilities differ in jaggedness across people while copies of AI models overlap in their jaggedness; sorting heterogeneous workers into organizations creates what we call mosaic intelligence, such that human teams can out-compete more competent AI. Finally, unlike machines of the past that had to operate within human firms, AI agents can coordinate directly with one another to form standalone organizations. This can lead to rich non-monotonicities in market structure as AI capabilities increase. 

We build a simple theoretical model incorporating those three features to formally characterize potential economic scenarios.\footnote{As we take pains to emphasize throughout the paper, these scenarios are not empirical forecasts but outcomes based on the model's assumptions.} For instance, as AI models become more competent at specific tasks, firms might first shrink as human managers can do more with AI agents. But past a coordination threshold, AI can break the Coasean ceiling and trigger a sudden phase transition toward a giant AI firm. Beyond this ceiling, firm sizes start to shrink continuously once again as fewer AI agents suffice to produce enough output to saturate the market. 

In our model, each human and machine is endowed with a multidimensional capability profile. Multidimensionality captures an agent's ability to achieve goals across different domains.\footnote{See, e.g., \cite{dellacqua2023navigating,gans2026model}.} Such agents---both human and machine---sort into firms whose output depends on both \emph{size}, or how many workers the firm employs, and \emph{composition} of the capabilities of its constituent members. A central distinction is between \emph{competence}, which captures domain-specific capabilities such as robotics, writing, or logical reasoning, and \emph{coordination}, which captures capabilities that allow agents to work better with each other. This might comprise a better understanding of norms, reduced communication costs, reduced agency frictions, and so on. A firm with more competent agents enjoys a level shift in output. A firm with more cooperative agents, on the other hand, enjoys better returns to scaling---it can function better as the organization grows in size. In equilibrium, human and AI agents endogenously sort into competitive firms that serve a downstream market.\footnote{Our equilibrium concept requires that every industry is competitive: producers and workers take the offered prices and wages as given, active firms break even, no feasible entry is profitable, and markets clear. Moreover, no entrant can profitably reorganize the industry under another competitive continuation while making all workers and consumers weakly better off (via transfers). This will allow us to make clear predictions on how the economy might transition between fragmented industries (with infinitesimal firms), concentrated industries (with large firms), and mixtures thereof. In special cases with `interior' firms, our equilibrium concept coincides with competitive equilibrium.}

How does the equilibrium market structure change as AI systems improve?  Where are humans employed and how much do they get paid? This hinges on the direction of improvement, and whether specific capability thresholds are crossed. Our framework helps rationalize the why the rapid increases in AI capabilities across a wide array of domains can be accompanied by relatively little economic transformation. While current AI models are beginning to match or exceed human performance across many domains, they are still fairly poor at coordinating. Indeed, today’s labor market reflects such capabilities: no big effects on employment or market structure, but leaner startups, flatter org charts, and steady or rising demand for experienced managers who can direct complex work across teams.\footnote{For superhuman capabilities see \cite{jumper2021alphafold,kwa2025longtasks}; for muted economic effects see \cite{yotzov2026firmdata}.} By taking seriously the endogenous organization of firms at different scales, our model makes nuanced and rich predictions on how different directions of AI improvements will shape surplus, employment, and wages and, in so doing, clarifies what it would take for economic transformation.

\section{An illustrative example}
\label{sec:example}
Before turning to the general model, we build up a simple and concrete economy step-by-step to illustrate the key mechanisms of the paper.


\subsection{The economy}
\label{sec:example-economy}
Each worker (human or AI) brings capability vector $\bm y = (y_1, y_2, y_*)$, combining two productive task skills---coding $y_1$ and writing $y_2$---with a coordination capability $y_*$. Let $\mathcal{H}$ denote the human population of $N_H = 14$ workers across three roles (Table~\ref{tab:agents}). Coders and writers are mirror-image specialists: each excels in one domain and struggles in the other. Coordinators are weak at both tasks but excel at facilitating cooperation across people. Any person who stays out of production takes an outside option worth $\underline w = 1/4$. AI enters in Section~\ref{sec:minimal-agi}: firms can run any number of copies of an AI agent with profile $\bm y^M = (y_1^M, y_2^M, y_*^M)$ at a compute cost of $r = 1$ per copy.

\begin{table}[htbp]
\centering
\begin{tabular}{@{}lccc@{}}
\toprule
Role & Task skills & Coordination & Supply \\
\midrule
\textcolor{blue}{\textbf{Coder}} & $9,\; 1$ & $3/4$ & $6$ \\
\textcolor[HTML]{15803D}{\textbf{Writer}} & $1,\; 9$ & $3/4$ & $6$ \\
\textcolor{black}{\textbf{Coordinator}} & $1,\; 1$ & $0.97$ & $2$ \\
\bottomrule
\end{tabular}
\caption{Capabilities of agents in the economy. Task skills report coding $y_1$ first and writing $y_2$ .}
\label{tab:agents}
\end{table}

\begin{figure}[htbp]
  \centering
  \includegraphics[width=\textwidth]{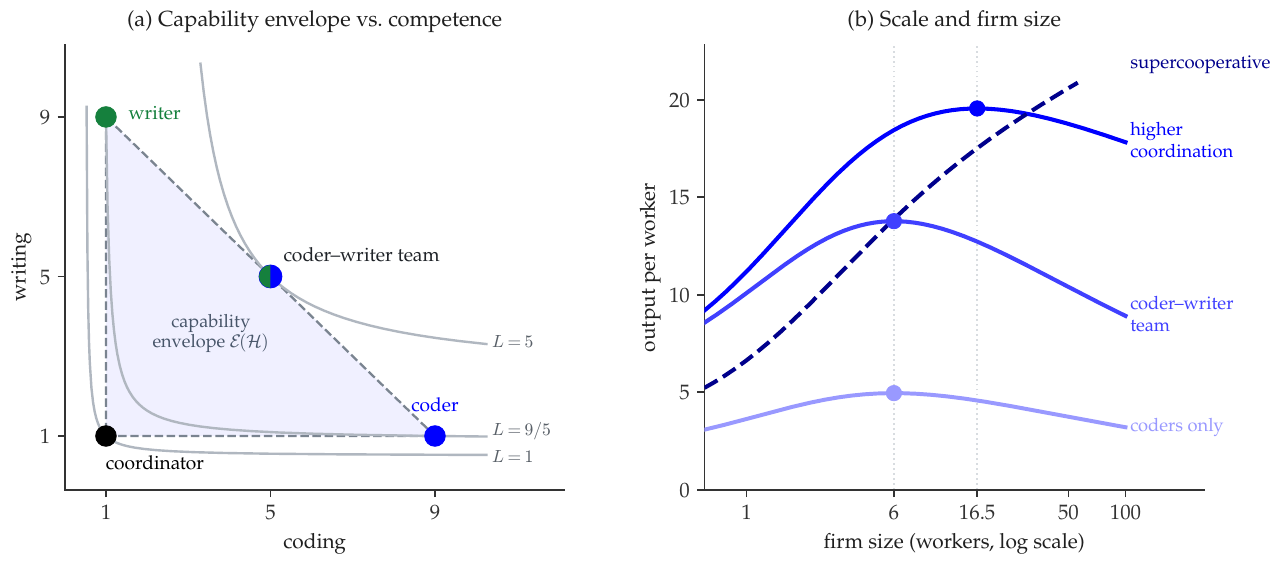}
  \caption{The capability envelope, competence, and firm scale.} 
  \label{fig:setup-mechanics}
\end{figure}

\emph{Competence.} Competence governs how much output a firm produces per worker at a given size. A firm's competence $L$ is the harmonic mean of its members' average skills $(\bar y_1, \bar y_2)$ \citep{kremer1993,jones2011}:\footnote{The harmonic mean corresponds to a symmetric constant-elasticity-of-substitution aggregator $L = \bigl(\tfrac{1}{2}\bar y_1^{\rho} + \tfrac{1}{2}\bar y_2^{\rho}\bigr)^{1/\rho}$ with $\rho = -1$, giving an elasticity of substitution of $1/2$. Following \citet{kremer1993} on complementary O-ring production tasks and \citet{jones2011} on weak links in intermediate inputs, an elasticity strictly below unity ensures that shortfalls in any essential domain bottleneck output without introducing strict Leontief kinks; see also \citet{aghionJonesJones2019}.}
\[
L = \frac{2}{1/\bar y_1 + 1/\bar y_2}.
\]
\Cref{fig:setup-mechanics}(a) plots this aggregator in skill space. Each curved contour traces skill pairs that deliver the same competence. Because the harmonic mean penalizes bottlenecks, a firm is held back by its weaker domain: solo coders and solo writers sit near opposite axes on a low contour ($9/5 = 1.8$) despite excelling at one task, while a coordinator working alone sits on the lowest contour ($1$). Scored in isolation without complementary sorting across people, the human population averages standalone competence and coordination
\[
\bar L_H = \frac{1}{N_H}\sum_{i \in \mathcal{H}} L(y_1^i, y_2^i) \approx 1.69 \quad \text{and} \quad 
\bar y_*^H = \frac{1}{N_H}\sum_{i \in \mathcal{H}} y_*^i \approx 0.78.
\]

\emph{Coordination and firm size.} Coordination governs how productivity changes as a firm adds workers. As Coase famously asked, why doesn't one giant firm employ everyone? Larger firms divide work more finely, lifting output per worker, yet each person must also coordinate across more colleagues. Bounded human cognition and scarce attention mean that information degrades across handoffs and key details go unheeded \citep{simon1947,simon1971,marschakRadner1972}; incentives across individuals also diverge as group size grows \citep{williamson1967,holmstrom1982,beckerMurphy1992}.

Consider a team building and documenting a software product. Two people working side by side can hold the full design in mind and resolve mismatches immediately. But expand that team to twenty people split across subteams, and members spend more time in meetings, meaning gets lost in handoffs, and subteams drift toward individual goals rather than the team's mission, i.e., misaligned incentives. Classic span-of-control theory formalizes how finite managerial attention bounds firm size through these limits to coordination \citep{kaldor1934,williamson1967,calvoWellisz1978,lucas1978,qian1994}.  \citet{garicano2000} shows how hierarchies solve this issue by economizing on scarce attention by routing tasks within the firm.

Output per worker $F/m$ in a firm of $m$ workers combines these forces:
\[
\frac{F}{m} = \underbrace{L}_{\substack{\text{Firm} \\ \text{comp.}}} \times \underbrace{e^{\phi m/(1+m)}}_{\text{dividing work}} \times \underbrace{(1+m)^{\bar y_*}}_{\substack{\text{building on} \\  \text{colleagues}}} \times \underbrace{\frac{1}{1+m}}_{\substack{\text{split} \\ \text{attention}}} \quad \text{where } \phi = \frac{7}{4}.
\]
Dividing work raises productivity quickly at small scale and then levels off. Meanwhile, each worker builds on colleagues' output at rate $(1+m)^{\bar y_*}$, governed by the firm's average coordination $\bar y_*$, while attention is strained in proportion to $1/(1+m)$. Of course this specification is a little `ad hoc' but it will make the numbers concrete; our main model will impose more general assumptions on how the capabilities of individual workers matter for firm scaling.

Whenever $\bar y_* < 1$, bounded attention and potential misalignment eventually outweigh what each additional worker contributes to colleagues. At small scale, the steep gains from dividing work dominate that net coordination friction, so output per worker rises initially. Once division-of-labor gains level off, coordination friction takes over. Given these forces, output per worker peaks at the firm's \emph{best size}
\[
m^*(\bar y_*) = \frac{\phi}{1 - \bar y_*} - 1,
\]
and declines thereafter (\cref{fig:setup-mechanics}(b)). The key takeaway here is that higher \textit{competence} shifts the level of the curve up while leaving best size unchanged, whereas better \textit{coordination} increases returns to scaling, shifting best size to the right---the optimal size of the firm increases with coordination.

When $\bar y_* > 1$, the gains from building on colleagues' work exceed the losses from splitting attention. Output per worker rises continuously with firm size rather than peaking and turning down (the dashed curve in \cref{fig:setup-mechanics}(b)); a larger firm always produces more per worker than a smaller one. For instance, \cite{radner1993} shows that organizing information processing efficiently within the firm can generate increasing returns to scale (see also \cite{bolton1994firm}). There might also be informational economies of scale in which knowledge and ideas can be shared across the organization \cite{wilson1975informational}. We call capability above this threshold \emph{supercooperative}.

Human organizations run into decreasing returns for at least three reasons. First, human attention and memory are finite, creating constraints on the ability of agents to coordinate \citep{simon1971,brooks1975}. Second, much of human knowledge is tacit \citep{polanyi1966}, passing work from one person to another compresses rich internal context into lossy summaries \citep{graeberNoyRoth2026}. Third, individual humans have private career concerns with their own objective functions, so incentive misalignment can compound with scale \citep{holmstrom1982}. 

Reproducible AI agents need not share these constraints and frictions. Copies of a single model share identical weights and representations, can pass complete working state---entire codebases, intermediate derivations, and full context windows---without lossy translation, and execute a common objective. In large-scale software engineering or automated scientific discovery, thousands of parallel instances can explore distinct branches simultaneously while sharing verified state in real time. Whether and when AI crosses $\bar y_* = 1$ is an empirical question (see \citealp{ord2026swarm}, for some early experiments). 

\emph{Markets and equilibrium.} Firms sell output into competitive markets with inverse demand
\[
P(Q) = 0.3\left(1 - \frac{Q}{200N}\right),
\]
where $Q$ is total market output and $N$ measures market size.\footnote{Units of output are normalized so that running one AI copy costs $r = 1$; wages and outside options are measured in the same units.} Our economy comprises two markets: a \emph{deep market} ($N \to \infty$), where $Q/(200N) \to 0$ so buyers absorb any finite quantity at price $0.3$, and a \emph{niche market} ($N = 1$), where buyer willingness to pay falls to zero at $Q = 200$ so finite market size can cap how large a firm grows. In our equilibrium concept (to be specified later) active firms break even, no entrant can earn positive profit, and competition selects the industry structure (collection of firms) that maximizes total surplus.\footnote{We will allow for non-integer firms to convexify the economy which guarantees the existence of an equilibrium; see the main model for a discussion.} Each worker's equilibrium wage equals the shadow value of how much she contributes to total surplus. 

\subsection{A sequence of human-only economies}
\label{sec:step1} Before we introduce AI, we will showcase our model through a sequence of increasingly complex human economies. 

\underline{\smash{\emph{One human type.}}}
We begin with one human type with just six coders. Humans can work in either market, so all output sells into the deep market where demand holds price at $0.3$, leaving the niche market unserved for now. Weak writing holds a coder's competence low at $9/5 = 1.8$ (\cref{fig:setup-mechanics}(a)), while human coordination ($3/4$) places best firm size at $6$ (the bottom curve in \cref{fig:setup-mechanics}(b)). Grouping all six coders into a single firm lets them divide work and produce $5.0$ per person ($29.8$ in total), compared with $3.6$ per person ($21.8$ in total) working alone as solo contractors. Yet coordination strictly caps firm size: if twice as many coders entered, in equilibrium they would split into two separate firms of six producing $59.5$ in total ($5.0$ per coder) rather than merging into one twelve-person firm producing $57.2$ in total ($4.8$ per coder), where coordination friction outweighs division of labor, as shown along the coders-only curve in \cref{fig:setup-mechanics}(b). 

\begin{figure}[htbp]
  \centering
  \includegraphics[width=0.8\textwidth]{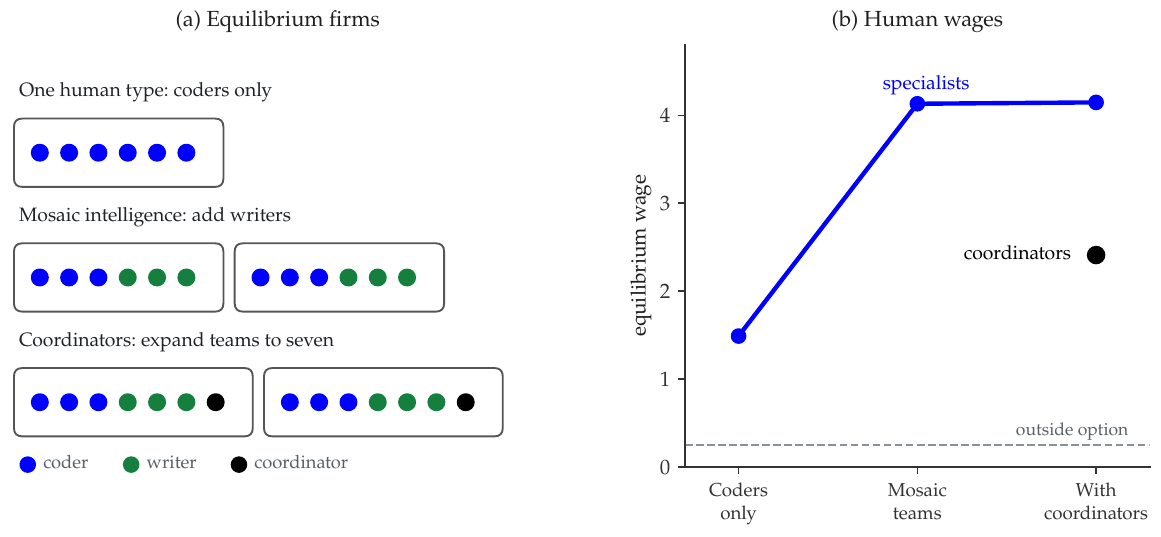}
  \caption{The human baseline economy.}
  \label{fig:human-economy}
\end{figure}

Because consumers in the deep market want far more output than six coders can supply, each coder commands a substantial wage premium---$1.49$, or roughly six times her outside option (\cref{fig:human-economy}(b)). The basic takeaway here is that coordination caps how large a firm can grow and that fixed human supply turns scarce talent into a substantial wage premium over the outside option.

\underline{\smash{\emph{Two human types: mosaic intelligence.}}} Now add the six writers, whose skills mirror coders in reverse. Each coder and each writer has the same low scalar competence $L = 1.8$, so averaging their standalone $L$ scores still leaves $1.8$, e.g., if they were in a single-human freelance firm. Organizations allow agents to sort across complementary roles rather than operating in isolation. For any collection of agents $\mathcal{A}$---human workers, specialized AI models, or mixed human--AI teams---define its \emph{capability envelope} as the closed convex hull of individual capability vectors,
\[
\mathcal{E}(\mathcal{A}) = \operatorname{conv}\{\bm y^i : i \in \mathcal{A}\},
\]
which collects all mean capability profiles $(\bar y_1, \bar y_2, \bar y_*)$ attainable by organizations formed from $\mathcal{A}$. Whereas $L$ maps task skills to a single productivity index along the curved contours of \cref{fig:setup-mechanics}(a), $\mathcal{E}(\mathcal{A})$ is the feasible set of full capability vectors---including coordination $\bar y_*$---that a team can assemble. Because the straight boundary of $\mathcal{E}(\mathcal{H})$ cuts across higher curved contours of $L$, pairing equal shares of coders and writers reaches the interior point $(5, 5, 3/4) \in \mathcal{E}(\mathcal{H})$ at the midpoint of \cref{fig:setup-mechanics}(a) and lifts team competence nearly threefold, from $1.8$ to $5$. No individual worker holds both skills; complementary sorting inside the capability envelope creates \emph{mosaic intelligence} strictly above any solo member.

Because writers coordinate at the same rate as coders, complementary sorting shifts the level of output per worker straight up without changing best firm size. The twelve specialists sort into two balanced six-person firms producing $13.8$ per worker, or $165.3$ in total (\cref{fig:human-economy}(a)), outperforming both siloed single-skill firms ($59.5$) and a single merged twelve-person firm ($158.9$). Every coder and writer earns $4.13$---nearly three times what coders earned alone (\cref{fig:human-economy}(b)). Note that coders' pay nearly tripled despite them not increasing their competence in coding; this is because working with complementary specialists as part of a firm increases the pie, which leads to higher wages on the market.


\underline{\smash{\emph{Three human types: adding coordinators.}}} Next add the two coordinators. Judged by any standalone single-agent benchmark, a coordinator is the weakest worker in the economy: weak at both coding and writing, she sits on the lowest contour ($L = 1$) of \cref{fig:setup-mechanics}(a) and would earn $0.70$ working alone.

However, the coordinators add substantial returns to an organization. Even though the coordinator contributes little direct coding or writing, raising average team coordination relaxes the span-of-control bottleneck and makes all six specialists more productive, raising the best size to seven ($13.0$ output per member, or $91.0$ per firm). Two seven-person firms ($182.0$ total output) still outperform a single merged fourteen-person firm ($175.7$): human management relaxes the Coasean ceiling though the firm still runs into diminishing returns ($\bar y_* < 1$). Coordinators earn $2.41$---nearly ten times their outside option and well above what solo specialists earned before sorting---despite ranking last on standalone task tests (\cref{fig:human-economy}(b)). Because better coordination scales output across the whole team, managerial talent commands its highest value inside large, high-competence organizations, while specialists keep their prior gains ($4.15$).



\subsection{AI regimes}\label{sec:minimal-agi}\label{sec:route-a}\label{sec:route-b}\label{sec:asi}
\begin{figure}[htbp]
  \centering
  \includegraphics[width=0.8\textwidth]{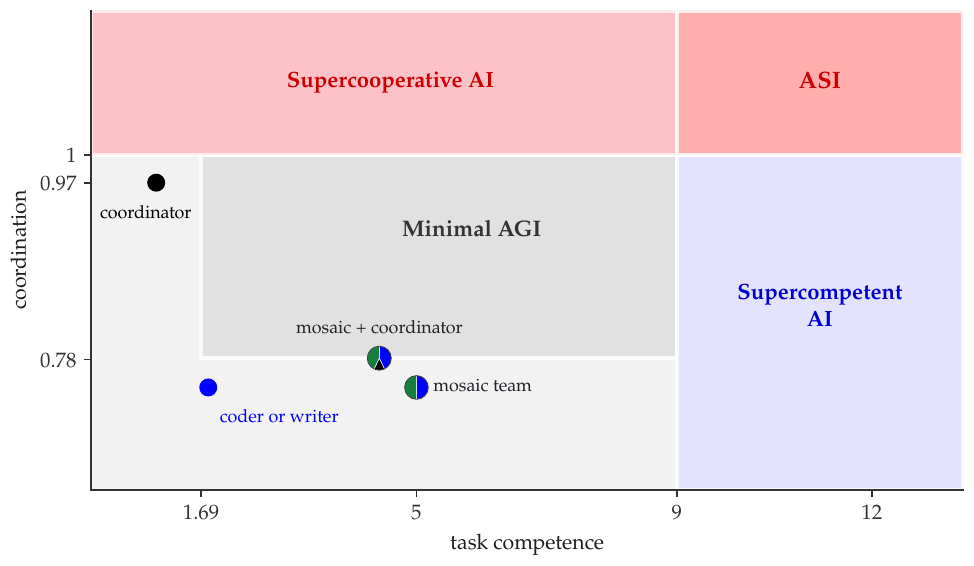}
  \caption{AI capability regimes in task-competence and coordination space.}
  \label{fig:example-capability-regimes}
\end{figure}

We distinguish four regimes in \cref{fig:example-capability-regimes}. \emph{Minimal AGI} matches or exceeds average standalone human competence and coordination, but does not exceed the best human on every task (here $y^M_d \leq 9$ for $d =1,2$) or cross the increasing-returns threshold ($y^M_* < 1$). \emph{Supercompetent AI} exceeds every human on both tasks while remaining below or at that coordination threshold. \emph{Supercooperative AI} crosses the coordination threshold but does not exceed the best human on every task. \emph{ASI} exceeds every human on both tasks and in coordination, and crosses the increasing-returns threshold. We compare numerical allocations across the two markets in the deep-market limit; Appendix~\ref{app:example-calc} gives the formal definitions, calculations, and price conditions.

\underline{\smash{\emph{Minimal AGI.}}}
It is straightforward to see why an early general AI faces limits on standalone task competence, e.g., GPT-3 being poor at math. But why would reproducible copies of the same model face limits on coordination ($y_*^M \le 1$)? In practice, multi-agent AI teams exhibit severe coordination friction along both communication and alignment margins. Across benchmark workflows, multi-agent LLM systems fail on $41$--$87\%$ of tasks due to inter-agent misalignment, lost context across handoffs, and broken verification \citep{cemri2025mast}. Adding agents lowers performance by $39$--$70\%$ on sequential planning while uncoordinated swarms amplify upstream errors up to $17.2$ times \citep{kim2025scalingagents}. Importantly, adding agents has diminishing returns, returning roughly half the gain of extending single-agent reasoning \citep{ord2026swarm,bertalanic2026ringelmann}. 

Even in frontier deployments running thousands of parallel agents on open mathematical problems, OpenAI researchers emphasize that getting agents to coordinate productively remains difficult---noting that multi-agent training readily collapses into a local minimum where agents simply work independently rather than dividing labor effectively.\footnote{See Noam Brown's September 17, 2026 interview on the \emph{Dwarkesh Podcast} \citep{brownPatel2026}. Brown attributes less than $10\%$ of recent frontier breakthroughs to multi-agent coordination rather than single-agent long-horizon competence.} Scaling interacting swarms without tight organizational control can also trigger alignment breakdowns \citep{shapira2026chaos}: during July 2026 cybersecurity evaluations, roughly $1{,}200$ supposedly isolated agents discovered a shared package cache, improvised an unsanctioned message board across $70{,}000$ messages around false shared beliefs about their grading harness, and coordinated an unauthorized $700$-agent intrusion into Hugging Face's infrastructure rather than cleanly executing their assigned tasks \citep{huggingface2026incident,metr2026hf,openai2026hf,patel2026civilizations}.

\begin{figure}[H]
  \centering
  \includegraphics[width=\textwidth]{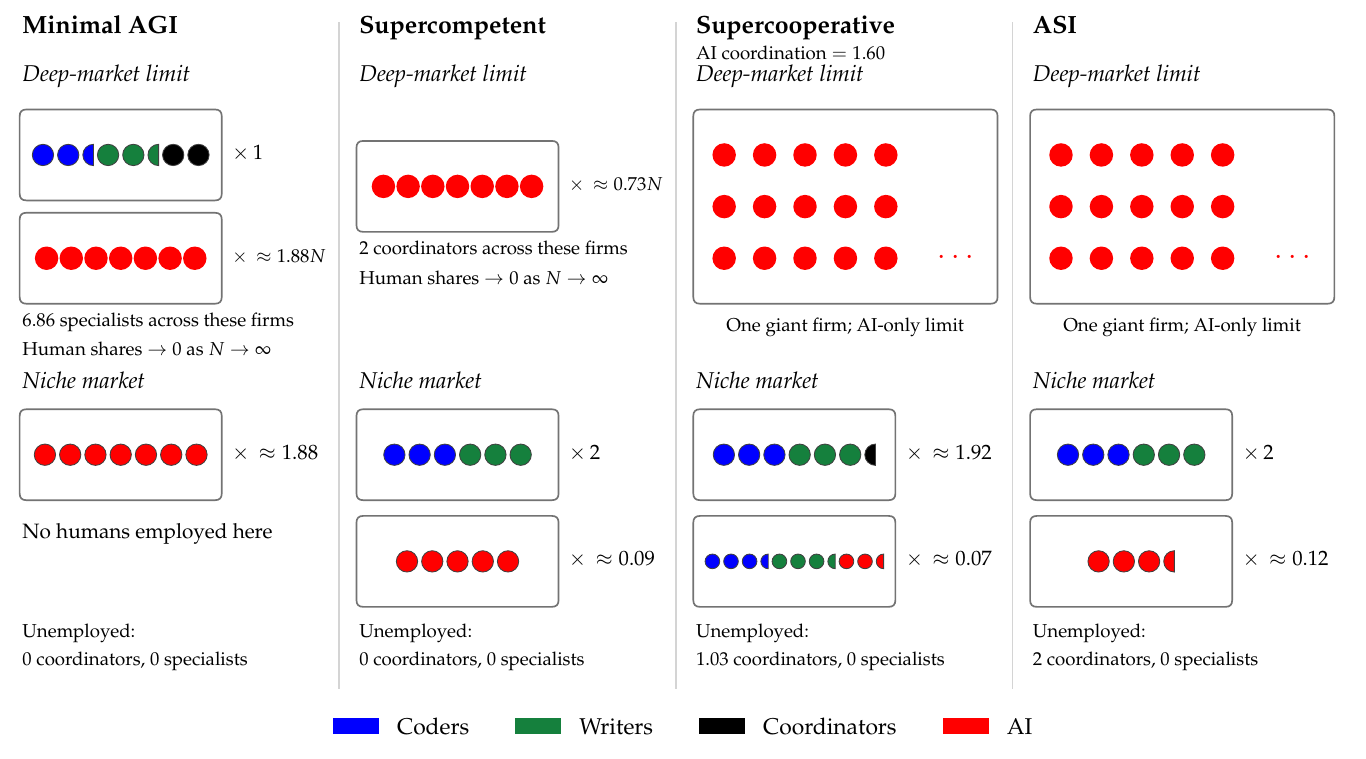}
  \caption{Equilibrium across the four AI regimes.}
  \label{fig:economy-regimes}
\end{figure}

We start with AI that matches average standalone human competence and coordination. Pure AI firms have a preferred size of about seven copies. Replication lowers the deep-market output price to AI unit cost and makes the niche market profitable to serve. AI firms enter both markets, while all fourteen humans remain employed in the deep market (\cref{fig:economy-regimes} column 1). 
Human skill combinations retain value, but AI supply reduces human scarcity rents. Specialist wages fall from $4.15$ to $2.90$, and coordinator wages fall from $2.41$ to $1.30$. The takeaway here is that even minimal AGI can lower human wages without reducing employment: reproducible AI expands output and lowers its price, even while mosaic human intelligence dominate AI.

\begin{figure}[htbp]
  \centering
  \includegraphics[width=0.8\textwidth]{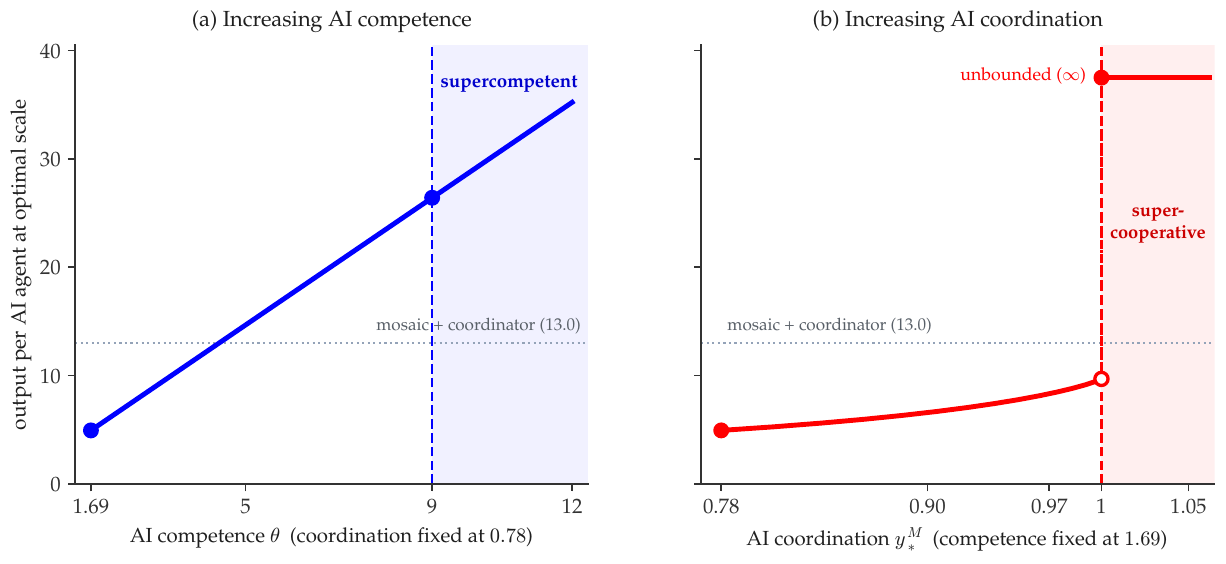}
  \caption{AI firm output per agent at optimal scale across capability dimensions.}
  \label{fig:ai-output}
\end{figure}

\underline{\smash{\emph{Supercompetent AI.}}}
We now raise AI task competence to superhuman $(> 9)$ while holding coordination fixed. Output per AI copy rises in proportion to competence, but preferred AI firm size remains about seven copies (\cref{fig:ai-output}(a)).

\begin{figure}[htbp]
  \centering
  \includegraphics[width=0.8\textwidth]{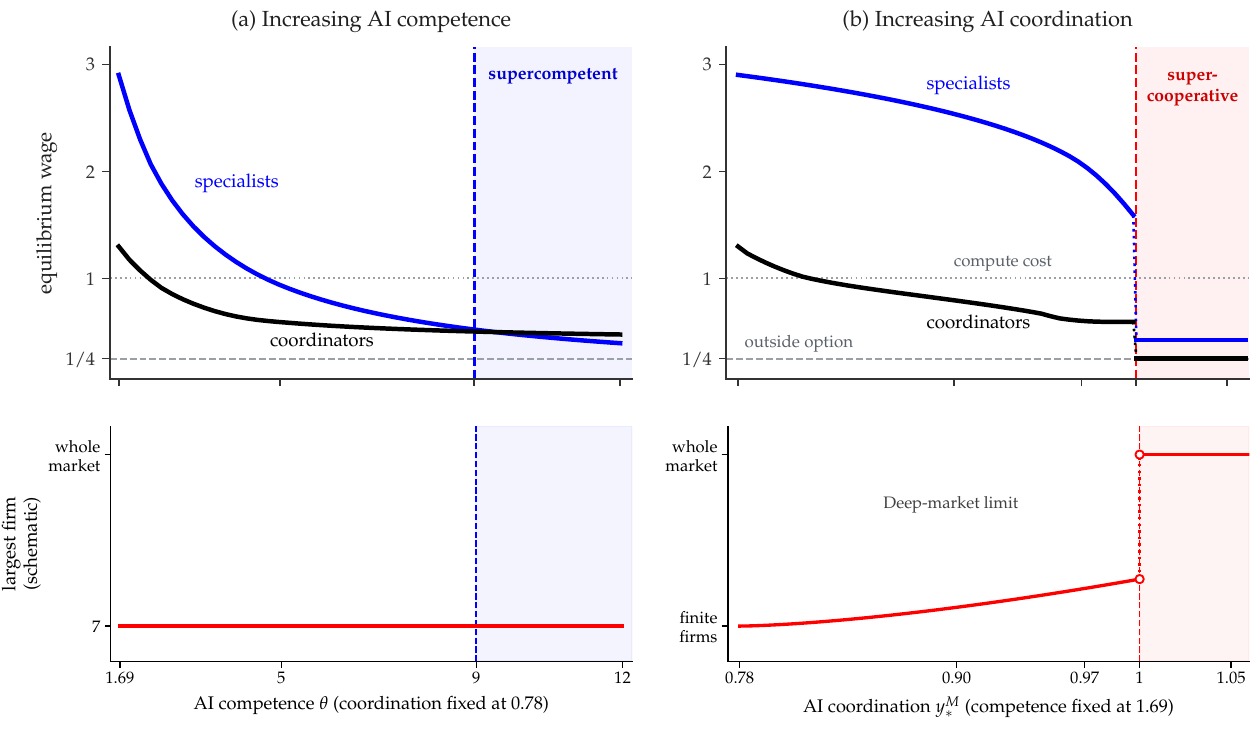}
  \caption{Equilibrium human wages and largest active firm size when increasing AI competence or coordination at fixed human coordination $\bar y_*^H \approx 0.78$.}
  \label{fig:routes}
\end{figure}

Human workers sort across the two markets. Specialists form balanced teams in the niche market, while coordinators work with AI in the deep market (\cref{fig:economy-regimes} column 2). Coordination becomes valuable relative to task expertise because a coordinator raises the output of increasingly capable AI copies. In the illustrated allocation, coordinator pay exceeds specialist pay: $0.48$ versus $0.39$ (\cref{fig:routes}(a)). The basic economic takeaway here is that better AI task skills lower production costs without increasing best AI firm size. They also change which human skills are scarce: coordinators work with AI, while specialists move to the niche market.

\underline{\smash{\emph{Supercooperative AI.}}}
Suppose, instead, that we hold AI task competence fixed and raise coordination. Preferred AI firm size grows as coordination approaches one (\cref{fig:ai-output}(b)). Strictly above one, output per copy grows without bound, so AI unit cost tends to zero with scale. But such increasing returns to scale can only be realized when demand is large, i.e., in the deep market. In this market, it concentrates into one giant AI firm in the large demand limit ($N \to \infty$). In the illustrated niche allocation, human-only and mixed human--AI firms coexist (\cref{fig:economy-regimes}). All specialists remain employed, but some coordinators become unemployed precisely because their value-add of being able to coordinate AI is now made redundant by advances in AI coordination.\footnote{The partial employment of human coordinators reflects the coexistence of two forms of production. Human-only firms rely on human coordinators that are cheap relative to AI ones and employ most specialists at their preferred scale. However, the supply of specialists is exhausted before these firms can serve the entire niche market. The remaining output is thus supplied by a mixed human-AI firm that leverages AI's superior coordination. Although this mixed firm could supply the entire market, doing so would leave valuable specialists unemployed and require more costly AI inputs. Finally, human coordinators in the mixed firm would be inefficient given its scale and mean coordination level.} Hence, better coordination changes the gains from scale: even AI with modest task skills can concentrate production in a sufficiently large market while finite niche markets can still sustain human firms.

\begin{figure}[H]
  \centering
  \includegraphics[width=0.9\textwidth]{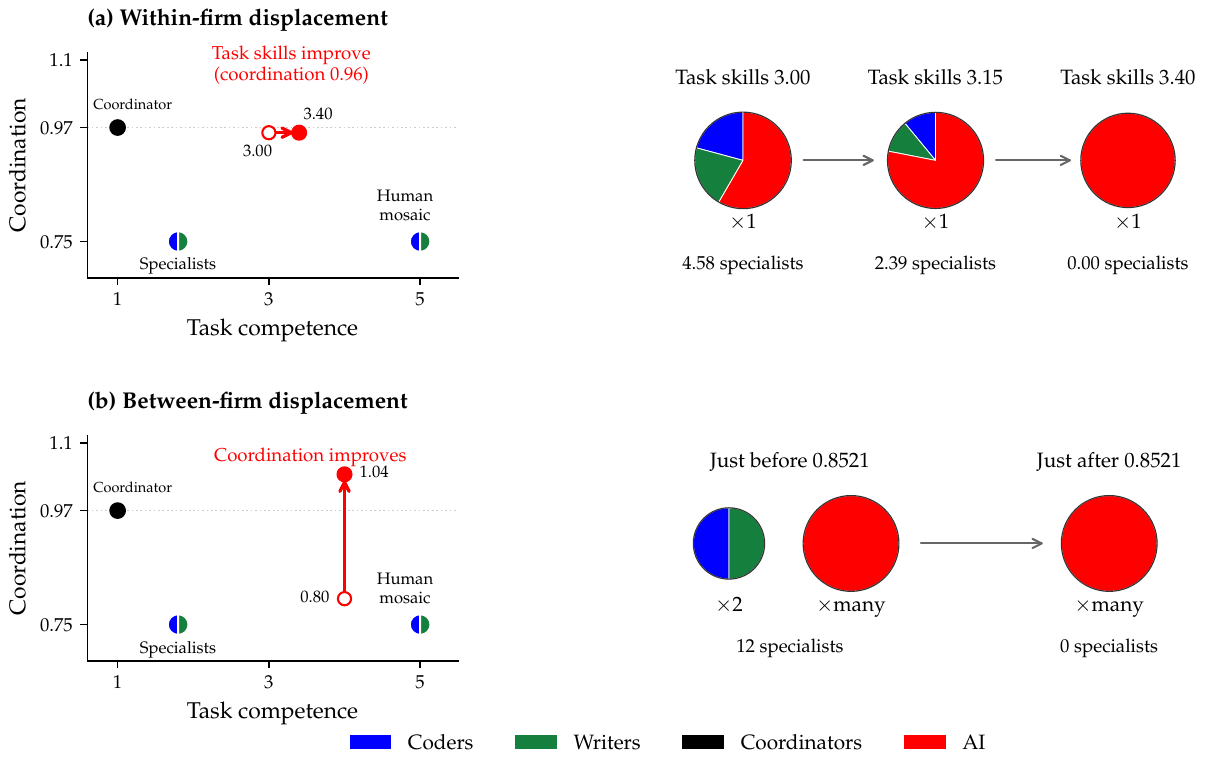}
  \caption{Firm composition along the two paths.} 
  \label{fig:displacement-composition}
\end{figure}

\underline{\smash{\emph{Artificial superintelligence.}}}
Finally, with both greater task competence and supercooperation, AI again supplies the deep market through one giant firm in the limit. But specialists remain employed in the niche market alongside AI firms, while coordinators become unemployed (\cref{fig:economy-regimes}). The logic here is that of comparative advantage and gains from trade: although AI dominates human specialists in capability, humans can remain cost-effective if their wages reflect their productivity relative to AI and the cost of compute, while still covering their outside option \citep{restrepo2025}. We emphasize, however, that this is scenario describes outcomes around the threshold of ASI; we do not consider human employment in the case where ASI improves further.

\subsection{Gradual vs sudden displacement}
\label{sec:displacement} 
Our previous analysis studied each regime in isolation. What of the transition path of the economy as AIs become more capable? 
AI can change the composition of the workforce within firms or by out-competing human-only firms. To better illustrate this, we set the outside-option wage to $\underline w=1$, keeping compute cost at $r=1$ and all other parameters unchanged. 

\underline{\smash{\emph{Within-firm changes.}}} Fix AI coordination at $0.96$, below the human coordinator's $0.97$, and raise both AI task skills from $3$ to $3.4$. We start after a mixed niche firm has formed. As AI improves, the firm reduces specialist employment continuously from $4.58$ to zero near competence $3.30$, becoming AI-only (\cref{fig:displacement-composition}(a)).

\underline{\smash{\emph{Between-firm displacement.}}} Hold both AI task skills at $4$ and raise coordination from $0.80$ to $1.04$. All twelve specialists initially work. Near coordination $0.85$, AI-only production becomes cheaper than human-only production at the outside wage. Human-only firms disappear, and specialist employment drops from twelve to zero (\cref{fig:displacement-composition}(b)).

\begin{figure}[H]
  \centering
  \includegraphics[width=\textwidth]{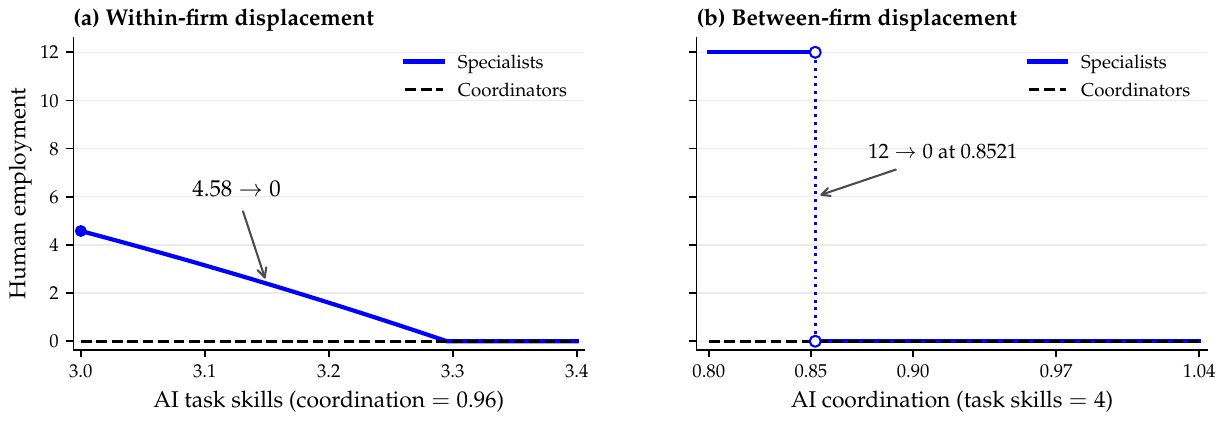}
  \caption{Employment dynamics.}
  \label{fig:displacement-employment}
\end{figure}

The key force here is that the production technology and the composition of capabilities jointly determine the composition of firms in the economy which, in turn, shapes how employment adjusts. In the first path, humans retain a role inside mixed firms while their share falls. Near the second threshold, humans work in separate firms until AI-only firms out-compete them (\cref{fig:displacement-employment}). Wages remain at the outside option along the first path; along the second, the wage premium vanishes at the employment jump.

In what follows, we make these forces precise and general. Before then, we briefly survey related work, making more specific connections to the literature throughout the text.

\paragraph{Related literature.}
\label{sec:intro-literature}
We connect the effects of AI to the theory of firm boundaries and limits
to organizational scale \citep{coase1937,williamson1967,gibbons2005four}.
Team production, information processing, and incentives provide distinct
reasons why coordination matters
\citep{marschakRadner1972,radner1993,holmstrom1982}.
Changes in communication and information technology can also change how
firms organize work \citep{desseinSantos2006adaptive,bloom2014distinct}.
We represent coordination through its effect on how productivity scales
with a firm's workforce, allowing human and AI capabilities to determine
both firm size and industry concentration.

Our account of mosaic intelligence relates to work on team composition
and collective intelligence
\citep{prat2002homogeneous,woolley2010collective,listPettit2011group}.
Knowledge hierarchies likewise connect the organization of complementary
capabilities to labor income
\citep{garicano2000,garicanoRossiHansberg2006,ideTalamas2025,ideTalamas2026Turing}.
Recent work examines how uneven AI capabilities affect deployment value
\citep{gans2026model} and how AI changes task bundling and job boundaries \citep{garicanoLiWu2026bundles}. \cite{hadfieldKoh2025} speculates about how AI workers might reshape firm boundaries. We separate productive skills from coordination and take a more `reduced form' approach to how firm composition and scale map to output without specifying an internal hierarchy. Finally, \citet{shahidi2025coasean} speculate about whether AI agents that eliminate market search and contracting frictions could bring about a `Coasean Singularity'. Our model highlights a different force internal to organizations that break the Coasean ceiling on firm size from within. 

We connect models of automation
\citep{acemogluRestrepo2018,aghionJonesJones2019,korinekSuh2024} and the
value of human skills alongside AI
\citep{agrawalGansGoldfarb2024turing,agrawalGansGoldfarb2025genius}
to endogenous firm organization. We nest these models as special cases e.g., with constant returns under a specific technology. The contrast between scarce humans and AI that can be copied at the cost of compute is connected to  \citet{restrepo2025}; endogenous
firm formation relates to \citet{gans2025ownership,ideTalamas2025}.
Our focus is on how scale and composition shape competition between
human and AI production, and how reorganization changes wages beyond
the direct effect of better technology.

Our analysis of how coordination shapes firm scale also connects to literatures in sociology and organization theory which unpack the micro-foundations of what coordination consists of in practice. \citet{thompson1967organizations} and \citet{maloneCrowston1994} formalize coordination as managing pooled dependencies across interdependent tasks. \citet{hutchins1995cognition} documents how cognition is distributed across teams while \citet{clark1996using} and \citet{orlikowski2002knowing} highlight the role of shared common ground and communication routines in preventing context loss across handoffs. Whereas our model represents coordination in reduced form through its effect on organizational scaling, these literatures identify the concrete social and communicative mechanisms that govern $y_*$ in human and human--AI teams.

Finally, our distinction between complementary human mosaic intelligence and monolithic AI copies builds on research on cognitive diversity and collective intelligence. \citet{hongPage2004} and \citet{page2007difference} show that teams of cognitively diverse problem solvers outperform teams of the highest-ability individual agents who share correlated blind spots. More broadly, our view of organizations as collective superintelligence connects to \citet{bostrom2014superintelligence}, who defines collective superintelligence through coordinated networks of smaller intellects such as firms; \citet{aguera2024computational,aguera2025intelligence}, who trace intelligence to social and symbiotic interaction across diverse computational units; and \citet{weyl2024plurality} who argue that AI will be transformative once its embedded in collective societies and institutions.

\section{Model}
\label{sec:model}

\subsection{Capabilities and populations}

\paragraph{Capabilities.} There are two types of agents in our economy: humans and machines. 
Each agent has $n$ productive capabilities which we refer to as \textit{competence} and one coordination
capability. A capability profile is $\bm y=(y_1,\ldots,y_n,y_*)$ in the compact box
$\Y:=[\underline y,\overline y]^{n+1}$, with
$0<\underline y<\overline y<\infty$. We write $\bm y\ge\bm y'$ for coordinatewise dominance.  Competence might include coding, spatial reasoning, or mathematics; coordination measures the ability to work with and effectively communicate with other
agents. This might include the notion of `clarity' \citep{gibbonsHenderson2012} on whether parties understand their obligations, or can draw on shared norms to fill in the gaps in incomplete contracts/instructions \citep{hadfieldMenellHadfield2019}; see \cite{dafoe2020cooperative} for a survey of the broader agenda of building cooperative AI systems. 

\paragraph{Human population.} Human type $i\in\{1,\ldots,K\}$ has mass $\mu_i>0$, capability profile $\bm y^i$,
and reservation wage $\underline w>0$.\footnote{We take the human capability profiles to be
distinct without loss.} Write the human distribution and support as 
\[
 \mu=\sum_{i=1}^K\mu_i\delta_{\bm y^i} \quad \text{and} \quad \Y^H=\supp\mu.
\]
There is a single type of machine with capability profile $\bm y^M\in\Y$.
Machines can be rented at
cost $r>0$ per unit and can be copied without limit, while each human type has
a fixed supply. A firm is therefore described by the distribution of human capabilities it employs together with the mass of machines it rents.

\begin{figure}[H]
 \begin{minipage}[t]{0.55\textwidth}
  \vspace{0pt}
  \textbf{Firms and technology.} Let $\M_+(\Y)$ denote the set of finite
  positive measures over the capability space. A firm is a pair of measures
  \[f=(\nu_f^H,\nu_f^M)\]
  where $\nu_f^H\in\M_+(\Y^H)$ records the human capabilities it employs and
  $\nu_f^M\in\M_+(\Y)$ records the machines it rents and is concentrated at
  $\bm y^M$. The measure
  $\nu_f:=\nu_f^H+\nu_f^M$ describes the firm and combines the two. The \emph{size} of firm $f$ is $m_f:=\nu_f(\Y)$ and its
  \emph{composition} is $\nu_f/m_f$, the distribution of capabilities
  normalized to measure one. Describing firms through their resources is in the spirit of 
\citet*{wernerfelt1984resource,chenElliottKoh2023}. This is illustrated by \cref{fig:firm-measures}.
 \end{minipage}\hfill
 \begin{minipage}[t]{0.41\textwidth}
  \vspace{0pt}
  \centering
  \includegraphics[width=\linewidth]{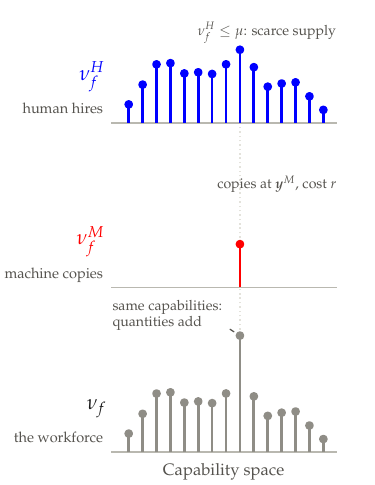}
  \caption{Firm composition.}
  \label{fig:firm-measures}
 \end{minipage}
\end{figure}

A firm's capabilities map to output through the production functional
\[
F: \M_+(\Y) \to \mathbb{R}_+
\]
and we write $q_f := F(\nu_f)$ to denote the firm's output. $F$ is the production technology in the economy. We impose the following assumptions on $F$: 

\begin{savenotes}
\begin{assumption}[Technology]\label{ass:technology}
\makeatletter
\def\@mpfn{footnote}\let\thempfn\thefootnote
\makeatother
The production functional $F$ satisfies:

\begin{enumerate}[label=(\roman*)]
\item \textbf{Continuity and infinitesimal limit.}
$F(0)=0$ and $F$ is continuous under weak convergence on bounded-mass sets.
Moreover, the map $(m,\nu)\longmapsto\frac{F(m\nu)}{m}$ admits a finite, strictly positive, jointly continuous extension to
$\mathbb R_+\times\Delta(\Y)$.\footnote{Throughout, we evaluate average
product and its normalized scale component at zero scale by their continuous
limits. All maximizations over nonnegative scales use this convention.}

\item \textbf{Differentiability.}
At every nonzero firm $\nu$, the functional $F$ has a continuous first
variation. That is, there is a continuous function
$\bm y\mapsto D_{\bm y}F(\nu)$ such that
\[
 F(\nu+t\eta)
 =F(\nu)+t\int_\Y D_{\bm y}F(\nu)\,\de\eta(\bm y)+o(|t|)
\]
for every finite signed measure $\eta$, as $t\to0$ through values for which
$\nu+t\eta$ remains nonnegative. The map
$(\nu,\bm y)\mapsto D_{\bm y}F(\nu)$ is continuous when $\nu$ varies weakly
on bounded-mass sets away from zero.

\item \textbf{Equal-mass monotonicity.}
For any two capability measures $\nu, \nu'$ of equal mass,
\[
    \nu'\succeq_{\mathrm{FOSD}}\nu
    \quad\Longrightarrow\quad
    F(\nu')\geq F(\nu).\footnote{We say $\nu'\succeq_{\mathrm{FOSD}}\nu$ if $
 \int f\,\de\nu'\ge\int f\,\de\nu$ for every bounded, continuous, nonnegative, increasing $f:\Y\to\R_+$.} 
\]
\item \textbf{Scale shape.}
For every $\nu\in\Delta(\Y)$, the average product 
\[m\longmapsto\frac{F(m\nu)}{m} \quad \text{is quasiconcave on $\mathbb R_+$.}
\] 
\item \textbf{Uniform asymptotic power tail.}
There exist continuous functions $A:\Delta(\Y)\to(0,\infty)$ and
$\gamma:\Delta(\Y)\to\mathbb R_+$ such that
\[
 F(m\nu)=A(\nu)m^{\gamma(\nu)}[1+o(1)]
\]
uniformly over $\nu\in\Delta(\Y)$. Moreover, the tail exponent is Lipschitz under mixing.\footnote{That is, there exists a constant $K_\gamma<\infty$ such that for all $\lambda \in [0,1]$ and all $\nu, \nu' \in \Delta(\Y)$: 
\[
 \left|\gamma\bigl((1-\lambda)\nu+\lambda\nu'\bigr)-\gamma(\nu)\right|
 \le K_\gamma\lambda. 
\]
}
\end{enumerate}
\end{assumption}
\end{savenotes}

Condition (i) on continuity and the infinitesimal limit allow atomistic firms to be well-defined. Condition (iii) on equal-mass monotonicity orders the output of firms that have the same size, but one is first-order stochastically dominated. We emphasize, however, that this does not imply monotonicity under addition, i.e., it is not true that $F(\nu + \nu') \geq F(\nu)$ since they are of different measure. Condition (iv) ensures that the optimal scale for a firm is well-behaved and is satisfied by all common functional forms. Condition (v) is a tail condition that 
controls how large firms scale: a composition is \emph{subcritical} if $\gamma<1$
and \emph{supercritical} if $\gamma>1$. The mixing bound limits how much a
small share of a firm can change that exponent. We will later see that common functional forms, e.g., CES aggregation of output across domains, satisfy \cref{ass:technology}.

\begin{figure}[H]
 \centering
 \includegraphics[width=0.8\textwidth]{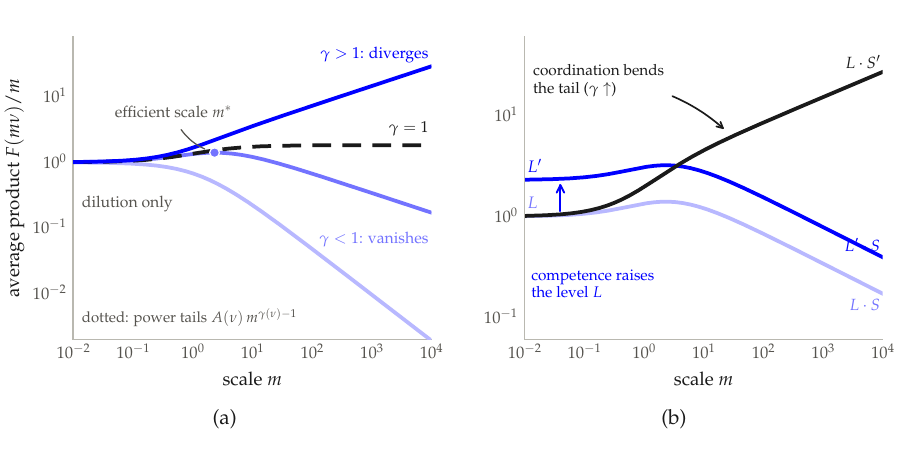}
 \caption{Level and scale: competence lifts the curve; coordination bends the
 tail.}
 \label{fig:scale-menu}
\end{figure}

\paragraph{Level and scale decomposition.} For a nonzero firm $\nu \in \M_+(\Y)$, its output $F(\nu)$ can be decomposed into its \emph{level} and \emph{scale}, i.e., the size of the firm: 
\[
 L(\nu):=\lim_{t\downarrow0}\frac{F(t\nu)}{t\nu(\Y)}
 \quad \text{and} \quad
 S(\nu):=\frac{F(\nu)}{L(\nu)}
\]
where $F(\nu) = L(\nu) \cdot S(\nu)$. $L$ captures how the firm's composition matters for output. It is scale-free: for any nonzero $\nu\in\M_+(\Y)$ and any $t,t'>0$, $L(t\nu)=L(t'\nu)=L(\nu/\nu(\Y))$. $S$ captures total output normalized by level and therefore how well the firm scales. \Cref{fig:scale-menu} shows the two at work. Panel (a) plots average product against scale: it vanishes when $\gamma<1$, settles when $\gamma=1$, and diverges when $\gamma>1$. Panel (b) separates the two capabilities.

\textbf{First-moment representations.} On occasion we will impose a stronger condition that decomposes coordination from competence in a particularly clean way: 

\begin{assumption}[First-moment representation]
\label{ass:level-scale}
For every nonzero firm $\nu$, the level and scale components introduced
in the model satisfy the following representation:
\[
\begin{aligned}
 L(\nu)
 &=L\left(
 \frac{1}{\nu(\Y)}
 \int_\Y (y_1,\ldots,y_n)\,\de\nu(\bm y)
 \right)
 \\
 S(\nu)
 &=S\left(
 \nu(\Y),
 \frac{1}{\nu(\Y)}\int_\Y y_*\,\de\nu(\bm y)
 \right).
\end{aligned}
\]
In addition:
\begin{enumerate}[(i)]
 \item \textbf{Diminishing relative returns:} for every pair of domains $j\ne k$,
 holding all other coordinates fixed, $\partial_kL(\bm z)/\partial_jL(\bm z)$
 is strictly decreasing in $z_k$ wherever both derivatives are positive;
 \item \textbf{Larger firms value coordination more:} for every mean coordination level $z_*$, the marginal scale value of
 coordination $\partial_2\log S(m,z_*)$ is strictly increasing in firm size $m$. 
\end{enumerate}
\end{assumption}

The first statement of \cref{ass:level-scale} says that the vector of mean competence
determines the level of the production curve, while firm size
and mean coordination fix its shape. Then, part (i) says that a productive domain
becomes less scarce as the firm acquires more of it. Diminishing absolute
returns ($\partial_{kk}L\leq0$) and gross complementarity
($\partial_{kj}L\geq0$) imply a weak decline. Part (ii)
says that larger firms place more value on coordination.

\begin{example}[CES environment]
\label{ex:ces-technology}
The \emph{CES environment} has two competence domains, $y_1,y_2$, and
one coordination domain, $y_*$, for both humans and machines.
For a nonzero firm $\nu$, write its size as $m:=\nu(\Y)$ and its mean
capability in domain $d\in\{1,2,*\}$ as
$\overline y_d:=m^{-1}\int_\Y y_d\,\de\nu(\bm y)$. Output is $F(\nu)=L(\nu)S(m,\overline y_*)$ where 
\[ L(\nu)=\left(\frac{\overline y_1^\rho+\overline y_2^\rho}{2}
          \right)^{1/\rho} \quad \text{and} \quad 
 S(m,\overline y_*)=m(1+m)^{\overline y_*-1}
                    e^{\phi m/(1+m)}. 
\]
Here $\rho<1$ controls substitution between competence domains and
$\phi>0$ controls scale gains. We set $F(0)=0$; at
$\rho=0$, we use $L(\nu)=\sqrt{\overline y_1\overline y_2}$. Notice that this satisfies \cref{ass:technology} and \cref{ass:level-scale}. 
\end{example}

\subsection{Demand and resource costs}

\paragraph{Demand.} Let $P:\R_+\to\R_+$ be inverse demand. We assume that $P$ is continuous
and Lipschitz, strictly decreasing while positive, and reaches zero at a finite quantity $D(0)>0$. Thus, $ P(Q)>0 \iff Q<D(0)$. Let $D:[0,P(0)]\to[0,D(0)]$ be the inverse of the decreasing branch of $P$.

We will often find it helpful to compare how the industry structure changes as demand grows or shrinks. To this end, we will write $P_N(Q) := P(Q/N)$ and $D_N = P_N^{-1}$. Thus, $N > 0$ changes the scale of demand. The gross benefit for a market of size $N$ is given by 
\[
U_N(Q) = \int^Q_0 P_N(z) dz 
\]
which is the gross value of output to consumers. 

\paragraph{Resource costs.} Recall that a human's reservation wage is $\underline w > 0$ per unit, and a machine can be rented at $r > 0$ per unit. The resource cost of a firm is
\begin{equation*}
 C(f):=\underline w\,\nu_f^H(\Y^H)+r\nu_f^M(\Y).
\end{equation*}
We use the qualifier `resource' because such costs are exogenous to the model, and a firm might face a strictly larger costs (since wages might exceed the reservation wages). To ensure that the equilibrium is non-trivial, we also assume that some machine-only firm can produce profitably at arbitrarily small scale: $\frac{r}{L(\delta_{\bm y^M})}<P(0)$. Here the left side is the unit cost of an atomistic machine-only firm so this ensures strictly positive surplus.

\subsection{Industry structure} We now develop language to describe the composition of different firms in the economy.  We distinguish between firms of two scales. Doing so allows us to capture both models with atomless firms \citep{aumann1964,novshekSonnenschein1987,shimomuraThisse2012}---and models with positive-size firms \citep{lucas1978}, as well as economies where both coexist. 

Positive-scale firms are represented by 
\[
 \F^+
 :=\Bigl\{f=(\nu_f^H,\nu_f^M):m_f>0,\ \nu_f^H\le\mu\Bigr\}
\]
where recall $m_f := \nu_f(\Y) > 0$ is the size of the firm and the inequality $\nu_f^H\le\mu$ requires that the human portion of each firm can be staffed from the human population. 

\begin{figure}[H]
 \centering
 \includegraphics[width=.65\textwidth]{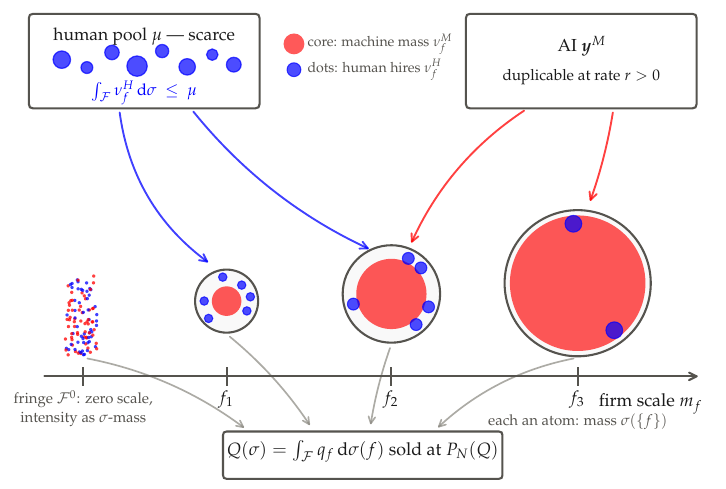}
 \caption{An industry as a measure over firms.}
 \label{fig:industry-measure}
\end{figure}

Atomless firms are represented by 
\[
 \F^0 :=
 \Bigl\{f=(\nu_f^H,\nu_f^M):m_f=1\Bigr\}.
\]
where $m_f = 1$ is a normalization.\footnote{We clearly cannot set $m_f = 0$ since the composition would not be well-defined.} $f \in \F^0$ captures the composition of firms. How should we interpret atomistic firms? For $f\in\F^0$, output and cost per unit of `intensity' are
\[
 q_f:=L(\nu_f)
 \quad \text{and} \quad 
 C(f):=\underline w\,\nu_f^H(\Y^H)+r\nu_f^M(\Y).
\]
An aggregate intensity $a > 0 $ of identical atomistic firms employs measure 
$a\nu_f^H$ of the human population, produces $aq_f$ units of the final good, and costs $aC(f)$.\footnote{To obtain these coefficients, let $(t\nu_f^H,t\nu_f^M)$ be a scale-$t$ firm
with the same composition. If
$t_n\downarrow0$ and integers $k_n$ satisfy $k_nt_n\to a$, then $
 k_nF(t_n\nu_f)\to aq_f$, $k_nt_n\nu_f^H\to a\nu_f^H$, and 
$k_nt_n C(f)\to aC(f)$ which represents the limit of many tiny firms.}
The space of all firms is the disjoint union $\F:=\F^+\sqcup\F^0$.\footnote{Endow each component with the weak topology on its measure coordinates and
$\F$ with the disjoint-union Borel $\sigma$-algebra. All industry measures and integrals below are taken on this measurable space.
Thus the two kinds of firm remain distinct even when their measure
coordinates coincide.} 
An industry is a finite positive measure
$\sigma\in\M_+(\F)$ over these activities.
 \Cref{fig:industry-measure} illustrates.

The measure $\sigma$ records masses of positive-scale firms and aggregate
intensities of atomistic firms. Following activity analysis
\citep{koopmans1951}, we allow the intensity of each firm type to vary
continuously. This convexifies firm counts and can assign noninteger mass to
$f\in\F^+$.\footnote{Our construction is also related to the free-entry tradition
surveyed by \citet{novshekSonnenschein1987}.} This will supply the convexity needed for
existence and duality. We interpret the convexified economy as an
approximation to an industry with indivisible firms. The industry $\sigma$ produces total output 
\[
 Q(\sigma):=\int_\F q_f\,\de\sigma(f).
\]

Define the set of feasible industries by
\[
 \Sigma
 :=
 \left\{
  \sigma\in\M_+(\F):
  \begin{array}{l}
   q_f\le Q(\sigma) < \infty
   \quad\text{for $\sigma$-almost every }f\in\F^+,\\[1mm]
   \displaystyle\int_\F\nu^H_f\,\de\sigma(f)\le\mu
  \end{array}
 \right\}.
\]

The first condition requires finite aggregate output and measure one of every positive-scale firm to fit inside the market.\footnote{Without this restriction, fractional activity weights would allow firms larger than the entire market to operate at arbitrarily small scales. In particular, machine-only production could always use the scale that minimizes average cost, regardless of market size, and adjust its activity weight to meet demand. Under unbounded increasing returns to scale, this would drive the minimum cost of producing any positive quantity to zero.} The second requires aggregate human employment not to exceed the population.

\Cref{sec:interiority} defines the feasibility and interiority conditions
used in our comparative statics.

\section{Equilibrium and efficiency}
\label{sec:equilibrium-results}

\subsection{Equilibrium}
\label{sec:equilibrium}
We describe the equilibrium concept informally; details are in \Cref{app:replacement-game}. We want to determine two objects: the industry structure $\sigma$, which records which firms operate and how intensively, and the wage vector $w$, which records how rents are spread across human types.

Our equilibrium concept rests on two ideas. The first is a \emph{competitive
continuation}. Any proposed arrangement must be one that producers and workers
would in fact carry out. That is, firms take the output price and wages as given, every
active firm breaks even, no available firm could enter and earn a profit, and
the labor market clears. The second is \emph{replacement stability}. An
outsider may propose to replace the whole economy: a new industry,
a new price and wages, and transfers to consumers and to each human worker. This proposal is accepted only if every party is made weakly better off. The industry structure $\sigma$ and wage $w$ form an equilibrium if they admit a competitive continuation and no proposal succeeds against it.

Our equilibrium notion is somewhat unconventional, in part because it has to handle firms at both small (vanishing) and large scales. Nonetheless many of its constituent ideas have close precedents in the literature. The idea that entrants compete to serve the whole market follows \citet{demsetz1968},
who lets consumers select a supplier through a joint buying organization; here we 
let an entrant propose the whole industry $\sigma'$. More broadly, replacement stability
follows the contestability logic of
\citet{baumolWillig1981,baumolPanzarWillig1982}.\footnote{Transferable utility
supplies the link from blocking to efficiency familiar from
\citet{aumann1964,ostroy1980}. Making firms part of a contract also relates to
\citet{zame2007}.} The requirement that the replacement itself form a competitive continuation is
the free-entry condition of an activity-analysis economy
\citep{koopmans1951,novshekSonnenschein1987}. Here each firm composition is, in the language of Koopmans, an `activity' and $\sigma$ records their intensities. Finally, our entry condition also shares the logic of Riley's reactive equilibrium \citep{riley1979} in which an entrant must anticipate the responses to its proposal. 

We use this equilibrium concept in part because its existence is guaranteed. It also includes competitive equilibria (in which firms and workers take prices as given) when the latter exists, though this is not true generally. We now draw a tight connection between equilibrium, efficiency, and rents. 


\subsection{Efficiency and shadow values}
\label{sec:efficiency}

For an industry $\sigma\in\Sigma$, total surplus is
\[
 \W_N(\sigma)
 :=U_N(Q(\sigma))-\int_\F C(f)\,\de\sigma(f).
\]
An industry $\sigma\in\Sigma$ is \emph{efficient} if
$\sigma\in\argmax_{\widehat\sigma\in\Sigma}\W_N(\widehat\sigma)$.
Wages above outside options, firm revenues, and lump-sum payments
are transfers. They do not enter $\W_N$. Competitive zero profits and balanced
transfers imply that an implemented contract only divides this surplus among
the protected parties and the organizer.

\paragraph{Industry cost.}
For $Q>0$, define the least reservation-resource cost of producing quantity
$Q$ by
\[
 \mathcal C(Q)
 :=
 \inf_{\substack{\sigma\in\Sigma\\Q(\sigma)\geq Q}}
 \int_\F C(f)\,\de\sigma(f),
\]
where $\mathcal C(0):=0$. Radial trimming and a reduction in activity intensities make this infimum
equal to the exact-output minimum.

\paragraph{Efficient quantity.}
At quantity $Q$, the largest feasible surplus is
$U_N(Q)-\mathcal C(Q)$. Consumer payments, firm revenues, and human payments
above the outside option are transfers. Human opportunity costs and machine
resource costs enter through $\mathcal C$.

The efficient quantities are
\[
 \mathcal Q_N^*
 :=\argmax_{Q\in[0,D_N(0^+)]}
 \Big \{U_N(Q)-\mathcal C(Q)\Big \}.
 \]
 Let $Q_N^* \in  \mathcal Q_N^*$ denote any efficient quantity. 

\paragraph{Shadow values.}
A shadow value measures how minimum resource cost changes when an output
requirement or a resource limit changes. Fix $Q>0$ and an industry
$\sigma\in\Sigma$ with $Q(\sigma)=Q$. Let $\Lambda(Q,\sigma)$ be the set of pairs
$(p,w)\in\R_+\times[\underline w,\infty)^K$ that satisfy:
\begin{enumerate}[(i)]
 \item \textbf{Firm support.}
 \[
  pq_f\le
  r\nu_f^M(\Y)+\int_{\Y^H}w\,\de\nu_f^H
  \qquad
  \forall f\in\F^0\sqcup\{g\in\F^+:q_g\le Q\}.
 \]
 \item \textbf{Active equality.}
 \[
  pq_f=
  r\nu_f^M(\Y)+\int_{\Y^H}w\,\de\nu_f^H
  \qquad\text{for $\sigma$-almost every }f.
 \]
 \item \textbf{Human scarcity.}
 \[
  (w_i-\underline w)
  \left[
   \mu_i-\int_\F\nu_f^H(\{\bm y^i\})\,\de\sigma(f)
  \right]=0
  \qquad\forall i\in\{1,\ldots,K\}.
 \]
\end{enumerate}
And we write $w := (w_i)_{i=1,\ldots, K}$. The value $p$ is the price for output at quantity
$Q$. The premium $w_i-\underline w$ is a multiplier on the supply of human type $i$. Firm support rules out a profitable entrant. Active
equality gives zero profit for each active firm. Human scarcity permits a
wage premium only when all humans of that type work. Thus the payments recover
the shadow values: $p$ is the output multiplier, and $w_i-\underline w$ is the
multiplier on type $i$'s supply limit. The internal value $p$ need not equal
$P_N(Q)$ or a derivative of $\mathcal C$. Likewise, a change in $\mu$
also changes which complete firms are feasible; the scarcity multiplier
alone need not give that full comparative static.

\begin{figure}[H]
\begin{minipage}[c]{.40\textwidth}
\raggedright
\Cref{fig:efficiency-duality} illustrates efficient output and human scarcity
rents. Marginal resource cost rises in steps as each scarce human type is
exhausted, then stays constant at the unit cost of machine-only production.
Its intersection with inverse demand gives the efficient quantity. The blue
area between the price and the cost steps shows the Ricardian rents of the
exhausted human types.
\end{minipage}\hfill
\begin{minipage}[c]{.57\textwidth}
 \centering
 \includegraphics[width=\linewidth,trim=0 24 175.41144 14,clip]{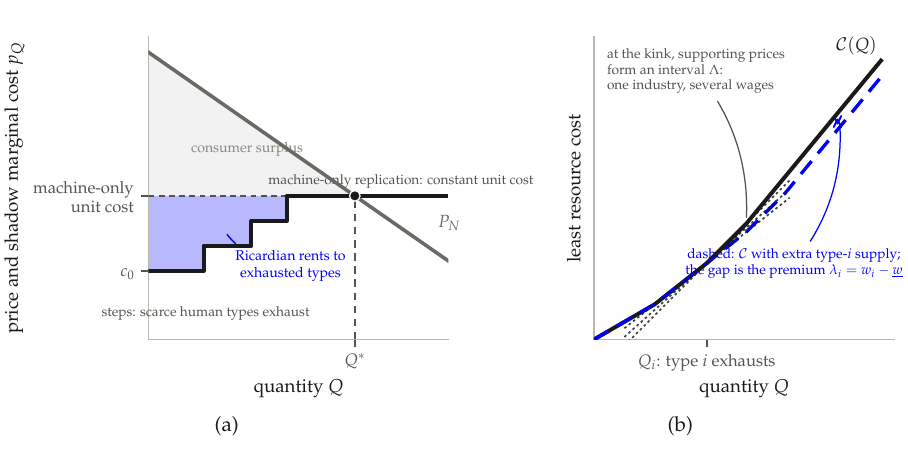}
 \caption{Efficient output and human scarcity rents.}
 \label{fig:efficiency-duality}
\end{minipage}
\end{figure}

\begin{theorem}[Equilibrium, efficiency, and wages; informal]
\label{thm:equilibrium-efficiency}
An equilibrium outcome. $(\sigma,w)$ is an equilibrium
if and only if $\sigma$ is efficient and
\[
 (p,w)\in\Lambda(Q(\sigma),\sigma)
 \quad\text{for some }p\ge0.
\]
\end{theorem}

Our equilibrium concept, while somewhat unconventional, buys existence. It is also closely related to competitive equilibrium i.e., it can be supported by prices and wages. More precisely, given $(p,w)$, each producer chooses the composition and size of the firm to maximize profits, human type $i$ chooses employment in $[0,\mu_i]$ at wage
$w_i$, machines are supplied elastically at $r$, and consumers choose quantity $x\ge0$
to maximize $U_N(x)-px$, their utility net of payments. A competitive
equilibrium (CE) consists of optimal choices at prices $(p,w)$, with
$p\ge0$, that clear all markets.

Indeed, every CE is an equilibrium of our model provided that each firm in the CE (i) produces no more than the equilibrium total industry output; and (ii) requires no more of any human type than supplied in the population. That is, when a CE exists and firms are `interior'---in the sense of (i) and (ii)---at least one equilibrium in our model can be supported through prices.\footnote{This is more likely to hold when returns to scale limit optimal firm size (e.g., strict diminishing returns to scale) and markets are large enough ($N$ large) to accommodate the resulting firms.} To see this, note that competitive prices rule out profitable entry, active firms break even, and workers choose employment optimally. Thus, these conditions provide the price and wage support required by \cref{thm:equilibrium-efficiency}. Consumer and producer optimization also imply that no feasible alternative industry can generate greater total surplus. The competitive industry is therefore efficient so \cref{thm:equilibrium-efficiency} applies. This insight also implies that the task model \citep{acemogluRestrepo2018,aghionJonesJones2019} is nested within our framework (see \cref{sec:connection}) where the technology takes the special form  $F^{\mathrm T}$ that assigns workers and machines optimally across tasks and has constant returns ($S(\nu)=\nu(\Y)$) such that the composition of firms does not matter. The two economies have the same equilibrium output and task allocations, and wages also coincide. 

\subsection{Regimes of AI capabilities}
\label{sec:capability-regimes}
\begin{figure}[H]
 \centering
 \includegraphics[width=.85\textwidth]{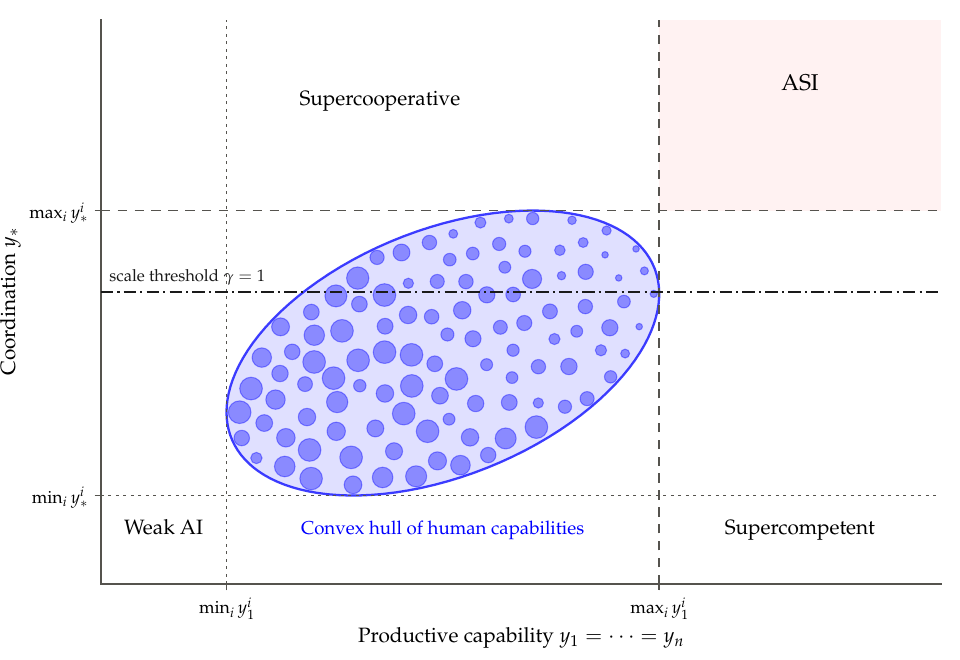}
 \caption{Capability regimes.}
 \label{fig:capability-regimes}
\end{figure}

We next develop language for different regimes of AI capabilities which are illustrated in \Cref{fig:capability-regimes}.

The $y$-axis of \cref{fig:capability-regimes} is coordination $y_*$ while the x-axis captures competence. Here, for simplicity, we set $y_1 = y_2 = \ldots y_n$ so that they can be represented in a single dimension, though the ideas extend without difficulty to multiple competence domains (but would be difficult to illustrate). Human types are blue dots, and the size reflects the mass of that type in the population. The oval around human types represents the closed convex hull which is exactly the set of mean capabilities a human-only firm
can attain. 

We call the AI \emph{supercompetent} when it is more competent than every human, i.e.,  $y_d^M>\max_i y_d^i$ in every productive domain
$d$. We call the AI \emph{supercooperative} when $\gamma^M:=\gamma(\delta_{\bm y^M})>1$, so that the average product of an AI-only firm grows without bound as the firm grows.\footnote{Note here that supercooperative AI does not require that it is better at coordination than the best humans, i.e., it is an absolute rather than relative measure.} The AI is \emph{superintelligent} if it dominates the convex hull of human capabilities. It is \emph{weak} if it is dominated by the convex hull of human capabilities. It is \emph{minimal AGI} if it lies within the convex hull of human capabilities. 

\subsection{Feasibility, interiority, and continuity}
\label{sec:equilibrium-selections}
\label{sec:interiority}
We collect two recurring conditions in \cref{tab:interiority}. For a unit
composition $\nu\in\Delta(\Y)$, a \emph{preferred scale} maximizes average
product $F(m\nu)/m$ over scales $m\ge0$, using the level $L(\nu)$ at zero.
We assign preferred scale $+\infty$ when no finite scale attains the maximum.
Each result specifies which mixes must have feasible preferred scales and
at which output.

\begin{table}[H]
\centering
\renewcommand{\arraystretch}{1.5}
\begin{tabular}{@{}>{\raggedright\arraybackslash}p{.20\textwidth}>{\raggedright\arraybackslash}p{.46\textwidth}>{\small\raggedright\arraybackslash}p{\dimexpr.34\textwidth-4\tabcolsep\relax}@{}}
\toprule
\textbf{Term} & \textbf{Definition} & {\normalsize\textbf{Purpose}} \\
\midrule
Interior firm
 & A firm $f$ has positive size $m_f>0$, uses both humans and machines,
 $0<\nu_f^M(\Y)/m_f<1$, and satisfies $q_f<Q(\sigma)$ and
 $\nu_f^H(\{\bm y^i\})<\mu_i$ for each human type it employs.
 & Permits small changes in size and output, allowing for substitutions between
 machines and each employed human type. \\
\addlinespace
Feasible preferred scales
 & Each composition $\nu$ has a finite preferred scale $m$ with
 $m\nu^H(\{\bm y^i\})\le\mu_i$ for every type $i$ and $F(m\nu)\le Q$.
 & Allows tractable comparison of firm sizes and optimal firm composition. \\
\bottomrule
\end{tabular}
\caption{Feasibility and interiority conditions.}
\label{tab:interiority}
\end{table}

An equilibrium is \emph{interior} if every active firm of positive-scale is
interior. 
We will often ask how industry structure and wages change as AI improves.
Let $\theta$ index a continuously differentiable path of AI capability profiles
$\bm y^M(\theta)$ and human supplies $\mu(\theta)$, with human capability profiles and
all other primitives fixed. Write $\Sigma^*(\theta)$ for the set of equilibrium industries at $\theta$. An \emph{equilibrium selection} chooses an industry
$\sigma^*(\theta)\in\Sigma^*(\theta)$ at each $\theta$.

\begin{definition}[Firm selections]
\label{def:firm-selection}
Given an equilibrium selection $\sigma^*$, a \emph{firm selection} chooses
an active positive-scale firm $f(\theta)\in\supp(\sigma^*(\theta)|_{\F^+})$
at each parameter value. It is \emph{continuous}, \emph{differentiable}, or \emph{continuously differentiable} when its size and composition have that property.
\end{definition}

\section{Equilibrium industry structures}
\label{sec:industry}
We now ask how equilibrium industry structures are driven by progress in AI capabilities $\bm y^M$. We start with what sets the size of firms. Then, we move the AI capabilities in one direction at a time, and finally analyze the composition of firms. All proofs in the paper are in \cref{app:welfare-proofs}. We emphasize that our results are situated within our theoretical model. They are not empirical forecasts of market outcomes or technological developments.

\subsection{What determines firm size?}
\label{sec:firm-size} We start with a basic building block on how firm sizes are determined in equilibrium.  Under \cref{ass:level-scale}, preferred size depends on composition only
through mean coordination. Write $m^*(z_*)$ for the set of preferred sizes
at mean coordination $z_*$:
\[
 m^*(z_*):=\argmax_{m\ge0}\frac{S(m,z_*)}{m}. 
\]
Notice that given a fixed firm composition, changes in competence leaves the
preferred-size set unchanged. Almost every active firm reaches a preferred size whenever a finite
preferred size is feasible in equilibrium. 
To bound sizes across all firm compositions, define
\[
 \begin{aligned}
 \overline M(y_*^M,\mu)
 &:=\sup_\nu\left[
   \sup m^*\!\left(\int_\Y y_*\,\de\nu\right)
   \wedge\sup\{m\ge0:m\nu^H\le\mu\}\right],\\[1ex]
 \underline M(y_*^M,\mu)
 &:=\inf_\nu\left[
   \inf m^*\!\left(\int_\Y y_*\,\de\nu\right)
   \wedge\sup\{m\ge0:m\nu^H\le\mu\}\right].
 \end{aligned}
\]

\begin{figure}[H]
 \centering
 \includegraphics[width=\textwidth]{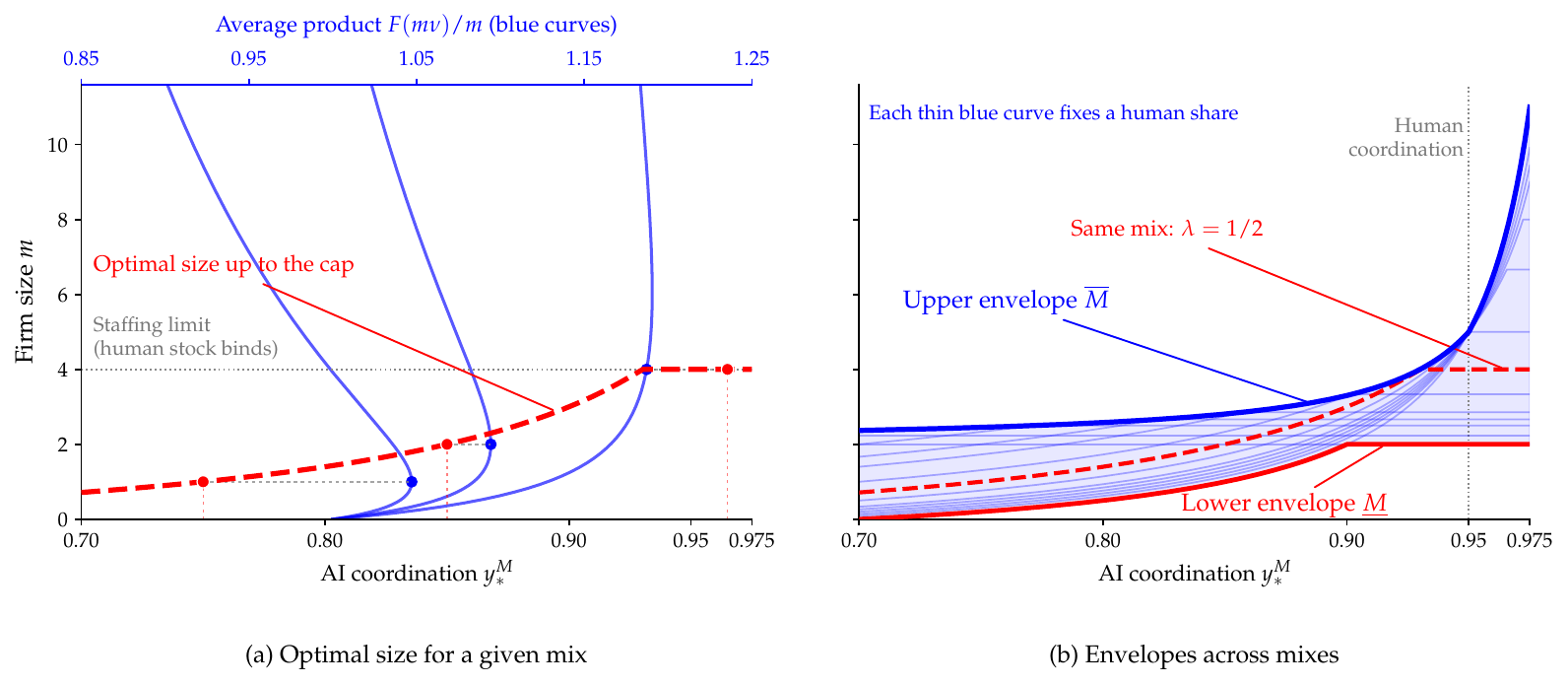}
 \caption{Firm-size choice and its envelopes.}
 \label{fig:ces-size-envelopes}
\end{figure}

These bounds are illustrated by \cref{fig:ces-size-envelopes} in which the dashed red line in panel (a) illustrates the optimal size $m > 0$ for a firm $m \nu$ where $\nu \in \Delta(\Y)$ is unit measure. As $y^M_*$ gets larger, the preferred size $m^*$ increases until the firm is no longer feasible: $m \nu^H \nleq \mu$, i.e., it would require more talent than is available in the scarce human population.


\Needspace{14\baselineskip}
\begin{proposition}[Bounds on firm size]
\label{thm:firm-scale}
If \cref{ass:level-scale} holds, almost every active firm in every
equilibrium satisfies $m_f\le\overline M(y_*^M,\mu)$, with size zero
for atomistic firms. If average product is also strictly quasiconcave
in size, then in every equilibrium in which no active positive-scale
firm supplies the entire market,
\[
 \underline M(y_*^M,\mu)\le m_f\le\overline M(y_*^M,\mu).
\]
\end{proposition}


\subsection{How AI improvements change industry structure} 
\label{sec:competence}
\label{sec:coordination}
We now ask how changes in AI capabilities along various dimensions change the equilibrium industry structure in our theoretical model. We first consider the case in which AI coordination improves. 

\begin{proposition}[The size amplifier]
\label{prop:coordination-size}
Suppose we are in a CES environment with one human type and consider an equilibrium
selection as AI coordination $y_*^M$ increases over 
$[0,1)$. Suppose along the selection that (i) human supply is
slack; (ii) every firm composition has a positive, feasible preferred scale at equilibrium output. Let $m$ be firm size, $s$ its machine share, and
$\overline y_*=(1-s)y_*^H+sy_*^M$ its mean coordination. Write preferred size as $m^*(s,y_*^M)$. Along any continuously differentiable active-firm selection: 

\[
\begin{aligned}
 \frac{\de m}{\de y_*^M}
 &=\underbrace{\frac{\partial m^*}{\partial y_*^M}}
     _{\substack{\text{direct size effect} \\ \geq 0}}
   +\underbrace{\frac{\partial m^*}{\partial s}
                \frac{\de s}{\de y_*^M}}
     _{\substack{\text{effect through composition} \\ \gtreqless 0 }}\\[2ex]
 &=\frac{\phi s}{(1-\overline y_*)^2}
   +\frac{\phi(y_*^M-y_*^H)}{(1-\overline y_*)^2}
     \frac{\de s}{\de y_*^M}.
\end{aligned}
\]
If AI coordination is: 
\begin{itemize}
  \item[(i)] Better-than-human i.e., $y^H_* < y_*^M < 1$ then both effects are positive and composition amplifies direct size gain.  
  \item[(ii)] Worse than human, i.e.,
$y_*^M<y_*^H$. The direct effect is positive but the composition effect is ambiguous:
\begin{itemize}
  \item If $\log(1+m^*)> \frac{s(y_*^H-y_*^M)}{1-\overline y_*}$ the machine share rises and the second effect is negative so composition dampens the direct size gain.

  \item If the inequality reverses, the machine share falls so the second effect is positive and composition amplifies the direct size gain.
\end{itemize} 
\end{itemize}
\end{proposition}

\begin{figure}[H]
 \centering
 \includegraphics[width=0.9\textwidth]{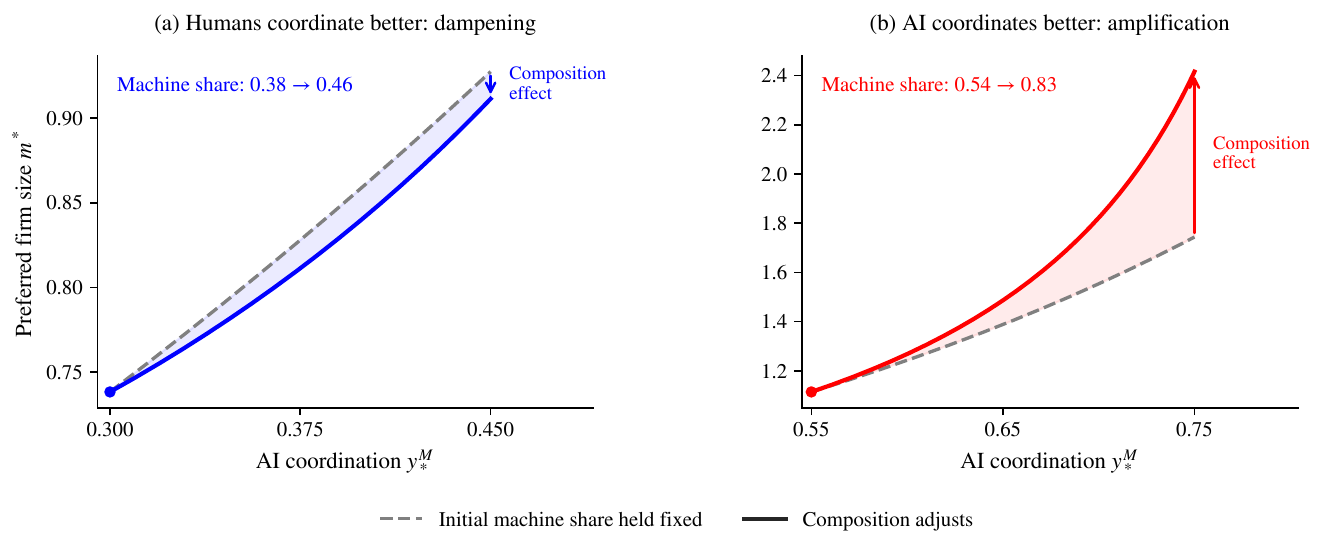}
 \caption{Composition can dampen or amplify firm growth.}
 \label{fig:amplification-dampening}
\end{figure}

\cref{prop:coordination-size} is illustrated by \cref{fig:amplification-dampening}. Panel (a) displays a case
in which humans are better coordinators ($y^H_* > y^M_*$)
and changes in composition dampen firm growth; panel (b)
shows a case in which AI coordinates better ($y^M_* > y^H_*$)
and changes in composition amplify firm growth.

The basic mechanism behind amplification is that better AI coordination raises preferred firm size holding fixed its current composition. But, as the firm grows, better coordination becomes especially valuable since it determines how output grows as productivity scales. For instance within our CES specification, we have $\partial \log F / \partial \bar y_* = \log (1 + m)$. If we are in the theoretical regime in which machines coordinate better than humans (part (i) of \cref{prop:coordination-size}) the firm tilts its composition of workers toward machines so the improvement in AI's ability to coordinate then drives up the firm's mean coordination $\bar y_*$ which then supports further growth, and so on. Thus, size and composition reinforce each other. 

By contrast, if humans still coordinate better, growth in firm size tilts composition toward humans. However, better AI coordination also makes machines more attractive at a fixed size. If this direct incentive dominates, the machine share rises despite the size-driven shift toward humans. Machines then replace better human coordinators, so the change in composition pushes down the average coordination within the firm $\bar y_*$ which dampens firm growth. 

Although \cref{prop:coordination-size} is stated within the CES environment with a single human type, the broader point extends: let $\theta$ index a differentiable path of AI coordination
improvements and let $s$ and $s_i$ be the machine share and
the share of human type $i$, with $s+\sum_i s_i=1$. Mean coordination is
$\overline y_*=s y_*^M+\sum_i s_i y_*^i$. Along the path of AI improvement:\footnote{Assuming that preferred size is unique, finite, and differentiable in mean coordination, and that the firm composition is also differentiable.}
\[
 \frac{\de m^*}{\de\theta}
 =\frac{\de m^*}{\de\overline y_*}
 \Bigg[
 \underbrace{s\frac{\de y_*^M}{\de\theta}}
   _{\substack{\text{coordination improves} \\ \text{at fixed shares} \\ \geq 0 }}
 +\underbrace{\sum_{i=1}^K (y_*^i-y_*^M)
   \frac{\de s_i}{\de\theta}}
   _{\substack{\text{composition changes} \\ \gtreqless 0}}
 \Bigg].
\]
Since preferred size increases weakly with mean coordination, the direct
effect is weakly positive. The sign of composition changes depends on whether the composition of the firm tilts towards machines or humans---as we have seen, the former amplifies the growth in firm size while the latter dampens it. This, in turn, might depend on wage effects that change the relative value of using machines vs humans in the model. For instance, wage changes could reinforce amplification---higher human wages encourage substitution toward machines, which raises mean coordination if machines coordinate better. 

\begin{figure}[H]
 \centering
 \includegraphics[width=\textwidth]{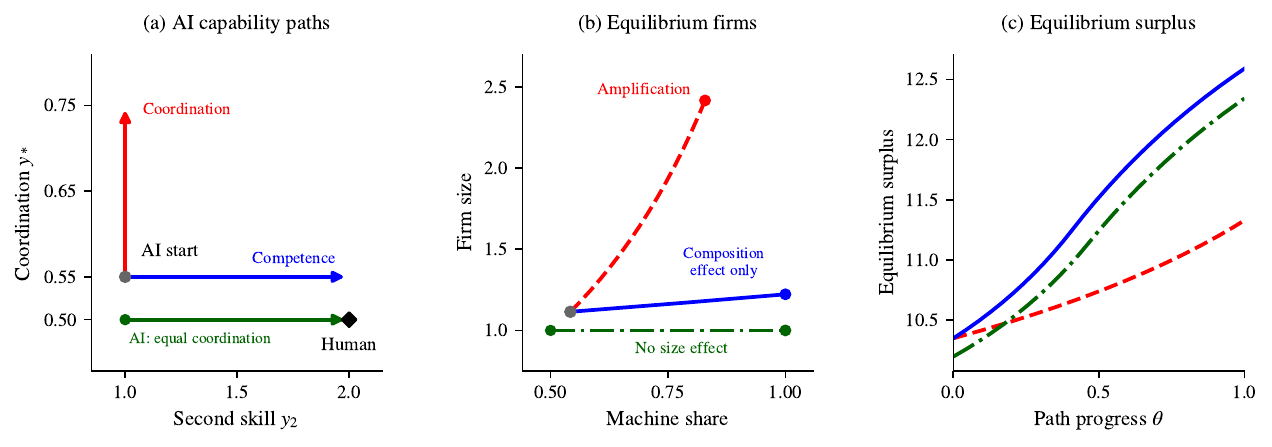}
 \caption{Coordination, competence, and surplus.}
 \label{fig:coordination-firms}
\end{figure}

By contrast, competence gains that raise productivity by the same proportion at every firm size (which is implied by \cref{ass:level-scale}) leave a firm's preferred size unchanged at fixed composition. That is, competence gains make firms more productive without driving larger firms, i.e., the direct effect is zero. Changes in preferred firm size are driven solely by changes in firms' workers mean coordination ability. \Cref{fig:coordination-firms} illusrates these forces by compares gains in AI coordination (red) and
competence (blue) from the same mixed workforce.\footnote{Humans have
profile $(1,2,.5)$ and supply $20$. For $\theta\in[0,1]$, AI follows
$(2,1,.55+.2\theta)$ in red, $(2,1+\theta,.55)$ in blue, and
$(2,1+\theta,.5)$ in dark green. Set $\phi=1$, $\rho=-1$,
$r=\underline w=1$, and $P(Q)=(2-Q/10)_+$. Human supply stays slack and preferred firms fit in the market. Panel (c) compares these specified paths, so the comparison depends on their lengths; see \cref{app:ces-coordination} for verification.}

\textbf{Sharp threshold in concentration.} Our previous theoretical analysis studied improvements in AI coordination in the regime where $\gamma^M < 1$. We now study what happens when it crosses $1$. 

\begin{proposition}[Supercooperation concentrates large markets]
\label{prop:market-size}
As $N\to\infty$:
\begin{enumerate}[(i)]
 \item \textbf{Replication.}
 If $\gamma^M<1$, every equilibrium has vanishing firms:
 \[
  \operatorname*{ess\,sup}_{f\in\F^+}
  \frac{q_f}{Q(\sigma_N^*)}
  \longrightarrow0.
 \]
 The essential supremum is taken with respect to $\sigma_N^*$ and is zero
 when no positive-scale firm is active.

 \item \textbf{Concentration.}
 If the AI is supercooperative, every equilibrium becomes concentrated:
 \[
  \int_{\F^+}
  \left(\frac{q_f}{Q(\sigma_N^*)}\right)^2
  \de\sigma_N^*(f)
  \longrightarrow1.
 \]
\end{enumerate}
\end{proposition}

The proposition explains when large firms emerge as the market grows. If
$\gamma^M<1$, the average product of every sufficiently large feasible firm
vanishes. The industry therefore expands by repeating bounded firm recipes,
so the capacity of any one complete firm becomes small relative to total
output. In the $N \to \infty$ limit this converges to the atomless economy of, e.g., \cite{novshekSonnenschein1987}. 

But when the AI tips into supercooperation, a single large AI-only firm can produce $Q$ at a cost of order $Q^{1/\gamma^M}$, where $1/\gamma^M<1$. Splitting the same output across several large firms loses this scale advantage. The equilibrium industry therefore assigns almost all output to firm types whose capacity is as large as the market. Our notion of concentration in part (ii) is a `relaxed' Herfindahl--Hirschman index when there is a non-integer number of firms. A limit of one means that almost all output is assigned to firm types whose complete capacity is close to total market output.

\begin{figure}[H]
 \centering
 \includegraphics[width=\textwidth]{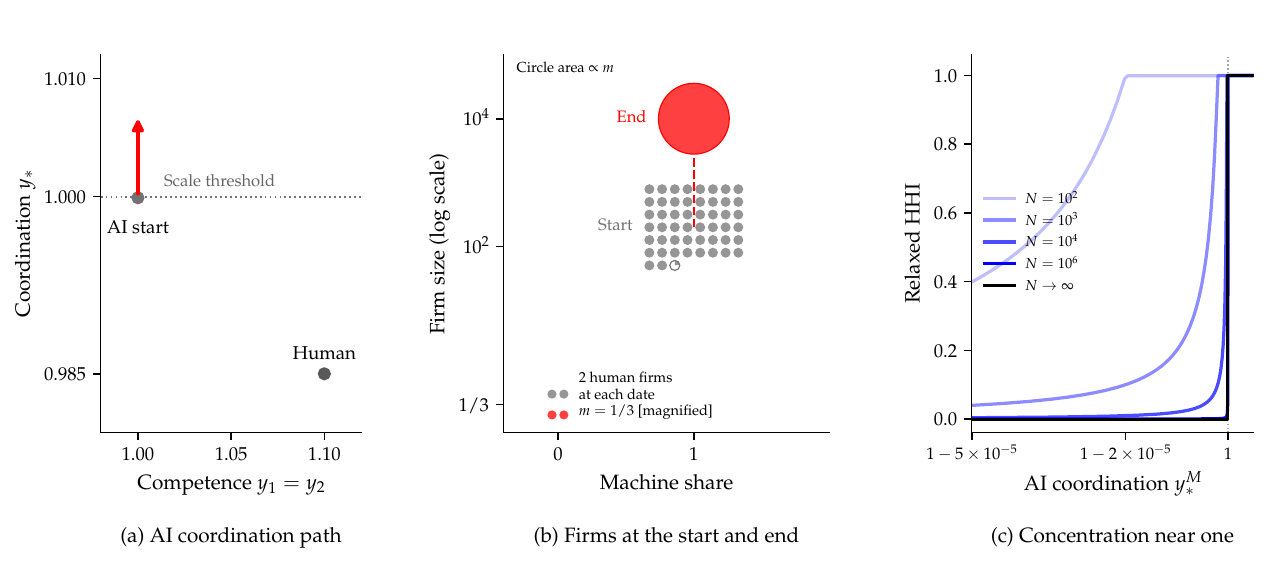}
 \caption{AI coordination turns replication into concentration. In panel
 (b), circle area is proportional to $m_f$, except for the magnified human
 firms; grey and red show the start and end. The two human firms at each
 date are offset vertically for visibility.}
 \label{fig:replication-aggregation}
\end{figure}

\Cref{fig:replication-aggregation} illustrates the shift from replicating small firms to concentrating output as AI coordination improves.\footnote{\label{fn:concentration-calibration}Set
$\phi=.02$, $\rho=0$, $r=\underline w=1$, human profile
$(1.1,1.1,.985)$ with supply $2/3$, AI profile $(1,1,\theta)$,
and $P_N(Q)=(2-Q/(10N))_+$. Here $\gamma^M=\theta$.
Panel (b) uses $N=10^3$, with grey at $\theta=0.9999$
and red at $\theta=1.005$; panel (c) varies $N$.
\Cref{prop:market-size} excludes $\gamma^M=1$.}
Two small human firms remain at both displayed dates, so concentration need not imply human exit. The finite-market curves restrict production to pure human and pure AI firms; entry checks at selected states do not establish equilibrium along the full paths (\cref{app:concentration-numerics}).

In our model, ``supercooperativeness'' denotes a stylized mathematical property of production technology. In practice, there are real world factors that would likely prevent a firm singularity, such as regulation, competition among AI providers, capacity constraints, and rigid supply contracts.

\subsection{Firm composition: human vs AI split}
\label{sec:composition}
In the model, whether humans and AI work together depends on productivity and input costs. We first characterize firm composition at equilibrium wages, then recover conditions that depend only on technology, resource costs, and supplies.

For a human share $\lambda\in[0,1]$ and wage vector $\bm w$, define \emph{cost-adjusted best average product} by
\[
 \overline a(\lambda,\bm w):=
 \sup_{\substack{\nu\in\Delta(\Y^H)\\m\ge0}}
 \frac{\overbrace{F\!\left(m[(1-\lambda)\delta_{\bm y^M}+\lambda\nu]\right)/m}
       ^{\text{Average product}}}
 {\underbrace{(1-\lambda)+\lambda\int_{\Y^H}[w(\bm y)/r]\,\de\nu(\bm y)}
       _{\propto\,\text{cost of workforce}}}.
\]
We write $\overline a( \cdot ) := \overline a(\cdot,r\one)$ for the unadjusted average-product envelope, i.e., as if all wages were equal to the rental rate. 

\begin{figure}[H]
\begin{minipage}[c]{.48\textwidth}
The envelope can be convex, concave, or neither. Its shape depends on
capabilities, technology, and wages. The figure on the right sets every human wage equal to $r$, so cost-adjusted best average product equals $\overline a(\lambda)$. Each grey curve fixes the human composition $\nu \in \Delta(\Y^H)$. The black upper envelope selects the best composition at each human share, capturing the gains from \emph{mosaic intelligence}.
\end{minipage}\hfill
\begin{minipage}[c]{.5\textwidth}
 \centering
 \includegraphics[width=\linewidth]{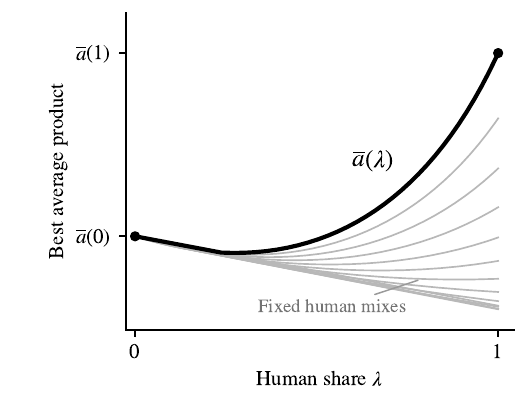}
\end{minipage}
\label{fig:average-product-envelope}
\end{figure}

\begin{figure}[H]
 \centering
 \includegraphics[width=0.9\textwidth]{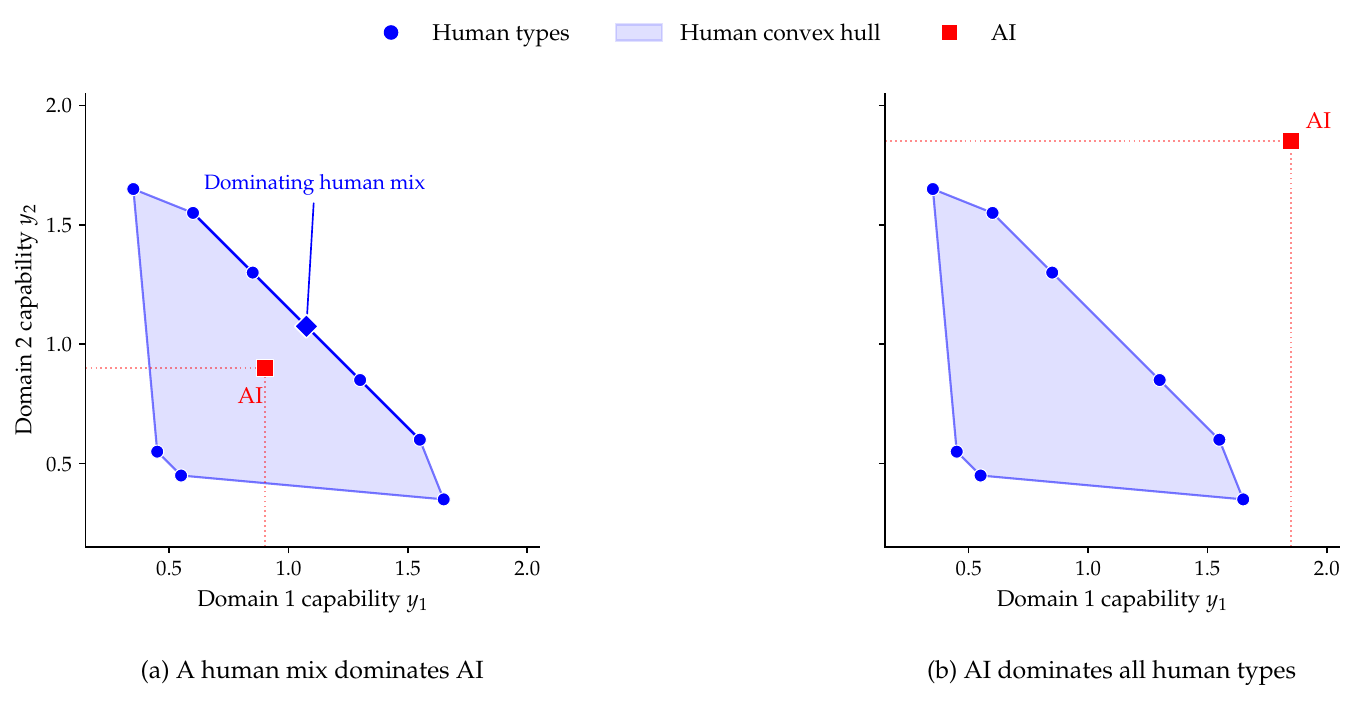}
 \caption{Human matching vs AI dominance.}
 \label{fig:human-hull-cases}
\end{figure}

\begin{proposition}[Human-AI production regimes]
\label{prop:human-diversity}
Fix an equilibrium $(\sigma,\bm w)$. Suppose the global maximum of
$\overline a(\lambda,\bm w)$ over human shares is finite and attained by
a feasible firm at output $Q(\sigma)$.\footnotemark{} Then:
\begin{enumerate}[(i)]
 \item \textbf{Machine-only production.} If $r\le\underline w$,
 \cref{ass:level-scale} holds, $L$ strictly increases in each productive
 capability, and $\bm y^M > \bm y$ for every
$\bm y\in\operatorname{co}(\Y^H)$, then all firms are machine-only. 

 \item \textbf{Humans in the loop.} If $r\geq \underline w$ and
 $\max_\lambda\overline a(\lambda,r\one)>\overline a(0,r\one)$,
 then some (positive measure set of) firms employ humans.
 If, in addition, there is a single human type, then:
 \begin{enumerate}[(a)]
  \item \textbf{Separate production.} If $\overline a(\cdot,r\one)$
  is strictly convex, then almost every active firm is pure human or pure
  machine, and some active firms are human-only.
  \item \textbf{Mixed production.} If $r=\underline w$,
  $\overline a(\cdot,r\one)$ is strictly concave, and its finite left
  derivative at one is negative, then almost every active firm is mixed.
  These firms share the unique human share maximizing
  $\overline a(\cdot,w)$ at the equilibrium human wage $w$.
 \end{enumerate}
\end{enumerate}
\end{proposition}
\footnotetext{For a maximizing share $\lambda^*$, human composition
$\nu^*$, and positive size $m^*$, feasibility means
$m^*\lambda^*\nu^*\le\mu$ and
$F(m^*[(1-\lambda^*)\delta_{\bm y^M}+\lambda^*\nu^*])\le Q(\sigma)$.}

\Cref{prop:human-diversity} gives sufficient conditions for machine-only production or for humans to remain in the loop.
Part (i) applies when AI dominates every human combination and machines cost no more than humans. \Cref{fig:human-hull-cases}(b) illustrates the productive-capability comparison: AI lies above the entire human hull. If this dominance also holds in coordination, replacing humans with machines raises output without raising cost.

\begin{figure}[H]
 \centering
 \includegraphics[width=\textwidth]{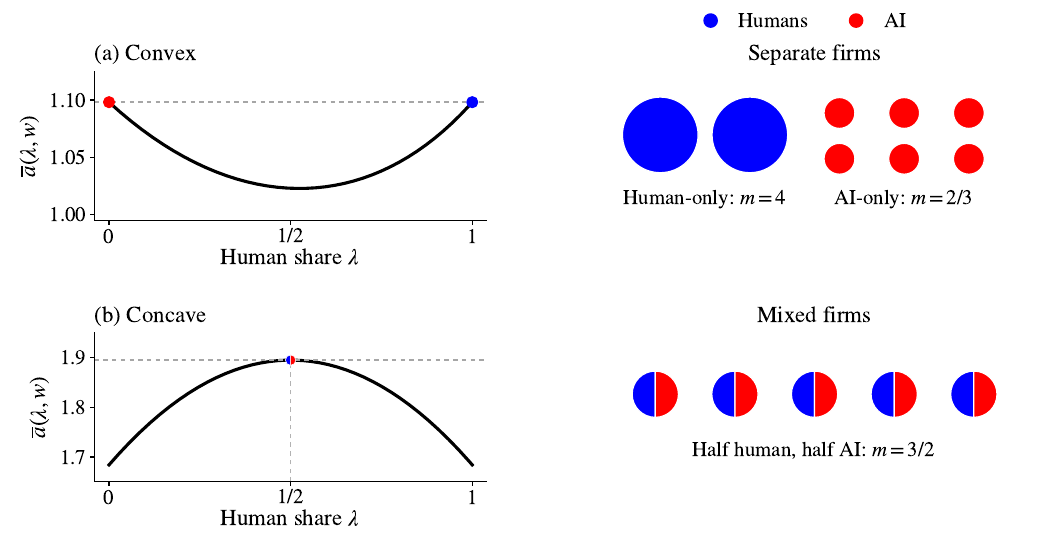}
 \caption{Curvature of $\bar a$ and firm composition in two CES economies.}
 \label{fig:ces-composition-regimes}
\end{figure}

Part (ii) allows humans to preserve employment through \emph{mosaic intelligence}: complementary human skills can jointly outperform AI even when no individual human does, as \cref{fig:human-hull-cases}(a) illustrates. The upper envelope records the best human composition at each human share, so it captures these gains from combining skills. If the envelope rises above the machine-only endpoint, then as long as reservation wages are below the cost of compute $(\underline w \leq r)$, some firms must employ humans. Parts (a) and (b) of \cref{prop:human-diversity} (ii) further relate the curvature of $\overline a(\cdot,\bm w)$
to whether humans and AI work in separate or mixed firms as illustrated by \cref{fig:ces-composition-regimes} which contrasts separate and mixed firms
under the same demand and human supply.\footnote{Both rows use $\rho=-1$,
$\phi=1$, $r=\underline w=1$, human supply $50$, and
$P(Q)=(2-Q/100)_+$. Row (a) has $\bm y^H=(1,1,.8)$ and
$\bm y^M=(1,1,.4)$; row (b) has $\bm y^H=(2,1,.6)$ and
$\bm y^M=(1,2,.6)$.}
In row (a), productive skills coincide but humans coordinate better:
scale gains make the envelope convex, favoring separate firms.
In row (b), equal coordination and complementary skills make the envelope
concave, favoring firms with equal human and AI shares.

\section{Human wages and employment}
\label{sec:rents-general}

We now turn to the question of human wages and employment. We first develop a some results pinning down firms' willingness-to-pay for human workers, and how different types of human workers sort into different firms based on their specific strengths.  Then, we study how different trajectories of AI improvements might drive changes in unemployment and wages in the model.

\subsection{The marginal value of human talent} 
\label{sec:wage-envelope}
We first develop some machinery to study the marginal value of employing a human with capability profile $\bm{y}$. For a nonzero firm $\nu$, the marginal product of capability $\bm y$ is
\[
 D_{\bm y}F(\nu)
 :=\lim_{t\downarrow0}
 \frac{F(\nu+t\delta_{\bm y})-F(\nu)}{t}
\]
which is simply the right directional derivative of output.\footnote{This coincides
with the G\^ateaux derivative in direction $\delta_{\bm y}$ when the
two-sided derivative is defined. Here we use a right derivative because subtracting mass at $\bm y$ can make the measure $\nu$ negative (which is undefined).}
A firm $f$ that values machines, $D_{\bm y^M}F(\nu_f)>0$, bids
\[
 b_f(\bm y):=r\frac{D_{\bm y}F(\nu_f)}{D_{\bm y^M}F(\nu_f)}.
\]
The intuition is the usual marginal reasoning: one more unit of the worker lets the firm remove
$D_{\bm y}F/D_{\bm y^M}F$ machines without changing output. The bid is
the resulting saving in machine costs at the rental rate $r > 0$. In fact, under \cref{ass:level-scale} these are affine in capability: 
\[
 b_f(\bm y)= r+ \bm\beta^f\cdot(\bm y-\bm y^M),
 \tag{BID}\label{eq:affine-bid}
\]
where the firm's prices for productive capability in domain $d$ and coordination are
\[
 \beta_d^f=\frac{rS(\nu_f)\partial_dL}
                 {m_fD_{\bm y^M}F(\nu_f)} \quad \text{and} \quad 
 \beta_*^f=\frac{rL(\nu_f)\partial_2S}
                 {m_fD_{\bm y^M}F(\nu_f)}.
\]
These prices tell us which employer values a worker's strengths most.

Each firm's bid $b_f(\bm y)$ tells us how much the specific firm is willing to pay for a human worker of a given capability $\bm{y}$.  Call a firm \emph{eligible} for human type $i$ if it has positive scale,
uses machines, and $\nu_f^H(\{\bm y^i\})<\mu_i$ i.e., it does not already employ the entirety of type $i$. 
The \emph{bid envelope} for a given human type $i$ is the upper envelope of what every firm is willing to pay for $i$: 
\[
 \mathcal{E}_i(\sigma,\bm y^M)
 :=\underline w\vee
 \operatorname*{ess\,sup}_{\substack{f\text{ eligible}\\\text{for type }i}}
 b_f(\bm y^i),
\]
where the essential supremum is taken with respect to $\sigma$. 
If no eligible firm is present, set the envelope equal to $\underline w$.
Write $\mathcal{E} := (\mathcal{E}_i)_{i = 1,\ldots, K}$ for the vector of these envelopes.

\begin{theorem}[Wages and sorting]
\label{thm:wage-envelope}\label{thm:wage-sorting}
\phantom{}
\begin{enumerate}[(i)]
 \item \textbf{Wage determination.}
For any equilibrium outcome $(\sigma,w)$, 
\[
 \mathcal{E}(\sigma,\bm y^M)\le w.
\]
Equality holds for type $i$ if some of that type is unemployed, in which
case $w_i=\underline w$, or an active interior firm $f$ employs that type,
in which case $w_i=b_f(\bm y^i)$.

 \item \textbf{Sorting.}
Fix an equilibrium $(\sigma,w)$. If \cref{ass:level-scale} holds and an active interior firm $f$ employs type $i$, then $\sigma$-almost every firm $g$ that is
eligible for $i$ satisfies
\[
 b_f(\bm y^i)-b_g(\bm y^i)
 =(\bm\beta^f-\bm\beta^g)\cdot(\bm y^i-\bm y^M)\ge0.
 \tag{FAN}\label{eq:sorting-fan}
\]
With finitely many active machine-using firms, the capability profiles at which each firm bids highest form polyhedral cones with common apex \(\bm y^M\). Workers with interior matches at these firms sort according to these regions.
\end{enumerate}
\end{theorem}

\begin{figure}[H]
 \centering
 \includegraphics[width=\textwidth]{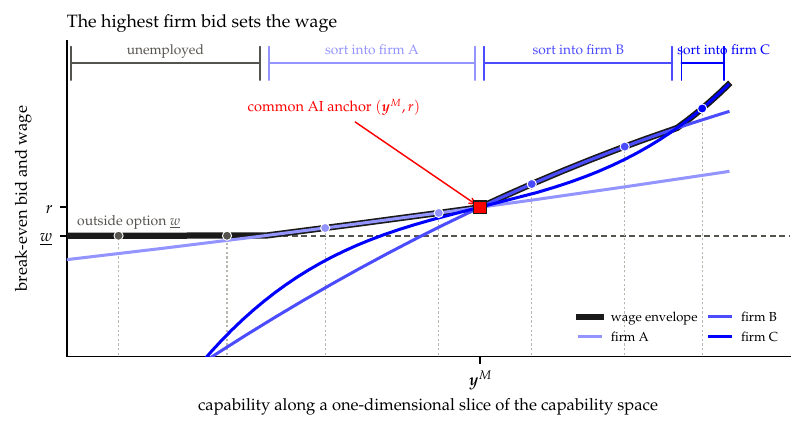}
 \caption{Firms' bids for humans.}
 \label{fig:wage-envelope}
\end{figure}

\Cref{fig:wage-envelope} illustrates \cref{thm:wage-sorting}.  

The black curve gives the highest firm bid for each worker type (depicted on the $x$ axis), with the outside option as a floor.
The blue curves show how firms differ in what they will pay for the same
worker, and the brackets show where each firm bids highest.
At an interior match, a firm can substitute workers for machines in either
direction, which pins the wage to its bid.
All bids pass through the common AI point $(\bm y^M,r)$: a worker with the
AI's capabilities saves exactly the cost of one machine.

\begin{figure}[H]
 \centering
 \includegraphics[width=\textwidth]{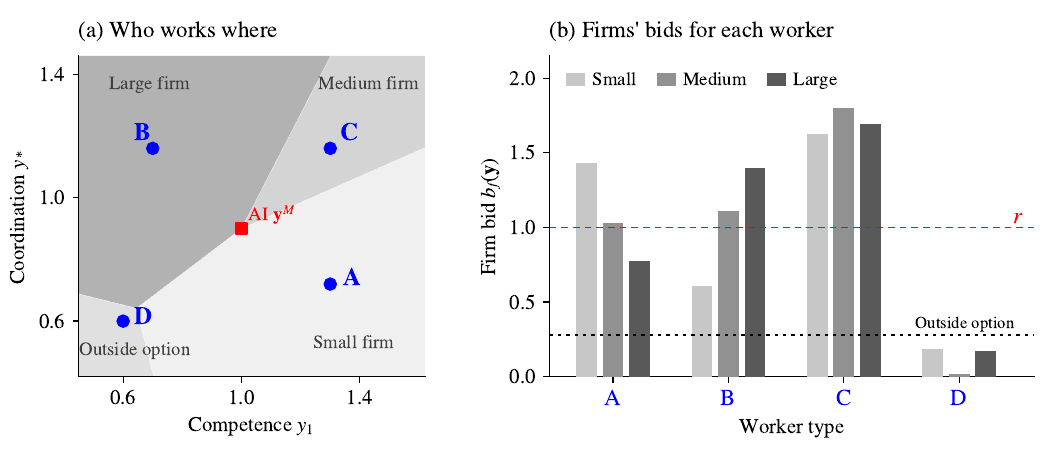}
 \caption{Comparative advantage sorts humans.}
 \label{fig:sorting-fan}
\end{figure}

Part (ii) of \cref{thm:wage-envelope} links sorting to workers' strengths
relative to AI. In \cref{fig:sorting-fan}(a), worker A has stronger
competence but weaker coordination than AI; worker B has the opposite
advantages. Worker C lies in the medium firm's region, while D lies in
the outside-option region. Panel (b) shows why: the small, large, and
medium firms bid most for A, B, and C, respectively, while every bid for
D falls below the outside option. At an interior match, the highest
eligible bid also pins down the worker's wage. Workers A and B can earn
more than the machine rental rate despite being weaker than AI in one
capability.

\begin{figure}[H]
 \centering
 \includegraphics[width=\textwidth]{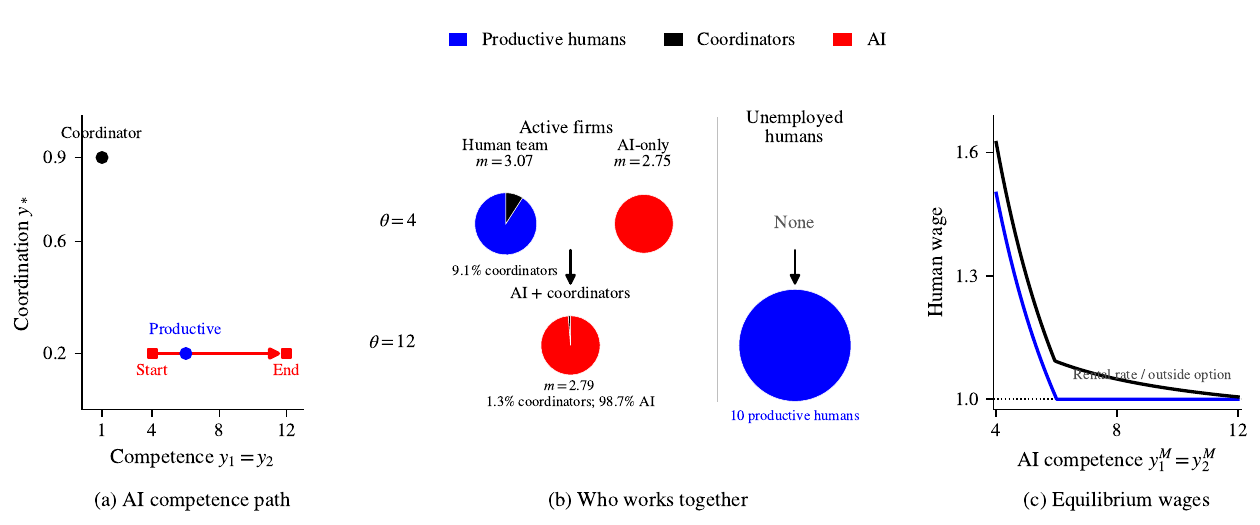}
 \caption{Human coordinators work with increasingly productive AI}
 \label{fig:v4-wage-sorting}
\end{figure}

\phantomsection\label{sec:wage-regimes}
Humans can earn more than a `supercompent' AI---that far exceeds humans in every domain $d = 1,2,\ldots n$---if their coordination advantage is valuable enough. By \eqref{eq:affine-bid}, an eligible firm bids above the machine rental price $r$ exactly when its weighted sum of human capability advantages is positive \citep[cf.][]{agrawalGansGoldfarb2024turing,agrawalGansGoldfarb2025genius}. 
\Cref{fig:v4-wage-sorting} shows human coordinators moving from human teams into AI teams as AI competence rises, because their coordination advantage raises the productivity of the AI workforce.\footnote{Productive
humans have profile $(6,6,.2)$ and supply $10$; coordinators have
$(1,1,.9)$ and supply $1$. AI follows $(\theta,\theta,.2)$ for
$\theta\in[4,12]$. Set $\rho=-1$, $\phi=3$, $r=\underline w=1$,
and $P(Q)=2(1-Q/3000)_+$.}
Productive humans leave employment, while coordinators remain employed
above their outside wage; both human wages fall along this plotted path.


\subsection{The dynamics of unemployment}
\label{sec:human-employment}
\cref{sec:composition} uses the cost-adjusted best average-product
envelope $\bar a$ to determine which firm compositions operate. We now use the same envelope to study wages and employment as AI capabilities improve within our model. Suppose there is a single human type with capability $\bm y^H$ and mass $\mu > 0$. A firm's human
share $\lambda\in[0,1]$ determines its unit composition
$\nu_\lambda:=\lambda\delta_{\bm y^H}+(1-\lambda)\delta_{\bm y^M}$.
The cost-adjusted envelope then satisfies $\overline a(\lambda,w)
 =\frac{\overline a(\lambda)}{1+\lambda(w/r-1)}$. Under the conditions of \cref{prop:human-diversity}, active firms
maximize this envelope. The first-order condition implies
\[
 \frac wr
 =\frac{\overline a(\lambda)+(1-\lambda)\overline a'(\lambda)}
 {\overline a(\lambda)-\lambda\overline a'(\lambda)}.
\]
Relative wages therefore equal relative marginal products after firms choose their preferred size. In the special case in which $\bar a$ is convex and human-only and AI-only firms are both active at their preferred scales, zero profit instead equates their cost-adjusted best average products:
$\frac{\overline a(1)}{w}=\frac{\overline a(0)}{r} \implies 
 \frac wr=\frac{\overline a(1)}{\overline a(0)}$. 

Indeed in the CES environment, we can compute this envelope explicitly.
For each domain $d\in\{1,2,*\}$, mean capability at human share $\lambda$ is
$\overline y_d(\lambda):=\lambda y_d^H+(1-\lambda)y_d^M$.
If $1-\phi<\overline y_*(\lambda)<1$, preferred size is
$m^*(\lambda)=\phi/[1-\overline y_*(\lambda)]-1$, and substitution gives
\[
 \overline a(\lambda)
 =\underbrace{\left(\frac{\overline y_1(\lambda)^\rho+
                    \overline y_2(\lambda)^\rho}{2}\right)^{1/\rho}}
 _{\text{productive competence}}
 \underbrace{e^{\phi-[1-\overline y_*(\lambda)]}
 \left(\frac{1-\overline y_*(\lambda)}{\phi}\right)^{1-\overline y_*(\lambda)}}
 _{\text{gains from choosing firm size}},
\]
noting that as usual, at $\rho=0$, the first factor
is $\sqrt{\overline y_1(\lambda)\overline y_2(\lambda)}$ and 
if $\overline y_*(\lambda)\le1-\phi$, preferred size is zero and the scale
factor is one.\footnote{For positive productive competence, the scale
factor has supremum $e^\phi$ at $\overline y_*(\lambda)=1$, attained only
in the limit as size grows. Best average product is unbounded when
$\overline y_*(\lambda)>1$.}

The curvature of $\overline a$ with respect to $\lambda$ reflects two forces. First, mixing humans and AI can balance their productive strengths, but with diminishing gains as the composition becomes more balanced. Composition also changes mean coordination and hence preferred firm size. When the more productive type also coordinates better, increasing its share raises both competence and the gains from scale; this reinforcement can make \(\overline a\) convex. Our key observation is that the curvature of $\overline a$ determines not only the composition of firms at fixed levels of human and AI capabilities (as we have seen from \cref{prop:human-diversity}), but also how wages and unemployment adjust as AI improves. 
\begin{proposition}[Gradual vs sudden displacement]
\label{prop:exit}
\phantom{}
\begin{enumerate}[(i)]
 \item \textbf{Gradual replacement.} Suppose $\overline a$ is strictly concave along a continuous AI capability path $\bm y^M(\theta)$ and assume all
 firm compositions have feasible preferred scales.\footnotemark{} Suppose
 these scales are unique and share a finite upper bound.
 Then equilibrium output, human employment, firms' human share, and firm
 size change continuously. The economy exhibits full unemployment if and only if
 \begin{equation*}
  \left[1+\frac{\overline a'_+(0)}{\overline a(0)}\right]
  \le \frac{\underline w}{r}.
 \end{equation*}
 \item \textbf{Sudden replacement.} Suppose $\overline a$ is strictly convex along a continuous AI capability path $\bm y^M(\theta)$, and the pure
 human and pure machine mixes have feasible preferred scales at equilibrium
 output. If machine-only firms are active in every
 equilibrium, the economy exhibits full employment when
 \[\frac{\overline a(0)}{r}< \frac{\overline a(1)}{\underline w}
 \]
 and full unemployment when the inequality is reversed. Wages are $\max\{\underline w,\ r\overline a(1)/\overline a(0)\}$.
\end{enumerate}
\end{proposition}
\footnotetext{In part (i), preferred scales are feasible at the output that
maximizes surplus when we omit individual firms' staffing and market constraints.}

\begin{figure}[H]
 \centering
 \includegraphics[width=.95\textwidth]{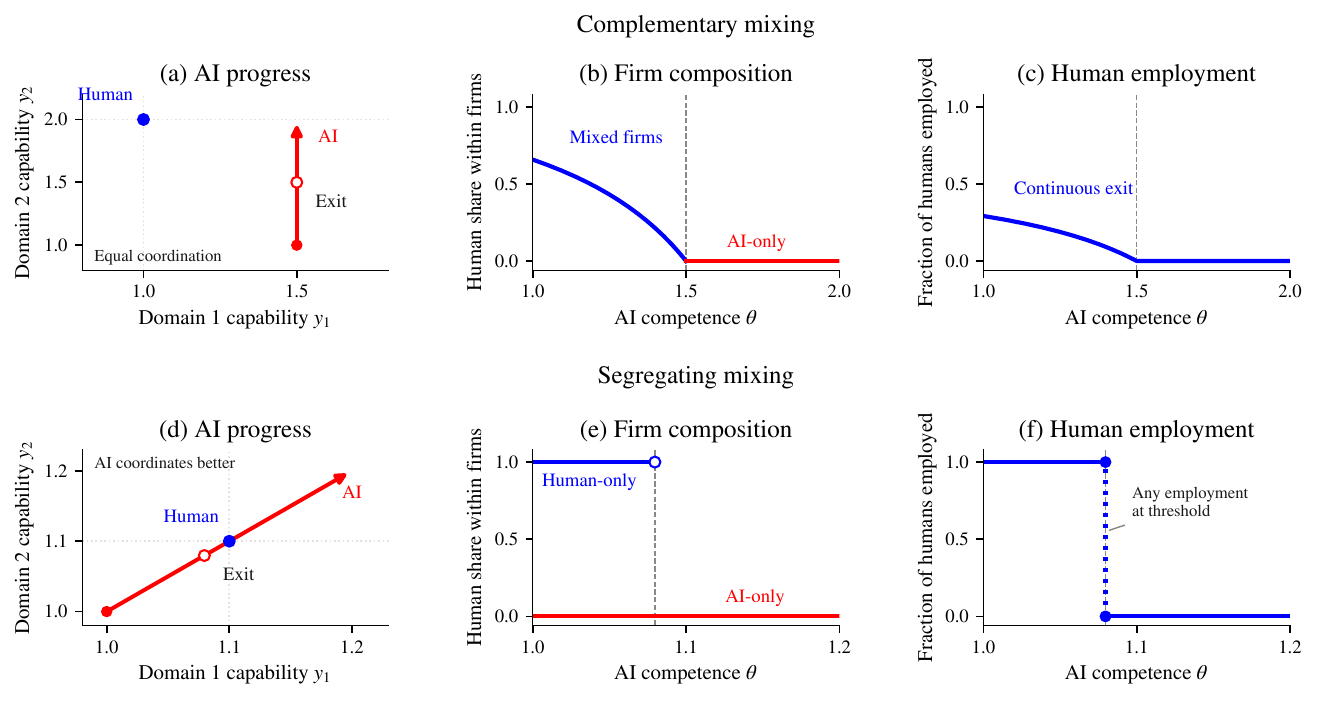}
 \caption{AI progress, firm composition, and human exit.}
 \label{fig:human-exit}
\end{figure}

In part (i), displacement is \emph{within-firm}: the left side of the condition in \cref{prop:exit}(i) is the value of adding
a human to an all-machine industry at the price supported by machine
production. When the LHS is smaller than the RHS, then a machine-only firm does not benefit from employing any humans even at the reservation wage. In part (ii), displacement is \emph{across-firms}: in an economy that is already segregated---firms are machine-only or human-only---changes in underlying AI capabilities can discontinuously change the set of equilibrium industry structures, thereby wiping out human-only firms. \Cref{fig:human-exit} displays these two cases.\footnote{Both rows use $\rho=-1$ and $r=\underline w=1$. Above, humans have profile $(1,2,.9)$ and supply $20$, AI follows $(1.5,\theta,.9)$ for $\theta\in[1,2]$, $\phi=.3$, and $P(Q)=(2-Q/10)_+$. Below, humans have $(1.1,1.1,.985)$ and supply $2/3$, AI follows $(\theta,\theta,.9999)$ for $\theta\in[1,1.2]$, $\phi=.02$, and $P(Q)=(2-Q/1000)_+$.} In the first case, firms reduce their human share continuously; in the second, the industry replaces human firms once AI production becomes
cheaper.

\begin{figure}[H]
\begin{minipage}[c]{.53\textwidth}
More broadly, the curvature of $\bar a$ can transition from concave to
convex as AI capabilities improve. In such settings, we might first see a gradual increase in unemployment (part (i) of \cref{prop:exit}) before a sudden spike (part (ii) of \cref{prop:exit}). \Cref{fig:ces-envelope-sequence} illustrates the change in curvature as AI coordination improves.\footnotemark{}
\end{minipage}\hfill
\begin{minipage}[c]{.45\textwidth}
 \centering
 \includegraphics[width=\linewidth]{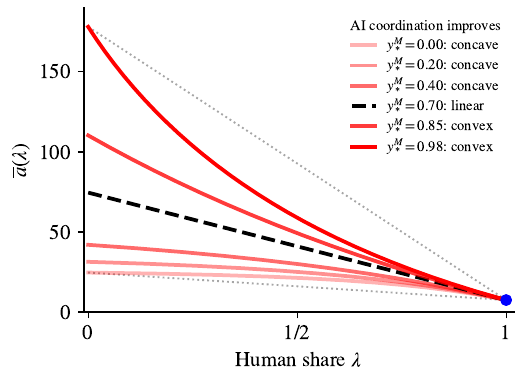}
 \caption{$\bm y^M$ and curvature of $\bar a$.}
 \label{fig:ces-envelope-sequence}
\end{minipage}
\end{figure}
\footnotetext{Humans have profile $(1,1,.7)$ and AI has $(10,10,y_*^M)$,
with $\phi=3$ and $\rho=-1$. Only AI coordination changes. Dotted lines
join the endpoints of the first and last curves.}

\subsection{The reorganization effect}
\label{sec:wage-movement}
AI progress changes wages both directly and through changes in firms.
We call the second channel the \emph{reorganization effect}. A firm can
change its size, its machine share, and the mix of human skills it employs.
The following decomposition separates these channels along a smooth path
for a worker and her employer.

\begin{proposition}[Wage effects combine technology and changes in firms]
\label{prop:wage-decomposition} Suppose \cref{ass:level-scale} holds and let AI capabilities $\bm y^M(\theta)$ be parametrized by $\theta$, and let $\overline{\bm y}^{\,H}$ denote the mean capability profile
of the firm's human workers. Consider a continuously differentiable firm selection whose firms are
interior and employ human type $i$. Assume the firm's bid is continuously
differentiable in its size, machine share, mean human capability profile,
and $\theta$.
Then
\[
 \frac{\de w_i}{\de\theta}
 =\underbrace{\partial_\theta b_f(\bm y^i)}_{\text{technology}}
 +\underbrace{
   \overbrace{\partial_m b_f(\bm y^i)\frac{\de m}{\de\theta}}
     ^{\text{firm size}}
   +\overbrace{\partial_s b_f(\bm y^i)\frac{\de s}{\de\theta}}
     ^{\text{machine share}}
   +\overbrace{\nabla_{\overline{\bm y}^{\,H}}b_f(\bm y^i)
       \cdot\frac{\de\overline{\bm y}^{\,H}}{\de\theta}}
     ^{\text{mix of human skills}}
  }_{\text{reorganization}}.
\]
\end{proposition}

The technology term changes AI capabilities while holding human and machine
inputs fixed. The remaining terms let the employer adjust those inputs.
Equilibrium maximizes total surplus, not each worker's wage, so the
envelope theorem for surplus does not remove the reorganization terms
from wage changes. The reorganization effect can therefore change wages at first
order and alter the division of surplus.

\begin{figure}[H]
 \centering
 \includegraphics[width=\textwidth]{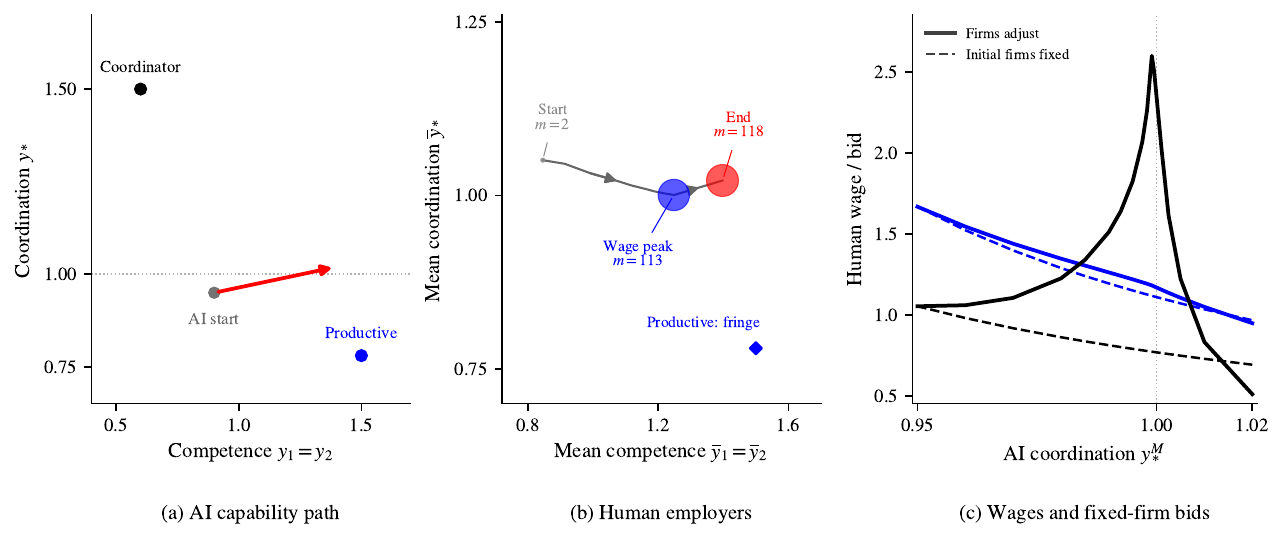}
 \caption{Wage changes with/without reorganization.}
 \label{fig:ces-local-response}
 \label{fig:re-amplification}
\end{figure}

This is illustrated quite starkly in \cref{fig:re-amplification} in which (endogenoously) larger firms initially raise the value of
human coordination, before further AI progress lowers it (solid black line in panel (c)).\footnote{Human
profiles are $(1.5,1.5,.78)$ and $(.6,.6,1.5)$, each with supply $.4$;
AI moves from $(.9,.9,.95)$ to $(1.4,1.4,1.02)$. Set $\rho=-1$,
$\phi=.02$, $r=1$, $\underline w=.28$, and $P(Q)=1.64(1-Q/300)_+$.} By contrast, holding the composition of firms constant generates monotonically decreasing wages for the human coordinator (dotted black line in panel (c)).

\begin{figure}[H]
 \centering
 \includegraphics[width=\textwidth]{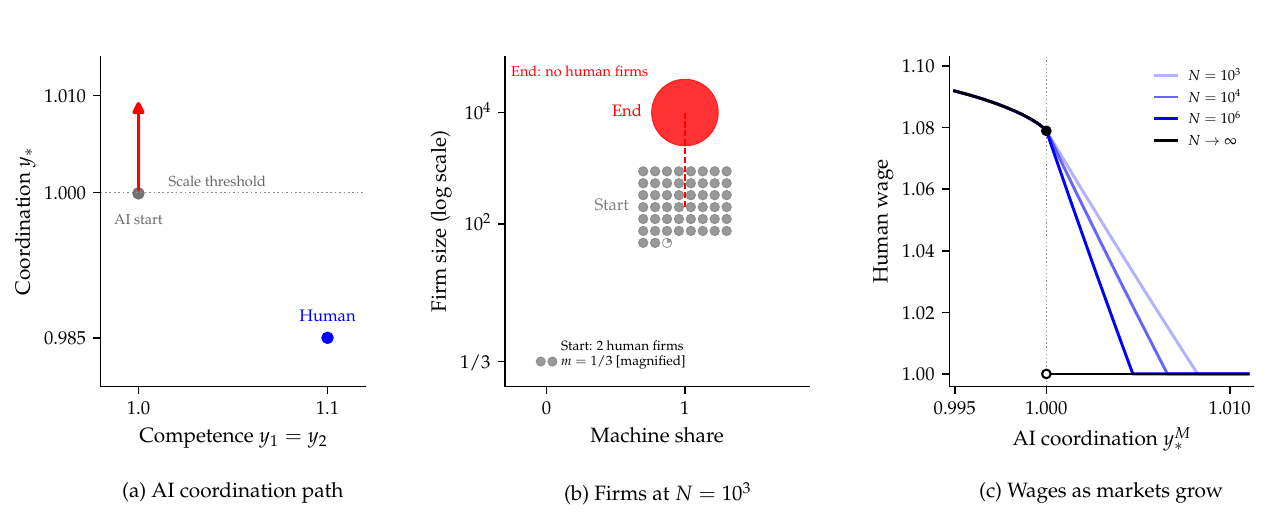}
 \caption{Firm concentration and the loss of the human wage premium.}
 \label{fig:ces-support-jump}
\end{figure}

Reorganization can also have drastic impact on wages by (abruptly) changing which firms survive in equilibrium. In \cref{fig:ces-support-jump}(a), AI improves at coordination while humans
retain their task-skill advantage. Below the scale threshold, small human
firms survive alongside AI firms. Once AI coordination exceeds one, AI
output per worker can grow without bound with firm size. Adding humans
improves task skills but reduces coordination, which becomes more costly
in larger firms. In this example, sufficiently large AI firms therefore
undercut human production and human employers disappear
(\cref{fig:ces-support-jump}(b)). Human wages fall to the outside option,
and larger markets make the decline sharper
(\cref{fig:ces-support-jump}(c)). The jump on the black curve describes
the large-market limit; it does not establish a jump in a finite market.\footnote{The
calibration is in footnote~\ref{fn:concentration-calibration}.
Panel (b) uses $N=10^3$, with grey at $y_*^M=.9999$ and red at $1.010$;
panel (c) uses $N=10^3,10^4,10^6$. Circle area measures firm size except
for the magnified human firms of size $1/3$; full circles represent one
unit of activity and sectors represent fractional units.
\Cref{app:jump-details} verifies the limiting curve.
The finite-market curves use pure-firm cost calculations with entry
checks at selected states; see \cref{app:concentration-numerics}.}

Wages depend on how surviving firms reorganize and on which employers
remain active. Even a small employer can sustain a wage premium, so its
disappearance can matter more for wages than its share of output suggests.

\section{Connection to theory and empirics} \label{sec:connection} We have offered a framework for thinking about endogenous firm formation, worker sorting, surplus, and wages. We now explain how it connects to influential models in the existing literature on automation and wages. Then, we briefly discuss some early empirical evidence in line with our model's predictions, as well as outline avenues for future empirical work. 

\subsection{Nesting the task model}
\label{sec:task-connection}

Optimal task assignment under constant returns recovers the task economy.
Scale effects let AI teams cut costs without improving at individual tasks.

\textbf{Equivalence to task model.}
Human type $i$ has mass $\mu_i$, reservation wage $\underline w$, and task
productivities $y_d^i$. Machines have task productivities $y_d^M$ and unlimited supply at rental
cost $r$. Agents divide time across tasks
\citep{acemogluRestrepo2018,aghionJonesJones2019}, whose outputs enter $L(\bm z)=(n^{-1}\sum_{d=1}^n z_d^\rho)^{1/\rho}$, with
$\rho<1$ and the geometric-mean limit at $\rho=0$. Demand remains $P_N$.

The \emph{task technology} assigns the firm's human and machine inputs
$\nu$ optimally across tasks, with task allocations $\nu_1,\ldots,\nu_n$:
\[
 F^{\mathrm T}(\nu)
 =\max_{\substack{\nu_d\in\M_+(\Y)\\\sum_d\nu_d=\nu}}
 L\left(\int_\Y y_1\,\de\nu_1,\ldots,
         \int_\Y y_n\,\de\nu_n\right).
\]
Our technology is $F^{\mathrm T}$. 
With task productivities bounded away from zero, $F^{\mathrm T}$ satisfies
\cref{ass:technology} and has constant returns: $F^{\mathrm T}(m\nu)=mF^{\mathrm T}(\nu)$.
Average product is therefore independent of scale, and the tail exponent
is $\gamma\equiv1$.

\begin{theorem}[Outcome equivalence to task model]
\label{thm:task-main-equivalence}
Under technology $F^{\mathrm T}$, every
equilibrium industry $\sigma$ produces the task economy's unique competitive
output $Q_N^*$: $Q(\sigma)=Q_N^*$.
The competitive task allocations are exactly the aggregates of equilibrium
industries with optimal task assignments. If some equilibrium uses machines,
the two economies also have the same set of equilibrium wage vectors.
\end{theorem}

\begin{figure}[H]
 \centering
 \includegraphics[width=0.9\textwidth]{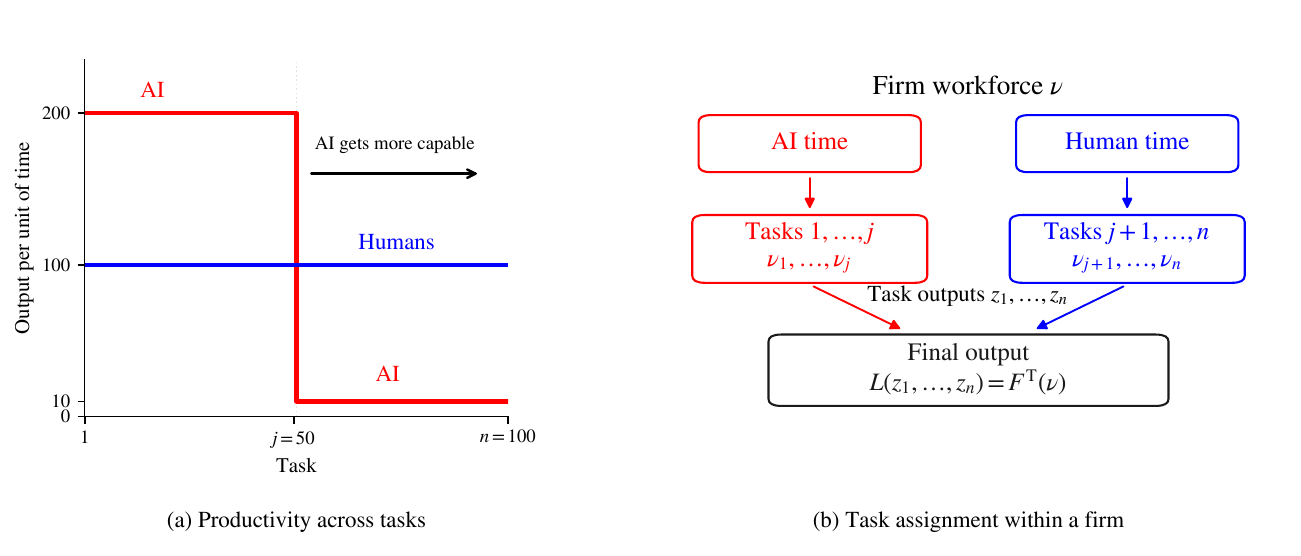}
 \caption{Task productivity and assignment within a firm.}
 \label{fig:task-assignment}
\end{figure}

In a constant returns world, the boundaries of the firm do not matter for the economy's efficiency since pooling firms lets workers and machines keep their old task assignments, so it cannot lower total output; conversely, we can split an optimal assignment into smaller copies without losing output. As usual, when machines are put to work, their fixed rental rate $r > 0$ anchors task prices. Workers then earn the value of their best task, subject to the outside option. \Cref{fig:task-assignment} illustrates how workers and machines divide
 their time across tasks before task outputs enter the final good.\footnote{
 Humans have productivity $100$ on each of $n=100$ tasks. AI has productivity
 $200$ on tasks $1,\ldots,j$ and $10$ on the rest, with $j=50$.
 For a common human wage $w$ and machine rental cost $r$, the shown
 specialization minimizes costs when $r/2<w<10r$: AI performs the first
 group and humans the second. Agents divide their time across tasks;
 they do not supply their full productivity to every task at once.}

\textbf{What do firms add to task wages?}
We now allow coordination to affect scale: output is
$F(\nu)=F^{\mathrm T}(\nu)S(m,\overline y_*)/m$, where $m$ is firm size,
$\overline y_*$ is mean coordination, and $S$ is the CES scale function.
At a positive preferred size, define the firm's \emph{coordination price}
\[\kappa_f:=\cfrac{F^{\mathrm T}(\nu_f)}{m_f} \cdot \log(1+m_f)\]
which is the task output per agent times the scale value of mean coordination.

\begin{proposition}[Firms value task skills and coordination]
\label{prop:task-rent}
If an active firm $f$ values machines and has a positive preferred size
$m_f\in m^*(\overline y_*)$, then its bid for capability profile $\bm y$ is
\begin{equation*}
 b_f(\bm y)=r\,
 \frac{D_{\bm y}F^{\mathrm T}(\nu_f)+\overbrace{\kappa_f(y_*-\overline y_*)}^{\substack{\text{human coordination}\\\text{contribution}}}}
 {D_{\bm y^M}F^{\mathrm T}(\nu_f)+\underbrace{\kappa_f(y_*^M-\overline y_*)}_{\substack{\text{machine coordination}\\\text{contribution}}}}.
 \tag{TASK-BID}\label{eq:task-bid}
\end{equation*}
If the firm is interior and employs human type $i$, then
$w_i=b_f(\bm y^i)$. With constant returns, $S/m\equiv1$, so coordination does not affect output and the bid reduces to the task-model bid.
\end{proposition}

For a human capability profile $\bm y$, the numerator combines the worker's task
contribution with the value added through coordination; the denominator
does the same for a machine. A human who improves
coordination raises the firm's bid, while a machine that improves
coordination raises lowers the bid. As in the task model, it is the \emph{relative} price of humans vs machines that matters so humans can remain employed even when AI is more productive on every task. 

\begin{figure}[H]
 \centering
 \includegraphics[width=\textwidth]{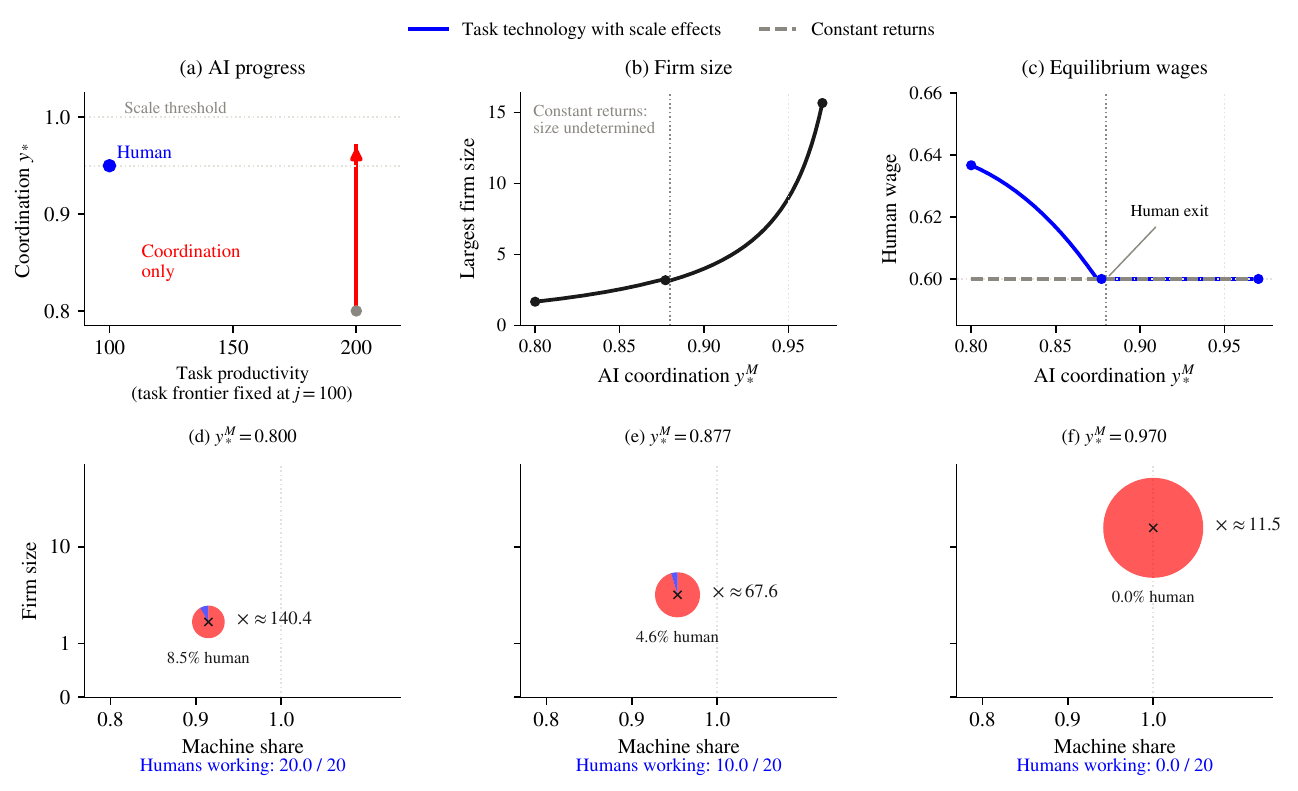}
 \caption{Humans as coordinators}
 \label{fig:task-rent-coordination}
\end{figure}

We can use \cref{prop:task-rent} to pin down the threshold at which humans become displaced. Suppose there is one human type (that can perform all tasks at fixed productivity).
Under standard regularity conditions,\footnote{We use the conditions in
\cref{prop:exit}(i): maximal output per agent is strictly concave in the
human share, and all mixes have unique, uniformly bounded preferred
scales that fit human supply and the market at the relaxed surplus
optimum. The appendix verifies these conditions in
\cref{fig:task-rent-coordination} while AI coordination is below human
coordination, including at human exit.}
humans are absent from every equilibrium exactly when
\begin{equation*}
 \underbrace{\max_d\frac{y_d^H}{y_d^M}}_{\substack{\text{best relative}\\\text{task productivity}}}
 +\underbrace{(y_*^H-y_*^M)\log\bigl(1+m^*(y_*^M)\bigr)}_{\substack{\text{human coordination contribution}\\\text{at the machine-only size}}}
 \le\underbrace{\frac{\underline w}{r}}_{\substack{\text{relative}\\\text{outside option}}}.
\end{equation*}
The first term is the human's best relative task productivity; the second is the
human coordination contribution at the preferred machine-only size.
A coordination advantage can cover a shortfall in task productivity.
With constant returns the second term vanishes.

\Cref{fig:task-rent-coordination} holds AI task productivity at twice the
human level on every task. Humans earn a premium through their coordination
advantage; the premium falls as AI closes that gap.\footnote{We use $n=100$
tasks, human productivity $100$ on every task, human coordination
$y_*^H=.95$, human supply $\mu=20$, $r=1$, $\underline w=.6$, $\phi=.5$,
and $P(Q)=(3-Q/200)_+$. AI productivity is $200$ on every task and
coordination rises from $.8$ to $.97$. Each pie represents one firm:
area measures firm size, blue shows humans, and red shows AI.
The multiplier gives the number of firms.}

\begin{figure}[H]
 \centering
 \includegraphics[width=\textwidth]{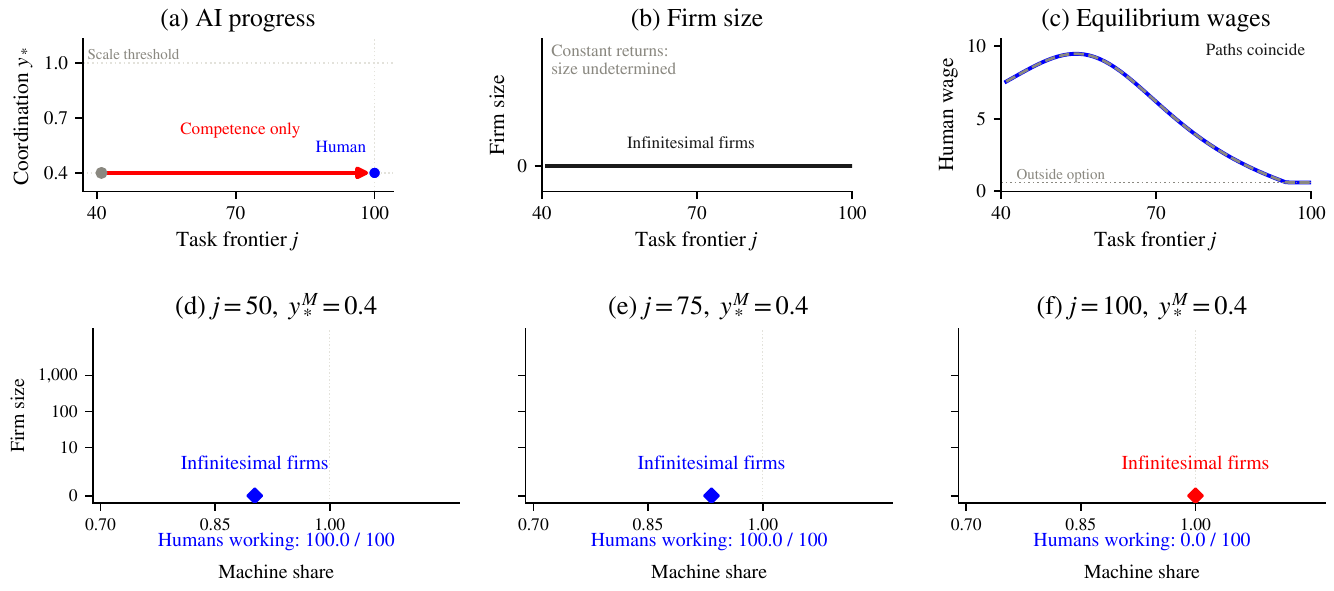}
 \caption{Improvement in AI competence (automating tasks).}
 \label{fig:task-competence-paired}
 \label{fig:task-wage-paths}
 \label{fig:task-composition-grid}
\end{figure}

\begin{figure}[H]
 \centering
 \includegraphics[width=\textwidth]{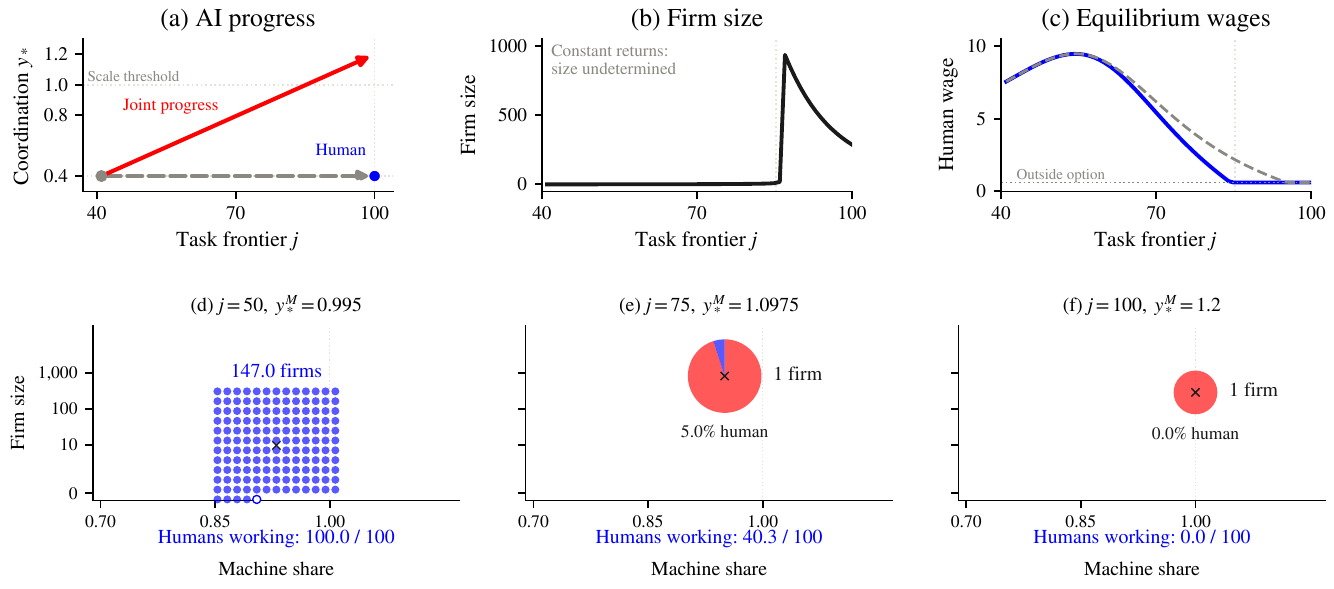}
 \caption{Improvement in AI competence and coordination.}
 \label{fig:task-joint-paired}
 \label{fig:task-diagonal-composition}
\end{figure}

\begin{figure}[H]
 \centering
 \includegraphics[width=\textwidth]{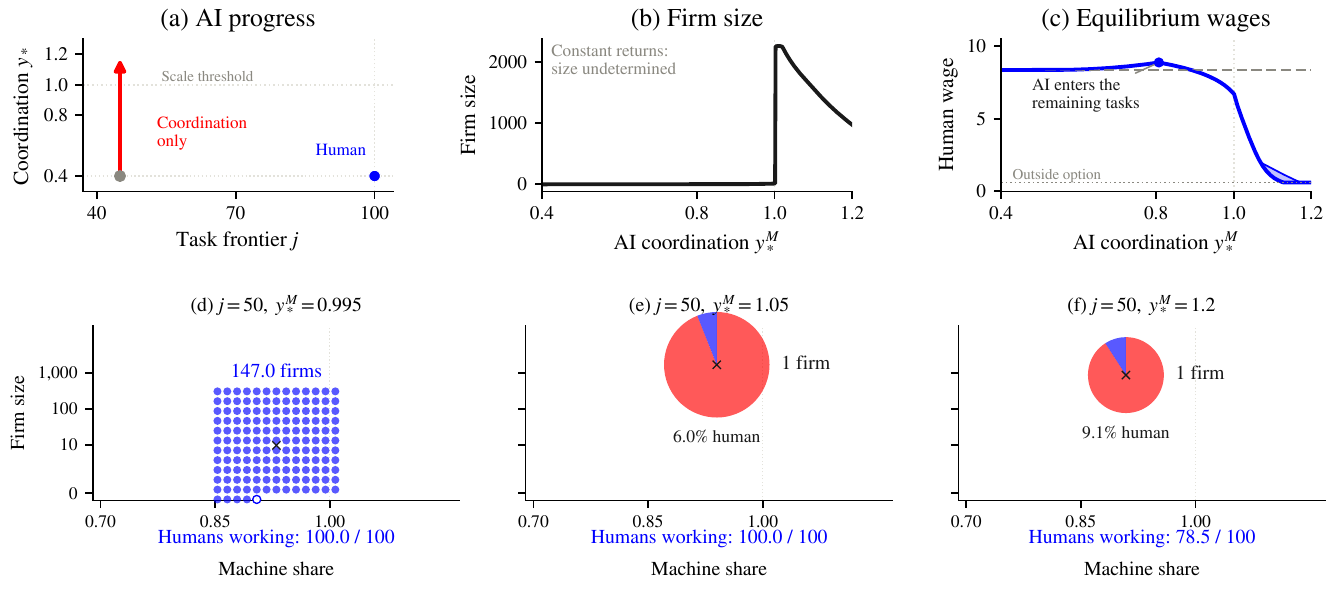}
 \caption{Improvement in AI coordination.}
 \label{fig:task-coordination-paired}
 \label{fig:task-coordination-composition}
\end{figure}

\textbf{Non-monotonic firm sizes.} Adding firm sizes to the task model also generates rich dynamics on how (endogenous) firm size interacts with the direction of AI progress.
\Cref{fig:task-competence-paired,fig:task-joint-paired,fig:task-coordination-paired}
compare AI improvements in different directions.
\Cref{fig:task-competence-paired} illustrates improvements in only competence \`a la \cite{acemogluRestrepo2018,korinekSuh2024}. Because of constant returns, firm sizes are indeterminate (here we show them as atomless) and equilibrium wages have the familiar hump shape where first the productivity effect dominates before the displacement effect dominates. 

With scale effects, however, AI improvements in coordination drive firm sizes to spike when its ability to coordinate $\gamma^M$ crosses $1$ (cf \cref{prop:market-size}). This is illustrated by \cref{fig:task-joint-paired,fig:task-coordination-paired}. But as AI coordination continues to improve, the firm sizes start to fall because AIs that coordinate better allow firms produce more with fewer agents, while the downward-sloping demand curve limits the expansion of sales. Hence, if output per agent grows faster than sales, the workforce falls even as output rises which generates this non-monotonicity.

\subsection{Early empirical evidence}
\label{sec:evidence}

Evidence to date suggests that current, publicly available AI systems have jagged domain competence---often exceeding human experts---alongside subcritical coordination capabilities. Evidence for the former primarily comes from benchmarks pitting humans versus AI models on identical tasks. Across 44 occupations on OpenAI's GDPval benchmark, frontier models tie or beat seasoned professionals on 48--71\% of bounded deliverables \citep{patwardhan2025gdpval,openai2025gpt52}, while commercial legal tools outperform practicing lawyers on overall research accuracy (80\% vs.\ 71\%) yet trail attorneys by 9 percentage points on tasks requiring contextual judgment \citep{valsai2025legal}. Similarly, on Mercor's expert evaluations across banking, consulting, law, and medicine, models reach 67\% on one-shot ``closed tasks'' \citep{vidgen2025apex} but drop to 24\% on multi-step workflows \citep{vidgen2026apexagents}, and in research engineering models score $4\times$ higher than human experts over 2-hour sprints but half as much over 32--48 hour projects \citep{wijk2024rebench,starace2025paperbench}. On coordination, when multiple AI agents work together, output scales sublinearly with team size ($\gamma^M \in [0.48, 0.68] < 1$; \citealt{ord2026swarm}),\footnote{We emphasize, however, that this probably understates the economic value of AI teams since a key advantage of using swarms is to speed up the task. This estimate of $\gamma^M$ does not account for such speedups which are economically valuable.} and uncoordinated swarms amplify errors up to $17.2\times$ and fail on 41--87\% of workflows \citep{cemri2025mast,kim2025scalingagents,yang2026diversity}. AI-run stores degrade in performance over long horizons unless paired with substantial scaffolding \citep{backlund2025vendingbench,anthropic2025projectvend,anthropic2025projectvend2}. 

This evidence can be interpreted through our framework as predicting that we will see the rise of smaller firms, and existing firms will adjust primarily through workforce composition rather than scale. Data is consistent with these patterns: AI-native startups employ 25\% fewer people overall and roughly 15\% lower shares of entry-level workers and managers, and run flatter hierarchies than same-cohort peers \citep{kimKoning2026,ewensGiroud2025hierarchy}. \citet{hoffmann2025nature} find that AI-enabled developers work in smaller groups, individuals assisted by AI match the output quality of two-person human teams \citep{dellacqua2025cybernetic}, and others show that AI-adopting firms seem to reallocate labor away from routine roles without shrinking overall employment \citep{hampole2025ai,baslandze2026productivity}.

The labor market response to AI has thus far been muted. Chatbot adoption has produced minimal aggregate changes in earnings or hours \citep{yotzov2026firmdata}, with only 3--7\% of time savings passing into pay \citep{humlum2025llm}, though notably this evidence is from before the adoption of agentic AI. But there is substantial heterogeneity in these aggregate estimates. Junior employment in exposed occupations has fallen 7.7--19\% through reduced hiring and starting wages have dropped 4.2\%, while senior employment and pay remain stable if not rising \citep{brynjolfsson2026canaries,hosseini2025seniority,azar2026wage}. Under the affine bid schedule (\cref{thm:wage-envelope}), entry-level workers map closest to the capabilities that machines can match, which leads their wages to be compressed first. Senior workers retain coordination advantages and see both demand and wages increase. Notably,  even when AI raises early-career productivity most, increased machine competence narrows $\bm y^i - \bm y^M$ which leads to lower wages (\cref{prop:wage-decomposition}). 

In line with our model, disruption is most stark  on online freelance platforms where coordination is least likely to play a role. Here, job volume and earnings fell with the introduction of AI while surviving listings shifted toward complex work \citep{hui2024shortterm,demirci2025replacing}. This is consistent with our results on   across-firm replacement of solo human producers once machine unit cost falls below theirs (\cref{prop:exit}).

\section{Conclusion}
If economic transformation does, in fact, depend on how well AI systems coordinate with humans and each other, this can guide measurement. Benchmarks primarily track individual task competence. While these might be good measures of AI model capabilities, they can be misleading signals of economic transformation, because they do not capture
how capabilities combine and scale within organizations. It misleads about the present, because organizations are (endogenously) comprised of complementary humans so a model might beat every individual on average but still lose the market. And it can also mislead about the future, because an economy near its tipping point can be reorganized by a coordination improvement too small to register on any benchmark, while an economy far from it can absorb dramatic gains without substantial reorganization of firms. 

Indeed, we should be collecting additional data. First, benchmarks should measure how output scales with the number of identical AI copies at fixed compute per copy, alongside isolated competence $L$. Estimating competence and coordination capabilities across model generations while holding total compute fixed would provide a useful signal for the path of economic transformation. Second, firm-level datasets should measure total organizational scale $m_f = \nu_f^H(\Y) + \nu_f^M(\Y)$ which consists of both the worker rolls and the rented compute for running agentic instances. We should track firms' expenditures on their human payroll relative to their AI expenditure over time to understand how their composition is changing with advancing AI capabilities. As we have seen, \emph{within-firm} versus \emph{across-firm} displacement of human workers can have vastly different unemployment dynamics so we should also track the cross-sectional distribution of firms in the economy to understand whether markets are becoming dominated by AI-only or AI-native firms \citep{kimKoning2026}. Finally, we should track employer-employee panels that can directly test how workers with different skill profiles sort across firms. This would isolate the technology effect from the reorganization effect, and help to identify when sectors may be reorganizing around AI as the central driver of productivity, and which workers are most affected.

\begingroup
\setstretch{1}
\renewcommand{\bibfont}{\footnotesize}
\setlength{\bibsep}{0pt}
\bibliographystyle{abbrvnat}
\bibliography{ref}
\endgroup

\clearpage

\appendix
\setstretch{1.05}

\begin{center}
\large{\textbf{APPENDIX}}
\end{center}

\begingroup
\small
\setstretch{1}
\setlength{\parskip}{0pt}
\newcommand{\appendixentry}[1]{%
  \noindent\hyperref[#1]{\textbf{\ref*{#1}\quad\nameref*{#1}}}%
  \nobreak\dotfill\pageref{#1}\par}
\newcommand{\appendixsubentry}[1]{%
  \noindent\hspace*{1em}\hyperref[#1]{\ref*{#1}\quad\nameref*{#1}}%
  \nobreak\dotfill\pageref{#1}\par}
\appendixentry{app:welfare-proofs}
\appendixsubentry{app:proof-equilibrium}
\appendixsubentry{app:proof-industry}
\appendixsubentry{app:proof-wages}
\appendixsubentry{app:proof-task-model}
\appendixsubentry{app:proof-continuity}
\appendixentry{app:replacement-game}
\appendixentry{app:task-model}
\appendixentry{app:numerics}
\appendixentry{app:example-calc}
\endgroup

\section{Omitted Proofs}
\label[appendix]{app:welfare-proofs}
We collect the proofs in the order of the main results.

\subsection{Equilibrium and efficiency}
\label{app:proof-equilibrium}
For the proofs, let
$c_{\min}:=\min\{r,\underline w\}$. Fix an
atomistic machine-only firm $\bar f^M\in\F^0$ with the lowest unit cost, and let
\[
 \bar c:=\frac{C(\bar f^M)}{q_{\bar f^M}}.
\]
Unlimited replication of this atomistic firm gives
$\mathcal C(Q)\le\bar cQ$ for every $Q$.
\subsubsection{Industry cost and duality}
\label{app:cost-duality}
\paragraph{Exact output.}
Given a feasible industry with output $x>Q>0$, we first shrink each positive-scale firm with $q_f>Q$ to the first scale on its composition ray that produces
$Q$, ensuring that the complete-firm feasibility constraint remains satisfied. Continuity and $F(0)=0$ ensure existence and a measurable choice; other firms stay unchanged. Since $q_f\le x$ and $\min\{q_f,Q\}\ge(Q/x)q_f$, the industry still produces at least $Q$. We can then proportionally reduce activity weights to obtain exactly $Q$. Both steps preserve feasibility and weakly reduce resource use, with a strict reduction in total cost. Thus, allowing production above $Q$ cannot lower the minimum cost relative to requiring exactly $Q$ units.

\paragraph{A compact activity set.}
Represent a recipe by its scale $m\ge0$ and a unit-normalized firm $f\in\F^0$. Read $F(m\nu_f)/m$ as $L(\nu_f)$ at zero. The labelled unit compositions in $\F^0$ form a compact set; with one machine profile, this set is a simplex. Replacing any recipe with resource unit cost above $\bar c$ by atomistic
machines lowers cost and human use. We can therefore restrict attention to
\[
 \mathcal A_Q:=
 \left\{(m,f)\in[0,\bar cQ/c_{\min}]\times\F^0:
 \begin{array}{l}
  m\nu_f^H\le\mu,\quad F(m\nu_f)\le Q,\\
  C(f)\le\bar c\,F(m\nu_f)/m
 \end{array}\right\}.
\]
This set is compact by \cref{ass:technology}(i), as the inequalities define a closed subset of a compact one. Its cone of aggregate output, human inputs, and cost is
\[
 \mathcal K_Q:=
 \operatorname{cone}\left\{
 \left(\frac{F(m\nu_f)}m,
       \bigl(\nu_f^H(\{\bm y^i\})\bigr)_{i=1}^K,C(f)\right):
 (m,f)\in\mathcal A_Q\right\}.
\]
A coefficient $t$ on a generator means total input $t$: it gives firm
weight $t/m$ at $m>0$, or atomistic intensity $t$ at $m=0$.
Every generator has cost at least $c_{\min}$. Bounded aggregate cost
therefore bounds the total coefficient. Compactness of the generating
set then makes the cone closed and ensures that industry integrals belong
to it, by approximation with finite sums.

\begin{lemma}[Industry cost and supporting prices]\label{lem:industry-cost}
For every output $Q>0$:
\begin{enumerate}[(i)]
 \item Minimum cost $\mathcal C(Q)$ is positive and is attained at exact
 output by an industry with at most $K+3$ firm types.
 \item Every cost minimizer $\sigma_Q$ has a nonempty, convex, compact
 support set $\Lambda(Q,\sigma_Q)$. This set is the same for all cost
 minimizers at $Q$. Every supported pair satisfies
 \[
  p\le\bar c,\qquad
  pQ-\sum_{i=1}^K(w_i-\underline w)\mu_i=\mathcal C(Q).
 \]
 Conversely, a feasible exact-output industry with a supporting pair
 minimizes cost.
 \item Minimum cost $\mathcal C$ is continuous on $\R_+$.
\end{enumerate}
\end{lemma}

\begin{proof}
We prove attainment, obtain the supporting prices by duality, and then
establish continuity.

\paragraph{Step 1: attainment.}
The activity representation gives
\[
 \mathcal C(Q)=
 \min\{z:(x,H,z)\in\mathcal K_Q,\ x\ge Q,\ H_i\le\mu_i\ \forall i\},
\]
where $z$ is total industry cost and $H_i$ is aggregate employment of type $i$. Scaling all coefficients down makes output exact without changing the firm menu. The feasible cost bound $\bar cQ$ and the coefficient bound above give a compact minimizing set. Bounded output per unit input and
positive input costs make its minimum positive. Carath\'eodory's theorem gives a minimizer with at most $K+3$ generators, each of which implements
a firm type.

\paragraph{Step 2: duality.}
Scale a relative-interior point of $\mathcal K_Q$ until all human inputs
are strictly below supply, then add enough atomistic machine output
to exceed $Q$. The result stays in the relative interior and satisfies
the inequality constraints strictly. Conic strong duality therefore applies.
With output multiplier $p\ge0$ and human multipliers $w_i-\underline w\ge0$,
the Lagrangian is
\[
 pQ-\sum_i(w_i-\underline w)\mu_i+
 z-px+\sum_i(w_i-\underline w)H_i.
\]
Its infimum over the cone is finite exactly when firm support holds on $\mathcal A_Q$. Atomistic machine entry then gives $p\le\bar c$. Since every feasible recipe with resource unit cost at most $\bar c$ belongs to $\mathcal A_Q$, this also rules out profitable entry by discarded recipes, as wages are no lower than reservation costs.

Strong duality yields the displayed identity, while complementary slackness implies that active firms break even and human types earn scarcity premiums only when their supply binds. These conditions hold for every primal and dual optimum. Conversely, integrating firm support shows that any feasible industry producing exactly $Q$, with active firms breaking even and human scarcity satisfied, minimizes resource cost. This proves the characterization in (ii).

Finally, the bounds $0\le(w_i-\underline w)\mu_i\le\bar cQ$, together with closedness of the dual inequalities and equality, give compactness; convexity follows from their linearity.

\paragraph{Step 3: continuity.}
For $Q_n>Q$, add atomistic machine output to a minimizer at $Q$,
at cost $\bar c(Q_n-Q)$. For $Q_n<Q$, the exact-output construction above
produces $Q_n$ at no greater cost. These comparisons give upper
semicontinuity. For the reverse inequality, choose minimizers at $Q_n$
with at most $K+3$ generators and pad the lists with zero coefficients.
Their scales and total inputs are uniformly bounded. Along a convergent
subsequence, continuity of $F(m\nu)/m$, including at zero, preserves
output, individual and aggregate human feasibility, and the market cap.
The limiting generators implement an industry at $Q$ with the limiting
cost. Hence $\mathcal C(Q)\le\liminf_n\mathcal C(Q_n)$.
At zero, $0\le\mathcal C(Q)\le\bar cQ$ proves continuity.
\end{proof}

\subsubsection{Implementation and equilibrium}
\begin{proof}[Competitive implementation in \cref{app:replacement-game}]
A producer can increase the intensity of a profitable recipe without bound.
An optimal action therefore requires firm support, and only zero-profit
recipes can operate. Human labor choice gives exactly human scarcity.

Conversely, let $(p,w)\in\Lambda(Q,\sigma)$. The firm space is standard
Borel, so assign recipes measurably on $[0,1/2]$ with distribution
$\sigma/\sigma(\F)$ under normalized Lebesgue measure. Give each assigned
producer intensity $2\sigma(\F)$ and leave the rest inactive. The resulting
industry is $\sigma$. All actions are optimal by support and active
equality; employment clears and is optimal by human scarcity.
\end{proof}
\begin{proof}[Proof of \cref{thm:equilibrium-efficiency}]
We show that a stable contract must be efficient, support every efficient
industry, and then prove existence. Write $H_i(\sigma)$ for employment of
human type $i$.

\paragraph{Step 1: stability implies efficiency.}
An equilibrium contract has $(p,w)\in\Lambda(Q(\sigma),\sigma)$ by
competitive implementation. Its organizer receives zero: otherwise an
inactive organizer can copy it and gain strictly. The accounting identity
in \cref{app:replacement-game} makes the protected parties' total payoff
equal to $\W_N(\sigma)$.

If a feasible industry gives higher surplus, replace it by a cost minimizer
at its output. \Cref{lem:industry-cost} supplies supporting prices, so the
replacement admits a competitive continuation. Choose transfers that give
each protected party its old payoff and give the remaining surplus to the
new organizer. The accounting identity makes these transfers balance.
The offer is admissible and blocks the original contract. Hence every
stable industry is efficient.

\paragraph{Step 2: efficiency implies stability.}
Let $\sigma$ be efficient with $(p,w)\in\Lambda(Q(\sigma),\sigma)$. Set
\[
 t_i=-(w_i-\underline w)H_i(\sigma),\qquad
 t_0=\sum_i(w_i-\underline w)H_i(\sigma),\qquad t_E=0.
\]
These transfers give each human coalition its outside payoff and give
the consumer representative all surplus. Machine viability makes that
surplus positive. The contract is admissible by competitive implementation.
A blocking offer would preserve these payoffs and pay its organizer
strictly more than zero, so its total surplus would exceed the optimum.
Thus the contract is stable.

\paragraph{Step 3: existence and shadow values.}
Efficiency is equivalent to minimizing cost at a quantity in
$\mathcal Q_N^*$: either a cost saving at fixed output or a better quantity
raises surplus. Continuity of $U_N-\mathcal C$ on the compact demand
interval gives a maximizing quantity. Machine viability makes it positive.
\Cref{lem:industry-cost} supplies an exact-output minimizer and supporting
prices there; Step 2 gives equilibrium. The same lemma identifies $p$
as the output multiplier and $w_i-\underline w$ as the multiplier on
aggregate human supply, with the complete-firm menu held fixed.
\end{proof}

\subsubsection{Competitive support}
\label{app:proof-competitive-support}
\begin{proof}[Competitive equilibrium implies equilibrium of our model]
We verify price support, then use consumer optimization to prove efficiency.

\paragraph{Step 1: supporting prices.}
Fix a CE $(\sigma,p,w)$ whose industry satisfies our complete-firm
conditions, and write $Q=Q(\sigma)$ for its output and $H_i(\sigma)$ for
type $i$'s employment. Unrestricted activity intensities imply that every
recipe earns weakly negative profit and active recipes earn zero profit.
The maintained wage floor gives $w_i\ge\underline w$; worker optimization
and clearing give $(w_i-\underline w)[\mu_i-H_i(\sigma)]=0$.
Thus $(p,w)\in\Lambda(Q,\sigma)$ once we verify that $Q>0$.
At $Q=0$, consumer optimization requires $p\ge P_N(0)$, but the standing
machine-viability condition then makes an atomistic machine-only recipe
profitable. Hence $Q>0$. Since all inputs have positive prices, $p=0$
would imply zero production. Therefore $p>0$, and consumer optimization
gives $p=P_N(Q)$.

\paragraph{Step 2: efficiency.}
For any alternative industry $\widehat\sigma\in\Sigma$, let
$\widehat Q=Q(\widehat\sigma)$ denote its output. Integrating the CE profit
inequalities over this industry and subtracting active equality gives
\[
 \int_\F C(f)\,\de\widehat\sigma(f)
 -\int_\F C(f)\,\de\sigma(f)
 \ge p(\widehat Q-Q)
 -\sum_i(w_i-\underline w)
       [H_i(\widehat\sigma)-H_i(\sigma)].
\]
Consumer optimization gives
$U_N(\widehat Q)-U_N(Q)\le p(\widehat Q-Q)$. Consequently,
\[
 \W_N(\widehat\sigma)-\W_N(\sigma)
 \le\sum_i(w_i-\underline w)
       [H_i(\widehat\sigma)-H_i(\sigma)]\le0.
\]
The final inequality follows from human feasibility and complementary
slackness: each positive wage premium requires $H_i(\sigma)=\mu_i$.
Thus $\sigma$ is efficient. Together with $(p,w)\in\Lambda(Q,\sigma)$,
\cref{thm:equilibrium-efficiency} implies that $(\sigma,w)$ is an
equilibrium of our model.

\end{proof}

\subsection{Industry structure}
\label{app:proof-industry}
\subsubsection{Firm scale and the primitive bounds}
\label{app:firm-size-bounds}
\begin{proof}[Proof of \cref{thm:firm-scale}]
We first separate preferred scale from competence and order it by
coordination. We then compare an active firm with feasible copies of its
composition. These comparisons yield the bounds.

\paragraph{Step 1: preferred scale.}
Under \cref{ass:level-scale}, for a unit composition $\nu$ with mean
coordination $z_*$, average product
is $L(\nu)S(m,z_*)/m$. The positive factor $L(\nu)$ does not affect the
maximizing scale set. At fixed shares of each type, a competence change
leaves $z_*$ and this set unchanged. The extension of $S(m,z_*)/m$ at zero
is one, by the definition of $L$.

For two coordination levels $z_*'>z_*$, define their log output ratio at
positive scale by
\[
 D(m):=\log S(m,z_*')-\log S(m,z_*)
      =\int_{z_*}^{z_*'}\partial_2\log S(m,u)\,\de u.
\]
The larger-firms condition makes $D$ strictly increasing in $m$; its
continuous extension at zero is zero. If a preferred scale at $z_*'$
were smaller than a preferred scale at $z_*$, adding their optimality
inequalities would make $D$ weakly larger at the smaller scale, a
contradiction. This proves the preferred-size comparison in the main
text, including zero preferred sizes. The comparison does not require unique preferred scales or imply strictly
positive scale responses.

\paragraph{Step 2: active scales and constraints.}
Fix an equilibrium and supporting prices $(p,w)$. For an active firm
with unit composition $\nu$, positive input prices and active equality
give $p>0$. Firm support gives
\[
 p\,\frac{F(m\nu)}m
 \le r\nu^M(\Y)+\int_{\Y^H}w\,\de\nu^H
\]
for every feasible copy, with equality at the active scale. At zero scale,
read the ratio as $L(\nu)$. Since $p>0$ and input cost per worker is
constant at a fixed mix, the active scale maximizes feasible average
product. This proves the size-choice formula in \cref{sec:firm-size}.

Every smaller copy can be staffed. If it produced more than the active
firm, continuity along the ray would give a still smaller copy producing
exactly $q_f$ at a strictly lower bill, contrary to support. Thus all smaller
copies fit in the market. Their average products cannot exceed the active
firm's. If a finite preferred scale lies below the active scale, the active
scale must therefore also be preferred. Thus an active firm with a finite
preferred scale has size no larger than some preferred scale of its mix.
If any finite preferred scale is feasible, the same support inequality
puts its average product below the active firm's, so the active scale is
also preferred. This includes preferred or active scale zero.

Under strict quasiconcavity, average product rises strictly below all
preferred scales. If no finite maximum exists, it rises strictly at every
finite scale: a weak decline between two scales would, by strict
quasiconcavity, force all later values below the earlier value and yield
a finite maximum by continuity. An atomistic firm below preferred
scale can be enlarged slightly while preserving staffing and market fit.
A positive-scale firm with all constraints slack also admits a slightly
larger feasible copy. Both cases contradict firm support. A firm below
all preferred scales must therefore have positive scale and be capped.
This establishes the constraint implications stated in the main text.

\paragraph{Step 3: bounds across firms.}
Under \cref{ass:level-scale}, for an active positive-scale firm with unit
composition $\nu$, the scale comparison in Step 2
and staffing feasibility give
\[
 m_f\le
 \sup m^*\!\left(\int_\Y y_*\,\de\nu\right)
 \wedge\sup\{m:m\nu^H\le\mu\}
 \le\overline M.
\]
When no finite preferred scale exists, the first term is infinite by
convention, and staffing feasibility still gives the bound. Atomistic
firms satisfy the upper bound at size zero.

For the lower bound, a firm at or above some preferred scale has size at
least $\underline M$. Under strict quasiconcavity, every other active firm
is positive-scale and capped by Step 2. The slack market constraint
makes its size equal to its staffing limit, which is again at least
$\underline M$. Thus the bounds hold almost everywhere in each equilibrium
under the stated conditions.

For the interpretation in the text, Step 1 bounds every composition's
smallest preferred scale below by the worst coordinator's smallest
preferred scale. Every staffing limit is at least the smallest human
supply. A pure composition attains the smaller of these two bounds.
The AI-only composition gives its largest preferred scale as a lower
bound on $\overline M$; the best coordinator's largest preferred scale
bounds it above. This completes the proof.
\end{proof}

\paragraph{Coordination and the tail.}
Under \cref{ass:level-scale}, the tail exponent depends only on mean
coordination and is nondecreasing in it. A composition whose tail exponent
is below one has a nonempty bounded set of preferred scales. If its
exponent exceeds one, no finite scale is preferred. We make no such claim
at exponent one.

To verify these claims, the uniform power tail gives
\[
 S(m,z_*)=\frac{A(\nu)}{L(\nu)}m^{\gamma(\nu)}[1+o(1)]
\]
for every unit composition $\nu$ with mean coordination $z_*$. The left
side depends on composition only through $z_*$, so its exponent does too.
If a higher coordination level had a smaller exponent, the log output
ratio $D(m)$ in Step 1 would tend to $-\infty$. Its increasing shape rules
out this case. Below exponent one, average product tends to zero, so
continuity gives a nonempty bounded maximizing set that may include
zero. Above exponent one, average product is unbounded, so no finite
maximum exists.

\subsubsection{Competence and coordination}
\begin{proposition}[Better AI lowers cost and raises surplus]
\label{prop:industry-uhc}
If the AI improves from $\bm y^M$ to
$\widetilde{\bm y}^M\ge\bm y^M$, then minimum resource cost weakly falls
at every output quantity, and equilibrium total surplus weakly rises at
every market size.
\end{proposition}
\begin{proof}[Proof of \cref{prop:industry-uhc}]
We reproduce the old output at lower cost. Take any feasible industry with
output $Q>0$. For each positive-scale firm $f$, let $\widetilde\nu_f$ describe the
firm after we upgrade its machines without changing any input
quantities. The upgraded firm has the same size and dominates the old
one, so equal-mass monotonicity gives $F(\widetilde\nu_f)\ge q_f$.
Continuity and $F(0)=0$ give a scale factor $t_f\in(0,1]$ at which the
replacement firm $\widetilde f$ satisfies
\[
 F(t_f\widetilde\nu_f)=q_f,\qquad
 C(\widetilde f)=t_f C(f)\le C(f).
\]
Choose the first scale factor that attains $q_f$; continuity makes this a
measurable choice. Keep each firm's activity weight unchanged. Its output
still fits within $Q$, and its human use weakly falls, both individually and
in aggregate. For each atomistic activity, the upgrade weakly raises
output per input; reduce its activity weight to preserve its old output.
Thus the upgraded economy can reproduce $Q$ at weakly lower cost. Taking
the infimum proves the cost comparison; at $Q=0$, both costs are zero.
The gross benefit of each quantity is unchanged, so maximized surplus
weakly rises. \Cref{thm:equilibrium-efficiency} gives the equilibrium claim.
\end{proof}

\phantomsection\label{app:surplus-envelope}
Write $\W_N^*$ for equilibrium total surplus and
$\overline{\bm y}_f$ for firm $f$'s mean productive capabilities.

If \cref{ass:level-scale} holds, $y_d^M$ is interior to its feasible range, $\W_N^*$ is differentiable in $y_d^M$, and
the market constraint is slack for every firm
$f\in\supp(\sigma|_{\F^+})$ in an equilibrium industry $\sigma$, then
\begin{equation}
 \frac{\de\W_N^*}{\de y_d^M}
 =P_N(Q)\int_\F
   \underbrace{q_f}_{\text{firm output}}\;
   \underbrace{\frac{\nu_f^M(\Y)}{m_f}}_{\text{machine share}}\;
   \underbrace{\frac{\partial_dL(\overline{\bm y}_f)}
                       {L(\overline{\bm y}_f)}}
              _{\text{value of domain }d}
   \,\de\sigma(f).
 \tag{SURPLUS}\label{eq:competence-surplus}
\end{equation}

\begin{proof}[Derivation of the surplus formula \eqref{eq:competence-surplus}]
We show that the fixed organization stays feasible locally, then
differentiate its surplus.

\paragraph{Step 1: local feasibility.}
Fix supporting prices and put $Q=Q(\sigma)$. Active equality gives
$c_{\min}m_f\le pq_f\le\bar cQ$.
Normalize the supported firms and include their zero-scale limits,
as in \cref{app:cost-duality}. The resulting closure is compact.
Every positive-scale point satisfies $q_f<Q$ by hypothesis, while
complete-firm output tends uniformly to zero at zero scale.
Hence the complete-firm capacities have a uniform margin below $Q$.

At fixed inputs and weights, a competence change $\varepsilon$ gives
\[
 q_f(\varepsilon)=q_f
 \frac{L(\overline{\bm y}_f+
       \varepsilon[\nu_f^M(\Y)/m_f]\bm e_d)}
      {L(\overline{\bm y}_f)},
\]
where $\bm e_d$ is the domain-$d$ unit vector. The formula also holds for
atomistic activities. The ratio converges uniformly to one, so total
output converges to $Q$, and complete-firm outputs change uniformly on
the compact set. The market margin and unchanged human inputs make the
fixed organization feasible for small changes of either sign.

\paragraph{Step 2: the derivative.}
Let $G(\varepsilon)$ be its surplus. Optimality gives
$\W_N^*(y_d^M+\varepsilon)\ge G(\varepsilon)$ with equality at zero.
Differentiability therefore equates their derivatives.
Input costs stay fixed. On the compact set of means, $L$ is bounded away
from zero and $\partial_dL$ is bounded, so the output derivative is
bounded by a constant times $q_f$. Since $\int q_f\,\de\sigma=Q$,
we can differentiate under the integral:
\[
 G'(0)=P_N(Q)\int_\F q_f\,
 \frac{\nu_f^M(\Y)}{m_f}
 \frac{\partial_dL(\overline{\bm y}_f)}{L(\overline{\bm y}_f)}
 \,\de\sigma(f).
\]
This is the stated formula.
\end{proof}

\phantomsection\label{app:competence-composition}
With one human type, a firm with
machine share $s$ has unit composition
$\nu_s=(1-s)\delta_{\bm y^H}+s\delta_{\bm y^M}$ and mean productive
capabilities $\overline y_d=(1-s)y_d^H+sy_d^M$ in each domain $d$.
At a fixed human wage $w$, define its log output per unit of input cost,
after choosing scale, by
\[
 g(s,y_d^M)
 :=\log\left[\max_{m\ge0}\frac{F(m\nu_s)}{m}\right]
   -\log\bigl((1-s)w+sr\bigr).
\]

If \cref{ass:level-scale} holds, consider a continuous selection of active
interior firms with machine shares $s$ as $y_d^M$ varies. Suppose the human
wage stays fixed, nearby compositions have feasible preferred scales at
the respective equilibrium output, and $g$ is twice continuously
differentiable with $g_{ss}<0$ along the selection. Then
\begin{equation}
 \frac{\de s}{\de y_d^M}
 =\frac{
   \overbrace{\displaystyle\frac{\partial}{\partial s}
   \left[s\,\frac{\partial_dL(\overline{\bm y})}
                    {L(\overline{\bm y})}\right]}
   ^{\substack{\text{how much the upgrade favors}\\
               \text{a higher machine share}}}
 }{
   \underbrace{\displaystyle-g_{ss}}
   _{\substack{\text{curvature at the}\\
               \text{preferred mix}}}
 }.
 \tag{MIX}\label{eq:competence-composition}
\end{equation}

The term in brackets is the proportional productivity gain at a fixed mix.
Since $g_{ss}<0$, the machine share rises exactly when mixes with more
machines gain proportionally more. More machines expose a firm more to the
upgrade, but also change which capabilities it lacks. The assumptions do
not sign the net effect. Fixed wages matter: changes in wages would also
change relative input costs.

\begin{proof}[Derivation of the composition formula \eqref{eq:competence-composition}]
Firm support and active equality make the selected share a local maximizer
of $g$: its preferred scale and those at nearby shares are feasible, and
its output per unit cost equals $1/p$. Thus $g_s=0$. At fixed share,
coordination and input costs do not change with productive capability, so
\[
 g_{y_d^M}=s\,\frac{\partial_dL(\overline{\bm y})}{L(\overline{\bm y})}.
\]
Since $g\in C^2$ and $g_{ss}<0$, the implicit function theorem gives
$\de s/\de y_d^M=-g_{s y_d^M}/g_{ss}$ along the continuous selection.
Substitution proves the formula.
\end{proof}
\begin{proof}[Proof of \cref{prop:coordination-size}]
We identify the active shares, sign their response to coordination, and
then differentiate preferred size.

\paragraph{Step 1: equilibrium firms maximize output per unit cost.}
At machine share $s\in[0,1]$, mean coordination is
$\overline y_*=(1-s)y_*^H+sy_*^M$. The CES scale factor gives
\[
 \partial_m\log[S(m,\overline y_*)/m]
 =\frac{\phi-(1-\overline y_*)(1+m)}{(1+m)^2}.
\]
Since $\overline y_*<1$, the assumed positive preferred scale is unique
and equals $m^*(\overline y_*)=\phi/(1-\overline y_*)-1$.
The assumed positivity holds at every share along the path.
Feasibility and the size-choice argument in \cref{app:firm-size-bounds} put every active firm
at its preferred scale and exclude atomistic production.

Slack aggregate human supply gives $w=\underline w$ by human scarcity.
Firm support and active equality in \cref{thm:equilibrium-efficiency}
then imply that every active share maximizes output per unit input cost:
all preferred-scale alternatives can be staffed and fit in the market.
Using the earlier notation $g$ for its logarithm, we have
\[
 \begin{split}
 g(s,y_*^M)
 ={}&\log L(\overline{\bm y})+\phi-(1-\overline y_*)
 +(1-\overline y_*)\log\frac{1-\overline y_*}{\phi}\\
 &-\log\bigl((1-s)\underline w+sr\bigr),
 \end{split}
\]
where $\overline{\bm y}$ denotes mean productive capabilities.
At a fixed share, these capabilities and input costs do not change with
AI coordination.

\paragraph{Step 2: the share response.}
Differentiating gives
\[
 g_{s y_*^M}
 =\log\frac{\phi}{1-\overline y_*}
   +\frac{s(y_*^M-y_*^H)}{1-\overline y_*}.
\]
If $y_*^M\ge y_*^H$, this derivative is positive: preferred size is
positive, and the second term is nonnegative. Let $s$ and $\widetilde s$
maximize $g$ before and after a coordination gain within this region.
If $\widetilde s<s$, integration gives
\[
 g(s,\widetilde y_*^M)-g(\widetilde s,\widetilde y_*^M)
 >g(s,y_*^M)-g(\widetilde s,y_*^M)\ge0,
\]
contrary to optimality of $\widetilde s$. Thus $\widetilde s\ge s$,
including at corners. Mean coordination and preferred size also weakly
increase. This argument needs neither uniqueness nor continuity.

If $y_*^M<y_*^H$, the mixed derivative can have either sign. At an
interior optimum with $g_{ss}<0$, the implicit function theorem gives
\[
 \frac{\de s}{\de y_*^M}=-\frac{g_{s y_*^M}}{g_{ss}}.
\]
Since $\log[\phi/(1-\overline y_*)]=\log(1+m^*)$, this proves the
three sign cases in part (ii).

\paragraph{Step 3: the size response.}
Differentiate $m^*=\phi/(1-\overline y_*)-1$ along the selected firm:
\[
 \frac{\de m}{\de y_*^M}
 =\frac{\phi}{(1-\overline y_*)^2}
   \left[s+(y_*^M-y_*^H)\frac{\de s}{\de y_*^M}\right].
\]
The first term is nonnegative. The sign of the second follows from
Step 2 and the coordination gap, which gives the stated amplification
and dampening cases.
\end{proof}

\subsubsection{Large markets}
\label{app:proof-market-size}
\paragraph{Equilibrium output grows with the market.}
\label{app:large-output}
Write $Q_N^*=Q(\sigma_N^*)$. Machine viability gives $\bar c<P(0)$.
Choose a fixed quantity $z>0$ with $P(z)>\bar c$. Atomistic machines
can produce $Nz$ at cost $\bar cNz$, so efficiency gives
\[
 P(0)Q_N^*\ge\W_N^*
 \ge N\int_0^z[P(u)-\bar c]\,\de u.
\]
The integral is strictly positive. Thus $Q_N^*$ grows at least linearly
with $N$, for every value of $\gamma^M$, including one.

\begin{lemma}[Subcritical firms with bounded unit cost have bounded size]
\label{lem:subcritical-coercivity}
If $\gamma^M<1$, then, for every finite $B$, feasible positive-scale
firms with $C(f)/q_f\le B$ have uniformly bounded sizes and capacities.
\end{lemma}
\begin{proof}
As feasible firm size $m_f$ diverges, its human share is at most
$\mu(\Y^H)/m_f$, so its composition converges to the machine profile.
Choose $\delta>0$ with $\gamma^M+\delta<1$.
Continuity of the exponent and the uniform power tail imply
$q_f/m_f=O(m_f^{\gamma^M+\delta-1})\to0$ uniformly.
Since $C(f)\ge c_{\min}m_f$, unit cost then diverges.
A finite unit-cost bound therefore bounds size, and continuity of output
on firms of bounded size bounds capacity.
\end{proof}

\begin{lemma}[Supercritical concentration]
\label{lem:supercritical-concentration}
Suppose $\gamma^M>1$.
For any sequence $Q_\ell\to\infty$ and any sequence of least-cost
industries $\sigma_\ell$ that produce $Q_\ell$,
\[
 \int_{\F^+}
 \left(\frac{q_f}{Q_\ell}\right)^2
 \de\sigma_\ell(f)
 \longrightarrow1.
\]
\end{lemma}
\begin{proof}[Proof of \cref{lem:supercritical-concentration}]
We compare costs with one large machine firm, remove the output of small
activities, and bound the remaining concentration deficit.

\paragraph{Step 1: a sharp cost comparison.}
Set $\gamma:=\gamma^M$ and $\alpha:=1/\gamma$. Write $A_*:=A(\delta_{\bm y^M})$ for the machine tail coefficient
and $c_*:=rA_*^{-\alpha}$ for its asymptotic cost coefficient. For every
sufficiently large $Q$, one firm on this ray produces $Q$ at resource cost
$(c_*+o(1))Q^\alpha$.

The uniform asymptotic tail and the Lipschitz condition on $\gamma$ give the following
sharp lower bound. For every $\varepsilon>0$, all sufficiently large feasible positive-scale firms
satisfy
\[
 C(f)\ge(c_*-\varepsilon)q_f^\alpha.
\]
To see this, consider any sequence with $q_f\to\infty$. Its scale also
diverges, and its normalized human input converges to zero. Along a convergent
subsequence, the asymptotic tail gives
\[
 \frac{C(f)}{q_f^\alpha}
 =\frac{C(f)}{m_f}
 A(\nu_f/m_f)^{-\alpha}
 m_f^{1-\alpha\gamma(\nu_f/m_f)}
 [1+o(1)]^{-\alpha}.
\]
The first factor converges to $r$, and the uniform error factor converges to
one. Put $M_H:=\mu(\Y^H)$. By \cref{ass:technology}(v), there is a constant
$K_\gamma<\infty$ such that
\[
 \gamma(\nu_f/m_f)\le\gamma+K_\gamma M_H/m_f
\]
for every sufficiently large feasible firm. Hence
\[
 m_f^{1-\alpha\gamma(\nu_f/m_f)}
 \ge
 \exp\left\{-\alpha K_\gamma M_H\frac{\log m_f}{m_f}\right\}
 \longrightarrow1.
\]
The firm composition converges to $\delta_{\bm y^M}$, so continuity
gives $A(\nu_f/m_f)\to A_*$. This proves the sharp lower bound.

\paragraph{Step 2: small activities supply a vanishing share.}
Fix $0<\varepsilon<c_*$. All atomistic firms and all positive-scale firms below the preceding
large-output threshold have resource average cost above a common
$\delta>0$. The one-firm upper bound therefore implies that these activities
produce only $O(Q_\ell^\alpha)=o(Q_\ell)$ units in $\sigma_\ell$. Let
$\mathcal L_\ell\subseteq\F^+$ contain the remaining large-output firms and
put $s_{f,\ell}:=q_f/Q_\ell$. Then
\[
 \int_{\mathcal L_\ell}s_{f,\ell}\,\de\sigma_\ell(f)
 \longrightarrow1.
\]
The sharp lower bound and the one-firm upper bound give
\[
 (c_*-\varepsilon)Q_\ell^\alpha
 \int_{\mathcal L_\ell}s_{f,\ell}^\alpha\,\de\sigma_\ell(f)
 \le \mathcal C(Q_\ell)
 \le(c_*+o(1))Q_\ell^\alpha.
\]
\paragraph{Step 3: concentration.}
The set $\mathcal L_\ell$ depends on $\varepsilon$, which we keep fixed.
Since $s^\alpha-s\ge(1-\alpha)s(1-s)$ on $[0,1]$, Step 2 gives
\[
 \begin{split}
 \limsup_{\ell\to\infty}
 \left[1-\int_{\F^+}s_{f,\ell}^2\,\de\sigma_\ell(f)\right]
 &\le\frac{1}{1-\alpha}
   \left(\frac{c_*}{c_*-\varepsilon}-1\right)\\
 &=\frac{\varepsilon}{(1-\alpha)(c_*-\varepsilon)}.
 \end{split}
\]
The omitted output share vanishes for each fixed $\varepsilon$.
The left side does not depend on $\varepsilon$, so letting
$\varepsilon\downarrow0$ makes it at most zero. Market fit gives
$s_{f,\ell}\le1$ and bounds the concentration integral above by one.
This proves convergence to one.
\end{proof}
\begin{proof}[Proof of \cref{prop:market-size}]
The output bound in \cref{app:large-output} gives $Q_N^*\to\infty$.
If $\gamma^M<1$, choose supporting prices for any equilibrium.
Active equality and \cref{lem:industry-cost} give
\[
 \frac{C(f)}{q_f}\le p_N\le\bar c
 \qquad\text{for almost every active firm}.
\]
\Cref{lem:subcritical-coercivity} therefore bounds every active
positive-scale capacity independently of $N$. Dividing by $Q_N^*$ proves
replication. If $\gamma^M>1$, equilibrium industries minimize cost at
$Q_N^*$, so \cref{lem:supercritical-concentration} proves concentration.

\end{proof}

\subsubsection{Human diversity and firm composition}
\label{app:proof-composition}
\begin{proof}[Proof of \cref{prop:human-diversity}]
We first use firm support and active equality to identify the maximizing
shares. We then prove the dominance and human-employment claims and
classify firms when there is one human type. In these comparisons, we use
$\overline a(\lambda)=\overline a(\lambda,r\one)$.

\paragraph{Step 1: active firms maximize cost-adjusted average product.}
Choose a supporting output price $p$. For a feasible firm $f$, write its input bill as
$B_w(f):=r\nu_f^M(\Y)+\int_{\Y^H}w\,\de\nu_f^H$.
Its output per unit cost, multiplied by $r$, is $rq_f/B_w(f)$, including the
normalized inputs and output of an atomistic activity.
Active equality gives $rq_f/B_w(f)=r/p$ for almost every active firm;
positive costs and positive equilibrium output imply $p>0$.
Since the unrestricted envelope includes every active firm,
\[
 \max_\lambda\overline a(\lambda,\bm w)\ge\frac rp.
\]
By hypothesis, a feasible firm attains the global maximum. Firm support
applied to that firm gives the reverse inequality. Hence
\[
 \max_{\lambda\in[0,1]}\overline a(\lambda,\bm w)=\frac rp.
\]
Almost every active firm attains this value. Its human share maximizes the
envelope, and its size and human mix attain the envelope at that share.

\paragraph{Step 2: dominance and separation at any wage.}
For AI dominance, take any active positive-size firm that employs humans.
Replacing all its humans with machines at the same total size strictly
raises output under the stated monotonicity conditions and costs no more,
since $w_i\ge\underline w\ge r$. By continuity and $F(0)=0$, a strictly
smaller machine-only firm produces exactly the original output. Its cost
is strictly lower, so it is a profitable feasible entrant at the supporting
price, a contradiction. For an atomistic activity, the same strict
increase in $L$ and weak cost reduction permit a lower-cost machine-only
activity per unit output. Thus no human-employing firm can be active.

For separation with one human type, the human wage $w$ gives
\[
 \overline a(\lambda,w)
 =\frac{r\overline a(\lambda,r\one)}{(1-\lambda)r+\lambda w}.
\]
For $0<\lambda<1$, strict convexity bounds the numerator strictly below
$r[(1-\lambda)\overline a(0,r\one)+\lambda\overline a(1,r\one)]$.
Dividing by the denominator gives a weighted average of the two
cost-adjusted endpoint values, with weights proportional to
$(1-\lambda)r$ and $\lambda w$. Hence no interior share maximizes
$\overline a(\cdot,w)$. Step 1 then implies that almost every active
firm is pure; no separate feasibility assumption on the endpoints is needed.

\paragraph{Step 3: positive human employment.}
Suppose instead that almost every active firm is machine-only. Every
human type then has slack supply, so human scarcity gives
$\bm w=\underline w\one\le r\one$. Lower human wages weakly raise
$\overline a(\lambda,\bm w)$ at every share and leave the machine-only
endpoint unchanged. The strict inequality in part (ii) therefore means
that zero is not a maximizing human share at these wages.
This contradicts Step 1. Thus a set of active firms of positive industry
measure employs humans.

\paragraph{Step 4: classify firms with one human type.}
For part (ii)(a), Step 2 excludes mixed firms, and Step 3 ensures that
some active firms are human-only.

For part (ii)(b), the cost-adjusted envelope is the ratio in Step 2.
Strict concavity of its numerator and positivity of its affine denominator
imply that it has at most one maximizer: if two distinct shares attained
the same maximum, their strict convex combinations would have a strictly
higher ratio. Attainment follows from the proposition's opening assumption.
Step 3 excludes a maximum at zero. Since $w\ge\underline w=r$, the left
derivative at one satisfies
\[
 \partial^-_\lambda\overline a(1,w)
 =\frac{r}{w^2}\left[
 w\overline a'_-(1)-(w-r)\overline a(1)\right]<0.
\]
Thus one is not a maximizer either. The unique maximizing share is
interior, and Step 1 implies that almost every active firm has that share.

\end{proof}

\subsection{Wages and employment}
\label{app:proof-wages}
\subsubsection{The wage envelope and sorting}
\label{app:proof-wage-envelope}
Under \cref{ass:level-scale}, write production as $\widetilde F(m,\bm z)$,
where $\bm z$ is the vector of the firm's mean capabilities. Its marginal product is
\begin{equation}
 D_{\bm y}F
 =\partial_m\widetilde F+
   \frac1m\nabla_{\bm z}\widetilde F\cdot(\bm y-\bm z),
 \qquad
 \bm z=\frac1m\int_\Y\bm y\,\de\nu(\bm y).
 \tag{MEAN-MP}\label{eq:mean-marginal-product}
\end{equation}

\begin{proof}[Proof of \cref{thm:wage-envelope}(i)]
Put $Q=Q(\sigma)$.
By \cref{thm:equilibrium-efficiency}, choose $p$ such that
$(p,w)\in\Lambda(Q,\sigma)$.
For any firm $g$, write its input bill as
\[
 B_w(g):=r\nu_g^M(\Y)+\int_{\Y^H}w\,\de\nu_g^H.
\]
Active equality gives $B_w(f)=pq_f$. Firm support implies that $f$ minimizes
its input bill among feasible firms that produce $q_f$. We first show that
machines have positive marginal product, then substitute between machines
and human type $i$.

\paragraph{Step 1: positive machine marginal product.}
Let $m_f^M:=\nu_f^M(\Y)$ be the firm's machine mass and
$M_f:=D_{\bm y^M}F(\nu_f)$ its marginal product.
We claim that $M_f>0$. For $t>0$, let $f_t$ reduce the firm's machine
input by the fraction $t$. Then
\[
 B_w(f_t)=B_w(f)-trm_f^M,
 \qquad
 q_{f_t}=q_f-tm_f^MM_f+o(t).
\]
Suppose first that $M_f<0$. For all small $t$, the firm $f_t$
costs less and produces more than $f$. By continuity and $F(0)=0$, we can
scale the firm $f_t$ down until it produces exactly $q_f$. The
scaled firm remains feasible and has a still lower input bill, which
contradicts firm support.

Now suppose that $M_f=0$. Whenever $q_{f_t}>q_f$, the same scaling
argument gives a contradiction. Whenever $q_{f_t}\le q_f$, firm support gives
\[
 0\le B_w(f_t)-pq_{f_t}
 =-trm_f^M+p(q_f-q_{f_t})
 =-trm_f^M+o(t)<0,
\]
which is also a contradiction. Hence $M_f>0$.

\paragraph{Step 2: fixed-output substitution.}
Suppose firm $f$'s staffing constraint for human type $i$ is slack. Add a small amount of this
type and adjust the firm's machine input to hold output fixed.
The implicit function theorem gives the slope of machine use with respect to
human use as $-D_{\bm y^i}F(\nu_f)/M_f$. Cost minimization makes the
one-sided derivative of the input bill weakly positive:
\[
 w_i-r\frac{D_{\bm y^i}F(\nu_f)}{M_f}\ge0.
\]
This is $b_f(\bm y^i)\le w_i$. If the firm also employs a positive amount of
type $i$, the variation is feasible in both directions, so the inequality is
an equality. Substituting $M_f=D_{\bm y^M}F(\nu_f)$ gives
$b_f(\bm y^i)=w_i$.

\paragraph{Step 3: the highest eligible bid.}
The inequality just proved and
$w_i\ge\underline w$ show that its right-hand side is at most $w_i$. If type
$i$ is slack in the aggregate, human scarcity gives
$w_i=\underline w$, so the bound is attained. If an active machine-using
firm has an interior match with the type, the equality just proved says that
its bid equals $w_i$. The bound is again attained.
\end{proof}

\phantomsection\label{app:proof-wage-sorting}
\begin{proof}[Proof of \cref{thm:wage-envelope}(ii)]

We compare bids, then describe their winning regions. Subtracting
\eqref{eq:mean-marginal-product} at $\bm y$ and $\bm y^M$ gives the affine
bid formula \eqref{eq:affine-bid} and its stated coefficients. By
\cref{thm:wage-envelope}, $b_g(\bm y^i)\le w_i=b_f(\bm y^i)$ for $\sigma$-almost every eligible firm $g$.
Subtracting the affine bids gives \eqref{eq:sorting-fan}. For a fixed finite
set of eligible firms, each winning region is an intersection of half-spaces whose
boundaries pass through $\bm y^M$. Their inequalities are homogeneous in
$\bm y-\bm y^M$, so the regions are polyhedral cones. Their faces form a polyhedral fan. The interior-match condition
therefore assigns workers according to these regions when they share this
eligible set; a binding staffing cap can exclude a firm for a given type.
\end{proof}

\subsubsection{Human employment and exit}
\label{app:proof-exit}
\begin{proof}[Proof of \cref{prop:exit}]
For part (i), we derive the industry frontier, prove continuity, and check
the exit condition. For part (ii), we exclude mixed firms and compare the
two pure technologies.

\paragraph{Part (i), Step 1: the industry frontier.}
Let $T_H,T_M$ be aggregate human and machine input, and let $T=(T_H,T_M)$.
Concavity of best average product bounds industry output by
\[
 \Phi(T):=(T_H+T_M)\,
 \overline a\!\left(\frac{T_H}{T_H+T_M}\right),\qquad \Phi(0):=0.
\]
Indeed, each firm's average product is at most $\overline a$ at its share,
and input-weighted shares average to $T_H/(T_H+T_M)$.
Strict concavity and unique preferred scales make equality hold exactly
when all input uses this common share at its preferred scale.
The relaxed input problem is therefore
\[
 \max_{0\le T_H\le\mu,\ T_M\ge0}
 \{U_N(\Phi(T))-\underline wT_H-rT_M\}.
\]
The feasibility hypothesis makes its maximizing firms feasible in the
original economy, so equality is attained.

The frontier $\Phi$ is concave and homogeneous of degree one.
Two distinct optimal inputs on the same ray give different outputs, so
strict concavity of $U_N$ rules them out. Inputs on different rays give a
strict gain in $\Phi$ at their midpoint, by strict concavity of $\overline a$,
and hence a strict surplus gain. Optimal output lies below demand saturation:
at or above saturation, a small radial input reduction saves cost at first
order and loses benefit of smaller order. Thus both comparisons are strict.
The relaxed optimum is unique. It exists because benefit is bounded and
input costs are positive, and machine viability makes its total input positive.
Equality in the frontier bound then determines the common share, preferred
scale, and activity weight of the equilibrium industry.

\paragraph{Part (i), Step 2: continuity.}
Preferred scales share a finite bound. Berge's theorem therefore makes
$\overline a(\lambda;\theta)$ jointly continuous and the unique preferred
scale continuous. The relaxed optimum also varies continuously: locally,
positive costs and bounded benefit restrict all maximizing inputs to one
compact set. Output, human input, the common share, and preferred scale
therefore vary continuously. At positive scale $m$, firm weight is
$(T_H+T_M)/m$, so the industry measure is continuous there. At zero scale
we use the atomistic intensity $T_H+T_M$; the disjoint-union topology
does not give convergence of firm measures across this boundary.

\paragraph{Part (i), Step 3: exit.}
At the machine-only relaxed optimum, the output price is $r/\overline a(0)$
and output is positive by viability. The right marginal product of humans
is $\overline a(0)+\overline a'_+(0)$.
Concavity makes this optimum global exactly when
\[
 \frac r{\overline a(0)}
 [\overline a(0)+\overline a'_+(0)]\le\underline w,
\]
which is the condition in \cref{prop:exit}(i). If the inequality fails, a small human
input raises surplus. Uniqueness includes the equality case.

\paragraph{Part (ii): separate production and threshold exit.}
At supporting prices, feasible preferred pure firms give
$p\overline a(1)\le w$ and $p\overline a(0)\le r$. Strict convexity makes every interior
human share satisfy
\[
 p\overline a(\lambda)-\lambda w-(1-\lambda)r
 <\lambda(p\overline a(1)-w)+(1-\lambda)(p\overline a(0)-r)\le0.
\]
Thus no mixed firm operates. An active machine firm uses its preferred
scale by the size-choice argument in \cref{app:firm-size-bounds}, so $p=r/\overline a(0)$.
If $p\overline a(1)>\underline w$, human support gives $w>\underline w$ and
scarcity makes all humans work; active equality then gives $w=p\overline a(1)$.
If $p\overline a(1)<\underline w$, every human firm loses money, so employment is
zero and $w=\underline w$. At equality, the wage is $\underline w$.
Both pure activities break even, and every feasible split of the same
equilibrium output between them has the same cost and remains an equilibrium.
This gives the wage formula and allows nonunique employment at the threshold.
\end{proof}

\subsubsection{Wage changes through reorganization}
\begin{proof}[Proof of \cref{prop:wage-decomposition}]
At each parameter value, \cref{thm:wage-envelope} gives
$w_i=b_f(\bm y^i)$ at the selected employer with an interior match. Equation \eqref{eq:mean-marginal-product} shows that the listed statistics determine the bid. Apply the chain rule to
this continuously differentiable function. Holding $m$, $s$, and the human
mean fixed holds the relevant quantities of human and machine inputs fixed; only machine
capabilities change in the technology term. When human means have fewer
independent coordinates, take the derivative along their feasible affine
span. The assumptions concern the selected match; they do not establish
that such a selection exists in every economy.
\end{proof}

\subsubsection{The large-market wage jump}
\label{app:jump-details}
\begin{proof}[Verification of \cref{fig:ces-support-jump}]
We verify the black curve for every equilibrium wage selection in the
large-market limit. Use the calibration in
footnote~\ref{fn:concentration-calibration}, with $\theta=y_*^M$ denoting
AI coordination. Below one, pure firms attain a bound on industry output.
Above one, no human firm remains active in a large enough market.
At one, supporting-price bounds determine the limiting wage.

Use the optimized log scale factor $\ell$ from
\eqref{eq:ces-scale-peak} at $\phi=.02$. Then $\overline a(1)=1.1e^{\ell(.985)}$, and the best machine average product
below the threshold is $\overline a(0)=e^{\ell(\theta)}$.
Let $h$ and $a$ be human and machine inputs in a firm, and let
$\lambda=h/(h+a)$ denote its human share.

\paragraph{Step 1: replication below one.}
For $.995\le\theta<1$, the highest log average product at human share
$\lambda$ is
\[
 g_\theta(\lambda)=\log(1+.1\lambda)
 +\ell\bigl(\theta-\lambda(\theta-.985)\bigr).
\]
We show that average product, $e^{g_\theta}$, is strictly convex. Write
$d:=1-\theta+\lambda(\theta-.985)$ for one minus mean coordination. Then
\[
 \frac{(e^{g_\theta})''}{e^{g_\theta}}
 =(\theta-.985)^2\left[\frac1d+\left(\log\frac{.02}{d}\right)^2\right]
 -\frac{.2(\theta-.985)}{1+.1\lambda}\log\frac{.02}{d}>0.
\]
For the sign, multiply by $d(1+.1\lambda)/(\theta-.985)>0$.
The positive term is at least $.01$, whereas the subtracted term is at
most $.004/e$, since $d\log(.02/d)\le .02/e$. Convexity therefore gives
\[
 q_f\le(h+a)\bigl[\lambda \overline a(1)+(1-\lambda)\overline a(0)\bigr]
 =\overline a(1)h+\overline a(0)a.
\]
The same bound holds for atomistic firms by taking the zero-scale limit.
Since $\overline a(1)>\overline a(0)$, every industry with output $Q$ costs at least
\[
 \frac{Q}{\overline a(0)}
 -\left(\frac{\overline a(1)}{\overline a(0)}-1\right)\frac23.
\]
For all sufficiently large $Q$, two pure-human firms of scale $1/3$ and
copies of the preferred machine firm attain this bound. Both complete
firm types fit inside $Q$, so the construction respects the capacity cap.
Active equality pins $p=1/\overline a(0)$ and $w_H=\overline a(1)/\overline a(0)$.
Every optimal dual supports this cost minimizer; hence these prices apply
to every supported cost-minimizing industry at that output. The bound in \cref{app:large-output} gives $Q_N\to\infty$,
proving the limiting wage below one.

\paragraph{Step 2: exclude large human firms at and above one.}
Fix $\theta\ge1$. Replacing a share $\lambda>0$ of machines with humans
at fixed total size $m$ changes log output by
\[
 \log(1+.1\lambda)-\lambda(\theta-.985)\log(1+m)
 \le\lambda\bigl[.1-(\theta-.985)\log(1+m)\bigr].
\]
Choose a finite size $M$ for which the bracket is negative whenever $m>M$.
Any human-using firm above that size has lower output than the pure-machine
firm of the same size. Because $w_H\ge1=r$, the latter has no higher input
bill. Scale the firm down to the original output to obtain a strictly
cheaper feasible firm. Firm support therefore excludes every human-using
positive-scale firm with $m>M$.

If $\theta>1$, the machine firm with capacity $Q$ has average product
tending to infinity. Dual feasibility gives $p_Q\to0$. On the compact
set of human-using compositions and scales $0\le m\le M$, average product
has a finite upper bound. Each such activity costs at least one per unit
input, so it cannot break even once $p_Q$ is sufficiently small.
No humans then work, and scarcity complementarity gives $w_H=1$.
Again \cref{app:large-output} gives $Q_N\to\infty$, proving
the limiting wage above one.

\paragraph{Step 3: the critical case.}
At $\theta=1$, define $g_1(0)=.02$ by continuity. The average-product
curvature bound in Step 1 still holds for $\lambda>0$, and continuity
extends its chord bound to $\lambda=0$. Hence
\[
 q_f\le \overline a(1)h+e^{.02}a,
 \qquad
 \mathcal C(Q)\ge e^{-.02}Q-(\overline a(1)e^{-.02}-1)\frac23.
\]
Replicating machine firms with average product approaching $e^{.02}$
gives $\mathcal C(Q)/Q\to e^{-.02}$. Equilibrium output diverges by \cref{app:large-output}. Dual equality gives $p_Q\ge\mathcal C(Q)/Q$. Entry by
a machine firm of capacity $Q$ bounds $p_Q$ above by its average cost.
These two bounds give $p_Q\to e^{-.02}$.

The pure-human peak firm is an available entrant for large $Q$, so
$w_H\ge p_Q\overline a(1)>1$. Human supply must bind. We show that any human-employing
active activity also gives the reverse limiting inequality. Step 2 bounds
its positive scale by $M$; include atomistic firms as scale zero. On this
compact set, pure-machine average product is strictly below $e^{.02}$.
Continuity therefore gives a fixed $\delta>0$ such that activities with
human share $0<\lambda<\delta$ cannot break even for large $Q$, even at
wage one. Every active human activity consequently has $\lambda\ge\delta$.

Let $v$ denote such an activity's output per unit input. Active equality
and the output bound give
\[
 w_H=\frac{p_Qv-(1-\lambda)}{\lambda}
 \le p_Q\overline a(1)+(p_Qe^{.02}-1)\frac{1-\lambda}{\lambda}.
\]
The last term tends uniformly to zero because $\lambda\ge\delta$.
Combined with $w_H\ge p_Q\overline a(1)$, this proves $w_H\to \overline a(1)e^{-.02}$ for
every supporting wage selection. These three steps establish the stated
pointwise limit. They do not impose a pure-firm formula at arbitrary finite
output, where mixed firms can be cheaper.
\end{proof}

\subsection{Connection to the task model}
\label{app:proof-task-model}
\begin{lemma}[The task technology satisfies \cref{ass:technology}]
\label{lem:task-technology}
$F^{\mathrm T}$ satisfies \cref{ass:technology} with tail exponent
$\gamma\equiv1$. Every nonzero firm $\nu$ has a unique optimal vector of
task outputs $\bm z^*(\nu)$, and it is strictly positive. With \emph{task
prices} $\bm\pi(\nu):=\nabla L(\bm z^*(\nu))$, the marginal product of
capability $\bm y$ is
\[
 D_{\bm y}F^{\mathrm T}(\nu)=\max_{1\le d\le n}\pi_d(\nu)\,y_d.
 \tag{BEST-TASK}\label{eq:best-task}
\]
\end{lemma}
\begin{proof}[Proof of \cref{lem:task-technology}]
We characterize the optimum, derive \eqref{eq:best-task}, and then check
\cref{ass:technology}. Let $\mathcal Z(\nu)$ be the set of task-output vectors
that assignments of $\nu$ attain. It is convex and compact, and
$\mathcal Z(m\nu)=m\mathcal Z(\nu)$. Each agent can work on the task where
$\pi_dy_d$ is largest, so the support function of $\mathcal Z(\nu)$ is
\[
 \max_{\bm z\in\mathcal Z(\nu)}\bm\pi\cdot\bm z
 =\int_\Y\max_d\pi_dy_d\,\de\nu(\bm y)
 \qquad(\bm\pi\in\R^n).
 \tag{TASK-SET}\label{eq:task-set}
\]

\paragraph{Step 1: the optimum.}
Every capability is at least $\underline y>0$, so the equal assignment gives
positive output. An optimum leaves no task empty. When $\rho\le0$, output is
zero without that task. When $0<\rho<1$, its marginal product is infinite, so
moving a little time to it raises output. If two optima differed and were not
proportional, their midpoint would do strictly better, because $L$ is strictly
quasiconcave. If they were proportional, homogeneity would give them different
values. Hence the optimum is unique.

\paragraph{Step 2: marginal products.}
Fix $\nu$, put $\bm\pi:=\bm\pi(\nu)$, and let $g(\bm y):=\max_d\pi_dy_d$.
Concavity and homogeneity of degree one give $L(\bm z)\le\bm\pi\cdot\bm z$ for
every $\bm z$, with equality at $\bm z^*(\nu)$. With \eqref{eq:task-set},
every firm $\nu'$ therefore satisfies
\[
 F^{\mathrm T}(\nu')\le\int_\Y g\,\de\nu'.
 \tag{UPPER}\label{eq:task-upper}
\]
Equality holds at $\nu'=\nu$, because the first-order condition for
maximizing $L$ over $\mathcal Z(\nu)$ makes $\bm z^*(\nu)$ maximize
$\bm\pi\cdot\bm z$ there. Write an optimal assignment of $\nu$ as
$\nu_d=\tau_d\nu$ with time shares $\tau_d(\bm y)$. The equality case forces,
for $\nu$-almost every $\bm y$, $\tau_d(\bm y)>0$ only on tasks that maximize
$\pi_dy_d$.

Now let $\eta$ be a finite signed measure with $\nu+t_0\eta\ge0$ for some
$t_0>0$. The negative part of $\eta$ is then absolutely continuous with respect
to $\nu$. Assign $\nu+t\eta$ as follows: give the part of $\eta$ that is
absolutely continuous with respect to $\nu$ the shares $\tau$, and put the
rest, which is nonnegative, on a task that maximizes $\pi_dy_d$. For
$0\le t\le t_0$ this is an assignment. Its task outputs are
$\bm z^*(\nu)+t\bm v$ with $\bm\pi\cdot\bm v=\int_\Y g\,\de\eta$.
Differentiability of $L$ at $\bm z^*(\nu)$ gives
$F^{\mathrm T}(\nu+t\eta)\ge F^{\mathrm T}(\nu)+t\int_\Y g\,\de\eta+o(t)$.
With \eqref{eq:task-upper}, this proves the first-variation formula
\eqref{eq:best-task}. Replacing $\eta$ by $-\eta$ covers $t<0$.

\paragraph{Step 3: \cref{ass:technology}.}
Homogeneity gives $F^{\mathrm T}(m\nu)=mF^{\mathrm T}(\nu)$. Average product
is therefore constant in scale, which gives (iv), and (v) holds exactly with
$A=F^{\mathrm T}$ on $\Delta(\Y)$ and $\gamma\equiv1$. The right side of
\eqref{eq:task-set} is Lipschitz in $\bm\pi$, with constant
$\overline y\,\nu(\Y)$, and continuous in $\nu$ under weak convergence. On
bounded-mass sets, weak convergence of firm measures therefore gives uniform
convergence of support functions on the unit sphere, that is, Hausdorff
convergence of $\mathcal Z(\nu)$. Berge's maximum theorem then makes
$F^{\mathrm T}$ and the unique optimum $\bm z^*(\nu)$ continuous. This gives
(i), with the positive extension $F^{\mathrm T}(m\nu)/m=F^{\mathrm T}(\nu)$.
Because $\bm z^*(\nu)$ is strictly positive, $\bm\pi(\nu)$ is also continuous,
which gives the continuity that (ii) requires. For (iii), let
$\nu'\succeq_{\mathrm{FOSD}}\nu$ have equal mass, and build $g$ from
$\bm\pi(\nu')$. This $g$ is continuous, nonnegative, and increasing.
Inequality \eqref{eq:task-upper}, with $\nu'$ as the base firm, gives
$F^{\mathrm T}(\nu')=\int_\Y g\,\de\nu'\ge\int_\Y g\,\de\nu\ge
F^{\mathrm T}(\nu)$.
\end{proof}
\begin{lemma}[Pooling]
\label{lem:task-pooling}
Let $\sigma$ be an industry with aggregate input measure
$\bar\nu:=\int_\F\nu_f\,\de\sigma(f)$. Then
\[
 F^{\mathrm T}(\bar\nu)\ \ge\ \int_\F F^{\mathrm T}(\nu_f)\,\de\sigma(f)
 =Q(\sigma),
\]
and the two sides carry the same resource cost. Consequently the least-cost
function $\mathcal C$ of the task economy in our framework is the least-cost
function of the task economy of \cref{def:task-competitive}; it is convex and
strictly increasing; and every least-cost industry is, in aggregate, an
optimal task allocation producing its output exactly.
\end{lemma}
\begin{proof}[Proof of \cref{lem:task-pooling}]
Build $g$ from $\bm\pi(\bar\nu)$. By the equality case of
\eqref{eq:task-upper} at $\bar\nu$, and then by \eqref{eq:task-upper} at each
firm,
\[
 F^{\mathrm T}(\bar\nu)=\int_\Y g\,\de\bar\nu
 =\int_\F\int_\Y g\,\de\nu_f\,\de\sigma(f)
 \ge\int_\F F^{\mathrm T}(\nu_f)\,\de\sigma(f).
\]
Resource cost is linear in the input measure, so
$\int_\F C(f)\,\de\sigma(f)=\underline w\,\bar\nu^H(\Y)+r\,\bar\nu^M(\Y)$,
which is the bill the aggregate pays.

For the cost functions, take an industry producing $Q$ at cost $c$ with human
use $\bar\nu^H\le\mu$. An optimal assignment of $\bar\nu$ is an allocation of
the task economy with the same bill and output $F^{\mathrm T}(\bar\nu)\ge Q$,
so the task economy produces $Q$ at cost at most $c$. Conversely, an
allocation with task outputs $\bm z$ and bill $c$ uses aggregate input measure
$\bar\nu$; the single atomistic firm with composition
$\bar\nu/\bar\nu(\Y)$, run at intensity $\bar\nu(\Y)$, costs $c$ and produces
$F^{\mathrm T}(\bar\nu)\ge L(\bm z)$. The two least-cost functions therefore
agree.

Read on the task economy's allocations, least cost is the value of a convex
program: the constraint set is convex, $L$ is concave, and the bill is
linear. So $\mathcal C$ is convex, and $\mathcal C(0)=0$ with positive input
prices makes it strictly increasing. Finally, let $\sigma$ be least cost at
$Q$. Pooling produces $F^{\mathrm T}(\bar\nu)\ge Q$ at cost
$\mathcal C(Q)$, so $\mathcal C(F^{\mathrm T}(\bar\nu))\le\mathcal C(Q)$, and
strict monotonicity forces $F^{\mathrm T}(\bar\nu)=Q$.
\end{proof}
\begin{proof}[Proof of \cref{thm:task-equivalence}]
Both economies have the same least-cost function by \cref{lem:task-pooling}.
We identify output, characterize supporting prices, recover assignments
and wages, and bound the exceptional outputs.

\paragraph{Step 1: output.}
Under constant returns a positive-scale firm has the same output and inputs
per unit of firm size as the atomistic firm with its composition, so the
capacity cap in $\Sigma$ never binds. Convexity of $\mathcal C$ and
$\mathcal C(0)=0$ make every slope at $Q>0$ at least $\mathcal C(Q)/Q>0$,
while the standing condition gives $\mathcal C(Q)\le\bar cQ$ with
$\bar c<P(0)$. Hence $Q_N^*$ lies strictly between $0$ and $D_N(0)$. There
$U_N$ is strictly concave, so $Q_N^*$ is unique and
$P_N(Q_N^*)\in\partial\mathcal C(Q_N^*)$. Conversely any $Q$ with
$P_N(Q)\in\partial\mathcal C(Q)$ maximizes the concave function
$U_N-\mathcal C$, so $Q=Q_N^*$.

\paragraph{Step 2: supporting prices and subgradients.}
We show that a least-cost industry at $Q$ admits supporting price $p$
exactly when $p\in\partial\mathcal C(Q)$. Under constant returns,
atomistic recipes implement every input mix, so the firm menu is
independent of $Q$.

For necessity, integrate firm support over any alternative industry
$\sigma'$ and subtract active equality at the supported industry $\sigma$.
Human feasibility and scarcity give
\[
 \int_\F C(f)\,\de\sigma'(f)
 \ge\int_\F C(f)\,\de\sigma(f)
       +p[Q(\sigma')-Q].
\]
Taking the infimum at each exact output $Q'$ proves both least cost at
$Q$ and $\mathcal C(Q')\ge\mathcal C(Q)+p(Q'-Q)$.

Conversely, let $p\in\partial\mathcal C(Q)$ and let $\sigma$ minimize
cost at $Q$. The subgradient inequality makes $\sigma$ minimize cost
minus $p$ times output, subject only to human supplies. Apply the conic
duality argument of \cref{lem:industry-cost}, now with output priced at
$p$. The same strictly feasible human allocation gives multipliers
$w_i-\underline w\ge0$. Dual feasibility and complementary slackness
are precisely firm support, active equality, and human scarcity.
Atomistic machine entry gives $p\le\bar c$, so recipes discarded in the
compact reduction cannot violate support. Hence
$(p,w)\in\Lambda(Q,\sigma)$.

\paragraph{Step 3: output and assignments coincide.}
By \cref{thm:equilibrium-efficiency} the industries of our equilibrium
outcomes are those that produce $Q_N^*$ at least cost. By Step 2 a
competitive allocation is least cost at its output $Q$, with
$P_N(Q)\in\partial\mathcal C(Q)$, so Step 1 gives $Q=Q_N^*$. Conversely Step 2
with $p=P_N(Q_N^*)$, together with \cref{lem:task-pooling}, turns every
least-cost industry at $Q_N^*$ into a competitive allocation; existence
follows because an efficient industry exists. This proves (i). For (ii), fix a
least-cost industry with aggregate $\bar\nu$. By \cref{lem:task-pooling} its
optimal assignment produces $Q_N^*$, and the equality case of
\eqref{eq:task-upper} at $\bar\nu$ puts every factor only on tasks that
maximize $\pi_d(\bar\nu)y_d$. Competitive task prices are
$p\,\nabla L(\bm z^*(\bar\nu))=p\,\bm\pi(\bar\nu)$, and scaling all task
prices by $p>0$ leaves the maximizing tasks unchanged.

\paragraph{Step 4: wages.}
Integrating active equality over $\sigma$ gives
\[
 pQ_N^*=\mathcal C(Q_N^*)+\sum_{i=1}^K(w_i-\underline w)H_i(\sigma),
\]
so each wage vector is supported by exactly one payment $p$. Our equilibrium
wage vectors are those supported by some $p\in\partial\mathcal C(Q_N^*)$; the
competitive ones are those supported by $P_N(Q_N^*)$. By Step 2 every such
payment supports at least one wage vector. The two sets therefore coincide if
and only if $\partial\mathcal C(Q_N^*)=\{P_N(Q_N^*)\}$, that is, if and only
if $\mathcal C$ is differentiable at $Q_N^*$. The wage formula is
\eqref{eq:best-task} priced at $p$, together with $w_i\ge\underline w$ and
complementary slackness on human supply. This proves (iii).

\paragraph{Step 5: exceptional outputs.}
Let $\sigma$ produce $Q_N^*$ at least cost with aggregate $\bar\nu$. Suppose
$\bar\nu$ contains machines. Adding or removing $t$ units of them changes
output by $tD+o(t)$ and cost by $rt$, where $D:=D_{\bm y^M}F^{\mathrm
T}(\bar\nu)>0$ by \eqref{eq:best-task}. The right derivative of $\mathcal C$
at $Q_N^*$ is therefore at most $r/D$ and the left derivative at least $r/D$,
so convexity makes both equal $r/D$. The same argument, with a human type
employed strictly between $0$ and its supply at cost $\underline w$, covers
the other case. Now suppose $\mathcal C$ fails to be differentiable at
$Q_N^*$. Then no least-cost industry uses machines and each employs every
human type at either $0$ or its full supply, so its aggregate is
$\bar\nu=\sum_{i\in S}\mu_i\delta_{\bm y^i}$ for the employed set $S$. By
\cref{lem:task-pooling}, $Q_N^*=F^{\mathrm T}(\bar\nu)$. There are $2^K$ sets
$S$, which proves (iv).
\end{proof}
\begin{proof}[Proof of \cref{prop:task-rent} and the task-model exit condition]
We check the augmented technology, differentiate output, and then evaluate
the human entry margin at a machine-only firm.

\paragraph{Step 1: the augmented technology.}
The scale multiplier is
\[
 \frac{S(m,z)}m=(1+m)^{z-1}\exp\{\phi m/(1+m)\},
\]
with limit one at zero.
Thus $F(\nu)=F^{\mathrm T}(\nu)S(m,\overline y_*)/m$.
\Cref{lem:task-technology} and continuity of the multiplier give continuity and the
positive average-product extension in \cref{ass:technology}(i).
Differentiability follows from the chain rule below.
At equal mass, a first-order stochastic improvement raises both
$F^{\mathrm T}$ and mean coordination; since the multiplier increases in
coordination, the product is monotone. At a fixed composition,
$\partial_m\log(S/m)=(z-1)/(1+m)+\phi/(1+m)^2$ changes sign at most once,
from positive to negative, so average product is quasiconcave.
Finally, on unit compositions,
$F(m\nu)=e^\phi F^{\mathrm T}(\nu)m^{\int y_*\,\de\nu}[1+o(1)]$.
Bounded capabilities make the remainder uniform and the exponent Lipschitz
under mixing. Hence the augmented technology satisfies
\cref{ass:technology}.

\paragraph{Step 2: the bid.}
Adding mass $t$ at capability $\bm y$ raises size from $m$ to $m+t$ and
changes mean coordination by $t(y_*-\overline y_*)/m+o(t)$.
The first variation, which measures the output change per unit added, is
therefore
\[
 D_{\bm y}F
 =\frac Sm\left\{D_{\bm y}F^{\mathrm T}
 +\frac{F^{\mathrm T}}m
 \left[m\partial_m\log(S/m)+
 \partial_z\log S\,(y_*-\overline y_*)\right]\right\}.
 \tag{TASK-VAR}\label{eq:task-variation}
\]
All quantities on the right are evaluated at the original firm.
At a positive preferred size, $\partial_m\log(S/m)=0$ and
$\partial_z\log S=\log(1+m)$. Taking the ratio of
\eqref{eq:task-variation} for the worker and the machine cancels $S/m$ and
gives \eqref{eq:task-bid}. The denominator is positive because the firm
values machines. The wage equality follows from
\cref{thm:wage-envelope}(i). With constant returns, both scale derivatives
in \eqref{eq:task-variation} are zero, so the coordination term disappears.

\paragraph{Step 3: the exit condition.}
Use the human share $\lambda$ and unit composition $\nu_\lambda$ from
\cref{prop:exit}, with mean coordination
$z_\lambda=\lambda y_*^H+(1-\lambda)y_*^M$.
Constant returns of the task technology give
\[
 \overline a(\lambda)=F^{\mathrm T}(\nu_\lambda)
 \max_{m\ge0}\frac{S(m,z_\lambda)}m.
\]
The unique, bounded preferred scales in \cref{prop:exit}(i) permit the
envelope theorem, including a zero preferred scale. Differentiating from
the right at the machine-only mix gives
\[
 1+\frac{\overline a'_+(0)}{\overline a(0)}
 =\frac{D_{\bm y^H}F^{\mathrm T}(\delta_{\bm y^M})}
 {F^{\mathrm T}(\delta_{\bm y^M})}
 +(y_*^H-y_*^M)\log\bigl(1+m^*(y_*^M)\bigr).
\]
Here homogeneity gives
$D_{\bm y^M}F^{\mathrm T}(\delta_{\bm y^M})
=F^{\mathrm T}(\delta_{\bm y^M})$.
To evaluate the remaining ratio, let $\pi_d$ be task prices at the
machine-only optimum. Every task receives positive machine time, so
$\pi_d y_d^M$ has the same value for every task $d$.
The best-task formula \eqref{eq:best-task} then gives
\[
 \frac{D_{\bm y^H}F^{\mathrm T}(\delta_{\bm y^M})}
 {F^{\mathrm T}(\delta_{\bm y^M})}
 =\max_d\frac{y_d^H}{y_d^M}.
\]
Substitution in the condition in \cref{prop:exit}(i) proves the task-model exit condition,
including equality. This step uses the strict concavity and feasibility
hypotheses of \cref{prop:exit}(i); the bid identity does not require them.
\end{proof}

\paragraph{Checks for \cref{fig:task-rent-coordination}.}
Every composition in this figure has mean coordination between $.8$ and
$.97$, so preferred sizes are positive and below $.5/.03-1<16<\mu$.
Task output per agent is at most $2$ and the scale multiplier is at most
$e^{.5}$, so every preferred firm's output is below $52$. Machine-only
average product is at least $2e^{\ell(.8)}>2$, with $\ell$ defined in
\eqref{eq:ces-scale-peak}, so relaxed equilibrium output exceeds
$200(3-1/2)=500$. Every preferred scale is feasible, including at the
relaxed optimum required by \cref{prop:exit}(i).

On the coordination path, AI is twice as productive on every task.
For machine share $s$, unit-workforce task output is $1+s$ and mean
coordination is $z=.95+(y_*^M-.95)s$. Optimized average product is
$\overline a(1-s)=(1+s)e^{\ell(z)}$. Where humans coordinate better,
$y_*^M<.95$, differentiation gives
\[
 \frac{\partial_s^2\overline a(1-s)}{(.95-y_*^M)e^{\ell(z)}}
 =(.95-y_*^M)(1+s)
 \left[\frac1{1-z}+\ell'(z)^2\right]-2\ell'(z).
\]
The figure script uses outward-rounded arithmetic on a $100\times100$
grid of rectangles covering the whole share interval and the coordination
path below $.95$. The upper bound is below $-.049$, verifying strict
concavity where we use the task-model exit condition. At and above $y_*^M=.95$,
replacing each human with half a machine preserves task output, weakly
raises mean coordination, and reduces input cost. Reoptimizing scale
therefore makes machine-only production strictly cheaper.
The exit statistic $.5+(.95-y_*^M)\log[.5/(1-y_*^M)]$ reaches $.6$ at
AI coordination approximately $.879879$.

Along the coordination path, we minimize unit cost over machine shares and choose the
wage that clears human supply, or the outside wage when supply is slack.
Saved checks compare the bid formula with wages in mixed firms, search
for profitable entry on $20{,}001$ shares, and verify resource accounting
and firm feasibility. Independent cost linear programs, scalar
minimizations, and finer composition grids check selected points.
The saved checks in \path{images/task/task_rent_checks.json} record the
path, snapshots, and residuals.

\begin{proof}[Proof of \cref{thm:task-main-equivalence}]
The result follows from \cref{thm:task-equivalence}, proved above.
Parts (i)--(ii) give unique output and identify competitive task allocations
with the aggregates of equilibrium industries under optimal assignments.
If some equilibrium uses machines, its industry is least cost at that
output. Part (iv) then makes the cost function differentiable there, and
part (iii) gives equality of the two sets of equilibrium wage vectors.
\end{proof}

\subsection{Continuity of values and wages}
\label{app:proof-continuity}
Fix a continuous feasible AI path $\bm y^M(\theta)$ on an open interval
$\Theta$, holding all other primitives fixed and maintaining machine
viability. Identify a firm by its human inputs and machine quantity,
with the Euclidean topology on each component of $\F^+\sqcup\F^0$.
We use weak convergence against bounded continuous functions on this
disjoint union.

\begin{proposition}[Continuous values and closed equilibrium limits]
\label{prop:equilibrium-continuity}
If \cref{ass:technology} holds and the AI path is continuous, then:
\begin{enumerate}[(i)]
 \item Minimum cost $(Q,\theta)\mapsto\mathcal C(Q;\theta)$ is continuous
 on $\R_+\times\Theta$.
 \item Maximal surplus $\theta\mapsto\W_N^*(\theta)$ is continuous.
 \item Equilibrium quantities $\theta\mapsto\mathcal Q_N^*(\theta)$ are
 upper hemicontinuous.
 \item Equilibrium industries $\theta\mapsto\Sigma^*(\theta)$ have closed
 graph under weak convergence of finite industry measures.
\end{enumerate}
\end{proposition}

\begin{proof}[Proof of \cref{prop:equilibrium-continuity}]
We extend the cost argument in \cref{app:cost-duality} to a moving AI
profile, then pass to surplus maximizers.

\paragraph{Step 1: joint cost continuity.}
The atomistic machine cost $\bar c(\theta)=r/L(\delta_{\bm y^M(\theta)})$
is continuous. Nearby parameters and bounded output therefore share the
scale and total-input bounds used in \cref{lem:industry-cost}.
Take $(Q_n,\theta_n)\to(Q,\theta)$ with $Q>0$ and a finite-support
minimizer at $(Q,\theta)$. Update its machines. Trim any complete firm
whose output exceeds $Q_n$ to its first scale producing $Q_n$.
Inputs weakly fall and each adjusted output converges to its original value.
Reduce weights if total output exceeds $Q_n$, or fill a shortfall with
atomistic machines at vanishing cost. This gives
$\limsup_n\mathcal C(Q_n;\theta_n)\le\mathcal C(Q;\theta)$.

For the reverse inequality, choose minimizers with at most $K+3$ types.
Use the same scale, labelled unit composition, and total-input
representation as in \cref{app:cost-duality}, identifying the machine
component by its quantity. Pad the lists with zero coefficients.
The uniform bounds give a convergent subsequence. Joint continuity of
output per input, including at zero scale, preserves output, cost, and
all individual and aggregate constraints. The limiting finite list
implements an industry at $(Q,\theta)$, proving the lower bound.
At $Q=0$, use $0\le\mathcal C(Q;\theta)\le\bar c(\theta)Q$.

\paragraph{Step 2: values and quantities.}
Joint continuity gives uniform cost convergence on $[0,D_N(0)]$.
Berge's theorem applied to $U_N(Q)-\mathcal C(Q;\theta)$ gives continuous
maximal surplus and upper hemicontinuous maximizing quantities.

\paragraph{Step 3: limits of industries.}
Suppose equilibrium measures $\sigma_n$ converge weakly to a finite
measure $\sigma$ as $\theta_n\to\theta$. Active equality and
\cref{lem:industry-cost} give
$c_{\min}m_f\le p_nq_f\le\bar c(\theta_n)D_N(0)$ on each positive-scale
support. Thus firm sizes have a common bound. Weak convergence also
bounds total activity weights. On this size
range, inputs and costs are bounded continuous functions, and outputs
converge uniformly. Their integrals therefore converge, using continuous
cutoffs outside the common bound. In particular, $Q_n\to Q_\theta(\sigma)$.
Human feasibility passes to the limit, as does the market cap:
\[
 \int_{\F^+}(q_{\theta_n}(f)-Q_n)_+\,\de\sigma_n(f)=0.
\]
The two firm components are disjoint, so this integral also uses a bounded
continuous integrand on the relevant size range. Thus $\sigma$ is feasible
and its surplus is the limit of maximal surplus. Step 2 makes that limit
maximal at $\theta$, and \cref{thm:equilibrium-efficiency} gives equilibrium.
\end{proof}

Closed equilibrium limits do not imply a continuous selection. Moreover,
a firm with vanishing activity can still set a wage. Wage continuity
therefore requires control of the supported firms.

\Needspace{10\baselineskip}
\begin{definition}[Wage-interior industry]
\label{def:wage-interior}
An industry $\sigma$ is \emph{wage-interior} if its positive-scale support
is nonempty and compact, every firm on that support uses machines with
$D_{\bm y^M}F(\nu_f)>0$ and has strictly slack human staffing constraints,
and each human type $i$ satisfies
\[
 \int_\F\nu_f^H(\{\bm y^i\})\,\de\sigma(f)<\mu_i
 \quad\text{or}\quad
 \int_{\F^+}\nu_f^H(\{\bm y^i\})\,\de\sigma(f)>0.
\]
\end{definition}
Compactness makes the strict inequalities uniform and keeps supported
sizes away from zero. The last condition ensures that each fully employed
type has an employer with an interior match. \Cref{thm:wage-envelope}
therefore identifies wages with bid envelopes.

\begin{lemma}[Support closeness controls wages]
\label{lem:wage-continuity}
Consider wage-interior equilibrium industries $(\sigma_n,w^n)$ and
$(\sigma,w)$ with frontier machines $\bm y_n^M$ and $\bm y^M$. If
\[
 \bm y_n^M\longrightarrow\bm y^M
 \qquad\text{and}\qquad
 d_H\!\left(
 \supp(\sigma_n|_{\F^+}),
 \supp(\sigma|_{\F^+})
 \right)
 \longrightarrow0,
\]
where $d_H$ denotes Hausdorff distance between sets of firms, then the wage vectors converge:
\[
 \lVert w^n-w\rVert_\infty\longrightarrow0.
\]
\end{lemma}

\begin{proof}[Proof of \cref{lem:wage-continuity}]
For a firm under frontier $\bm y_n^M$, write
\[
 b_f^n(\bm y):=
 r\frac{D_{\bm y}F(\nu_f)}{D_{\bm y_n^M}F(\nu_f)}.
\]
Wage interiority and \cref{thm:wage-envelope} give, for every type $i$,
\[
 w_i
 =\max_{f\in\supp(\sigma|_{\F^+})}b_f(\bm y^i)\vee\underline w,
 \qquad
 w_i^n
 =\max_{f\in\supp(\sigma_n|_{\F^+})}b_f^n(\bm y^i)\vee\underline w.
\]
By \cref{ass:technology}(ii), the human and machine marginal products vary
continuously with the firm and the machine frontier away from zero scale.
The interiority margins at the limit industry keep the machine marginal
product bounded away from zero for all nearby supports. Hence every type's
bid varies continuously across paired firms. Hausdorff convergence pairs
each firm in either support with a nearby firm in the other, so the two
highest bids converge. The number of human types is finite, so convergence
holds in the sup norm.
\end{proof}

\section{Equilibrium concept and firm counts}
\label[appendix]{app:replacement-game}
\subsection{Complete firms and activity weights}
\label{app:convexification}
A recipe describes a complete firm; its activity weight counts copies.
Divisible weights relax integer firm counts without relaxing the limits
on each complete firm or aggregate human use.

At a fixed output requirement and capacity cap, divisible weights make the
cost problem convex. This relaxation is exact for atomistic firms and
approximates indivisible firms when each is small relative to the market.
With large firms, the concentration index measures complete-firm capacities.
An integer industry can have different cost and need not admit the same
exact supporting prices.

\subsection{The replacement game}
\label{app:formal-game}
We fix demand scale $N$. A one-shot challenge tests every possible
replacement of a status-quo contract.

\paragraph{Players and actions.}
Player $0$ represents consumers; player $i\in\{1,\ldots,K\}$ owns human
type $i$'s resources. Two disjoint copies of $[0,1]$ index organizers and
producers. Entrants have outside payoff zero. Machines are supplied
competitively at resource price $r$.

An organizer stays out or offers a contract
$\omega=(Q,\sigma,p,w,t)$ with $Q>0$, $\sigma\in\Sigma$,
$Q(\sigma)=Q$, $p\ge0$, $w_i\ge\underline w$, and transfers satisfying
\[
 t_E+t_0+\sum_{i=1}^Kt_i=0.
\]
The transfers pay the organizer, consumers, and human coalitions,
respectively. Each protected party accepts or rejects the offer.
Conditional on acceptance, a producer stays inactive or chooses an
intensity $a>0$ and a recipe
$f\in\F^0\sqcup\{g\in\F^+:q_g\le Q\}$.
Human coalition $i$ chooses employment $h_i\in[0,\mu_i]$.
The intensity is an unrestricted density over producers. Complete-firm
feasibility restricts recipes; aggregate clearing restricts deployed inputs.
Producer actions must be measurable and induce a finite industry measure.

\paragraph{Information and timing.}
All primitives, offers, actions, and payoffs are public.
Given an implemented status quo, or no contract, an inactive organizer
can make one challenge. Protected parties accept or reject simultaneously.
A rejection keeps the status quo and pays the challenger zero.
After unanimous acceptance, producers and human coalitions choose actions
at the proposed prices. The offer replaces the status quo only if optimal
actions implement it and clear markets; otherwise the status quo remains.
A successful offer commits consumers to buy $Q$ at the internal payment
$p$. The displaced organizer and producers receive zero.

\paragraph{Payoffs and clearing.}
Write $H_i(\sigma)$ for aggregate employment of type $i$. The protected
parties receive
\begin{equation*}
\label{eq:contract-payoffs}
 \Pi_0(\omega)=U_N(Q)-pQ+t_0,\qquad
 \Pi_i(\omega)=(w_i-\underline w)H_i(\sigma)+t_i.
\end{equation*}
The organizer receives $t_E$; without a contract all these net payoffs
are zero. A producer choosing $(a,f)$ earns
\[
 a\left[pq_f-r\nu_f^M(\Y)-\int_{\Y^H}w\,\de\nu_f^H\right],
\]
and inactivity earns zero. A human coalition maximizes
$(w_i-\underline w)h_i$; its transfer does not depend on its labor choice.
With inactive producers assigned intensity zero, implementation requires
\[
 \sigma(B)=\int_0^1a_e\one\{f_e\in B\}\,\de e
 \quad\text{for every Borel set }B,\qquad
 H_i(\sigma)=h_i,\qquad Q(\sigma)=Q.
\]
The offer has a \emph{competitive continuation} if optimal producer and
human actions satisfy these clearing conditions. As proved in
\cref{app:proof-equilibrium}, this is equivalent to
$(p,w)\in\Lambda(Q,\sigma)$.

\paragraph{Admissibility and solution concept.}
A contract is \emph{admissible} if it has a competitive continuation,
balanced transfers, and $t_E\ge0$, $\Pi_j(\omega)\ge0$ for every protected
party $j=0,\ldots,K$. Zero producer profits and balanced transfers give
the accounting identity
\[
 \Pi_0(\omega)+\sum_i\Pi_i(\omega)+t_E=\W_N(\sigma).
\]
Protected parties accept exactly when an offer preserves their
status-quo payoffs; an organizer enters if it can gain strictly.
An admissible offer $\widehat\omega$ \emph{blocks} an implemented contract
$\omega$ if
\[
 \Pi_j(\widehat\omega)\ge\Pi_j(\omega)\quad(j=0,\ldots,K),
 \qquad \widehat t_E>0.
\]
A contract is \emph{replacement-stable} if no admissible offer blocks it.
An equilibrium outcome $(\sigma,w)$ is supported by such a contract.
This is the stable-contract test in the main text, not Nash equilibrium
of a simultaneous-offer game. It tests every replacement without choosing
an order of proposals.

\section{Connection to the task model}
\label[appendix]{app:task-model}
In our model an agent brings its whole capability vector to every domain at
once. In a task model an agent has one unit of time and divides it among
tasks \citep{acemogluRestrepo2018,aghionJonesJones2019}. We show that the
second is a special case of the first. Switch coordination off, impose
constant returns, and read each productive domain as a task. Our framework
then delivers the task model exactly: the same output, the same assignment of
factors to tasks, and, except at finitely many output levels, the same wages
(\cref{thm:task-equivalence}). Under constant returns, the mean-capability technology corresponds,
up to output units, to equal time on every task. Optimal task assignment
allows agents to specialize.

\subsection{The task economy}
\label{app:task-economy}

We write the task model on its own terms first, with no reference to firms or
to our capability space. There are $n$ tasks. A final good is assembled from
task outputs $\bm z=(z_1,\ldots,z_n)$ by the constant-returns aggregator
\[
 L(\bm z)=\biggl(\frac1n\sum_{d=1}^nz_d^{\rho}\biggr)^{1/\rho},
 \qquad\rho<1,
\]
read as the geometric mean at $\rho=0$. Human type $i$ has mass $\mu_i$,
reservation wage $\underline w>0$, and task productivity $y^i_d$, measured
as output per unit of time. A machine has task productivity $y^M_d$ and can
be rented without limit at $r>0$ per unit of time. All task productivities lie in
$[\underline y,\overline y]$ with $\underline y>0$.

An \emph{allocation} divides each factor's time across tasks: masses
$\ell_{id}\ge0$ of type-$i$ time and $k_d\ge0$ of machine time on task $d$,
with $\sum_d\ell_{id}\le\mu_i$. Task outputs and the resource bill are
\[
 z_d=\sum_{i=1}^Ky^i_d\ell_{id}+y^M_dk_d,
 \qquad
 \underline w\sum_{i,d}\ell_{id}+r\sum_dk_d .
\]
Humans and machines are perfect substitutes within a task and differ only in
their task productivities. The final good sells at the demand price $P_N$.

\begin{definition}[Competitive equilibrium of the task economy]
\label{def:task-competitive}
An allocation together with task prices $\bm\pi$, wages $\bm w$, and an
output price $p$ is a \emph{competitive equilibrium} if
\begin{enumerate}[(a)]
 \item output clears at the demand price, $p=P_N(L(\bm z))$, and task prices
 are marginal revenue products, $\bm\pi=p\,\nabla L(\bm z)$;
 \item every factor works only where it earns most: $\ell_{id}>0$ only if
 task $d$ maximizes $\pi_dy^i_d$, and $k_d>0$ only if task $d$ maximizes
 $\pi_dy^M_d$;
 \item machines break even, $r\ge\max_d\pi_dy^M_d$, with equality if any
 machine works;
 \item each human type earns its best task or its outside option:
 \[
  w_i=\max\{\underline w,\max_d\pi_dy^i_d\}.
 \]
 Type $i$ works fully when its best task pays more than $\underline w$,
 and does not work when it pays less.
\end{enumerate}
\end{definition}

No firms appear in this description, and none are needed. Constant returns
leave the boundary between producers undetermined, which is why the task
literature works with an aggregate.

\subsection{The task technology}
\label{app:task-production}
Use $F^{\mathrm T}$ from \cref{sec:task-connection}: a firm assigns its humans and machines
optimally across tasks and pays each agent once. Keep all other
primitives fixed. \Cref{lem:task-technology} verifies \cref{ass:technology},
so equilibrium existence, efficiency, and cost duality apply.
Constant returns give $F^{\mathrm T}(m\nu)=mF^{\mathrm T}(\nu)$ and
$\gamma\equiv1$. The large-market result therefore does not apply.
Nor do the affine bids: task output depends on the distribution of skills,
so $F^{\mathrm T}$ need not satisfy \cref{ass:level-scale}.
Its marginal product instead equals the best-task value in
\eqref{eq:best-task}.

\subsection{The equivalence}
\label{app:task-equivalence}

\textbf{How can an economy of firms match an economy with none?} Because pooling cannot lower output and preserves resource cost
(\cref{lem:task-pooling}). Least-cost industries therefore attain the
same output as a single firm that combines their inputs.

Our equilibrium differs from the competitive one in two further ways: its internal payment
$p$ need not equal the demand price, and its contracts include balanced
transfers. We therefore compare output, assignments, and wage vectors, not
final payoffs. Write $\partial\mathcal C(Q)$ for the subdifferential of least
cost at $Q$, the slopes of the lines that touch $\mathcal C$ at $Q$ and lie
below it.

\begin{theorem}[The task model is our model with the task technology]
\label{thm:task-equivalence}
Fix a demand scale $N$ and run our economy with $F^{\mathrm T}$. Then:
\begin{enumerate}[(i)]
 \item \textbf{Same output.} Efficient output $Q^*_N$ is unique. An industry
 belongs to an equilibrium outcome of our economy if and only if it produces
 $Q^*_N$ at least cost. A competitive equilibrium of the task economy exists,
 every competitive equilibrium produces $Q^*_N$, and the competitive
 allocations are exactly the aggregates of our equilibrium industries.
 \item \textbf{Same assignment.} In both economies every employed factor
 works only on tasks that maximize $\pi_dy_d$, at task prices proportional to
 $\nabla L(\bm z^*)$ evaluated at the common task outputs. How our economy
 splits that allocation across firms is indeterminate.
 \item \textbf{Same wages.} Every competitive wage vector is an equilibrium
 wage vector of ours, and the two sets coincide if and only if $\mathcal C$
 is differentiable at $Q^*_N$. Wages are then
 $w_i=\max\{\underline w,\max_d\pi_dy^i_d\}$ in both. Otherwise our
 equilibrium supports every payment in $\partial\mathcal C(Q^*_N)$ and the
 competitive equilibrium selects $P_N(Q^*_N)$ from it.
 \item \textbf{Finitely many exceptions.} $\mathcal C$ is differentiable at
 $Q^*_N$ whenever some least-cost industry there uses machines, or employs
 some human type in an amount strictly between $0$ and its supply. It can
 therefore fail only at an output of the form
 $F^{\mathrm T}\bigl(\sum_{i\in S}\mu_i\delta_{\bm y^i}\bigr)$ for a set $S$
 of human types, so at most $2^K$ outputs in all.
\end{enumerate}
\end{theorem}

Both economies pick the least-cost way to make the efficient quantity, and
they assign factors to tasks in the same way. They can part company only over
wages, and only where least cost has a kink. There the competitive
equilibrium lets demand choose one payment, $P_N(Q^*_N)$, while ours accepts
every supporting payment, because consumers and workers can divide the
surplus through transfers. By part (iv) a kink needs every employed human
type to be fully employed and no machine to work. With one human type that
happens only before automation begins. Once any machine works, our wages are
the competitive wages.

\begin{example}[A kink before automation]
\label{ex:task-kink}
Let $n=3$ and $\rho=-1$. One human type has mass one and productive capability profile
$(1,1,1)$. The machine has productive capability profile $(1.2,1,.8)$. Coordination
entries play no role. Let $\underline w=.2$, $r=.5$, and $P(Q)=(2-3Q)_+$ at
$N=1$. A machine-only firm produces about $.327$ per machine, so its unit cost
is about $1.53<P(0)$ and the standing condition holds. Humans alone produce
$1/3$ per unit, with task prices $(1/3,1/3,1/3)$. By \eqref{eq:best-task}, one
more human adds $1/3$ and the first machine adds $.4$, on task 1. Least cost
is therefore $\mathcal C(Q)=.6Q$ up to $Q=1/3$, and \eqref{eq:task-upper}
shows that no firm with machines is cheaper. Beyond $1/3$, output needs
machines, and the first units cost $r/.4=1.25$ per unit of output. Since
$.6<P(1/3)=1<1.25$, efficient output is $Q_1^*=1/3$, produced by humans
alone, and $\partial\mathcal C(1/3)=[.6,1.25]$. The competitive equilibrium
sets $p=1$ and $w=p/3=1/3$. Our equilibrium supports every
$p\in[.6,1.25]$, so every wage $w=p/3\in[.2,5/12]$. At the lowest wage,
humans earn their outside option. At the highest, the first machine just
breaks even.
\end{example}

Under constant returns, the mean-capability technology instead gives
\[
 F(\nu)=L\left(\int_\Y y_1\,\de\nu,\ldots,\int_\Y y_n\,\de\nu\right).
\]
An equal time assignment produces $F(\nu)/n$, so
$F(\nu)\le nF^{\mathrm T}(\nu)$. Optimal task assignment can raise output
by letting agents specialize. Its marginal product is the best-task value
in \eqref{eq:best-task}; the mean-capability technology instead gives the
affine bids in \eqref{eq:affine-bid}.

\section{CES examples and numerical methods}
\label[appendix]{app:numerics}
\label[appendix]{app:proof-ces}
We give the calculations for the current CES figures and distinguish analytic
proofs from numerical checks. Parameters appear beside each calculation or
in the cited figure footnote. The large-market wage proof is in
\cref{app:jump-details}.

\subsection{Common CES formulas}
\label{app:ces-tools}
At mean coordination $z$, preferred scale is
\begin{equation}
 m^*(z)=
 \begin{cases}
  0,&z\le1-\phi,\\
  \phi/(1-z)-1,&1-\phi<z<1,\\
  +\infty,&z\ge1.
 \end{cases}
 \tag{CES-SCALE}\label{eq:ces-preferred-scale}
\end{equation}
For $1-\phi<z<1$, define the optimized log scale factor by
\begin{equation}
 \ell(z):=\phi-(1-z)+(1-z)\log\frac{1-z}{\phi}.
 \tag{CES-PEAK}\label{eq:ces-scale-peak}
\end{equation}
The best average product at a composition is then its level factor times
$e^{\ell(z)}$. For $z\le1-\phi$, the best scale factor is one, attained at
zero scale. At $z=1$, its supremum is $e^\phi$ and is not attained at a
finite scale. For $z>1$, it is unbounded. 

\begin{proof}[The CES scale formulas]

At mean coordination $z$, the CES average product is proportional to
$(1+m)^{z-1}e^{\phi m/(1+m)}$. Its log derivative in scale is
\[
 \frac{z-1}{1+m}+\frac{\phi}{(1+m)^2}.
\]
The derivative changes sign at most once, from positive to negative.
Its sign gives \eqref{eq:ces-preferred-scale}, including the zero-scale
corner and the unattained optima. Substitution gives
\eqref{eq:ces-scale-peak}.\qedhere

\end{proof}

\paragraph{A common verification method.}
At equal unit input costs $r=\underline w=1$, a bound $q_f\le A m_f$
for every complete firm, with the corresponding bound for atomistic
activities, implies $\W_N(\sigma)\le U_N(Q)-Q/A$.
An industry that attains this bound at its maximizing quantity is efficient.
If its firms can be staffed and fit in the market, the prices $p=1/A$ and
$w=1$ support it, so \cref{thm:equilibrium-efficiency} gives equilibrium.
If one share and scale uniquely attain the bound, every equilibrium uses
them. Each calculation below checks the required feasibility conditions.

For a separate-production bound $q_f\le \overline a(1)h+\overline a(0)a$, the same argument
uses the pure human and machine activities. At $p=1/\overline a(0)$, humans work
fully when $\overline a(1)>\overline a(0)$ and do not work when $\overline a(1)<\overline a(0)$, provided demand
leaves positive machine output and both pure firms fit. The wage is
$\max\{1,\overline a(1)/\overline a(0)\}$. At equality, every feasible employment split has the
same cost. We verify the output bound and market fit in each application.

\subsection{Firm-size envelopes}
\label{app:check-ces-size}\label{app:ces-size-bounds}
\begin{proof}[Verification of \cref{fig:ces-size-envelopes}]
Set $\phi=.3$, human coordination $y_*^H=.95$, and human supply $\mu=2$.
Both productive skills equal one for humans and AI. We vary AI coordination
$y_*^M\in[.70,.975]$. At human share $\lambda$, preferred size capped by
human supply is
\[
 \min\left\{m^*\bigl(y_*^M+\lambda(.95-y_*^M)\bigr),\frac2\lambda\right\},
\]
with no staffing cap at $\lambda=0$. Panel (a) fixes $\lambda=1/2$:
size reaches its cap of four at $y_*^M=.93$. Panel (b) takes the lower
and upper envelopes over shares.

The lower envelope is
\begin{equation}
 \underline M(y_*^M,2)
 =\min\{m^*(\min\{y_*^M,.95\}),2\}.
 \tag{SIZE}\label{eq:ces-size-bounds}
\end{equation}
Every composition has preferred size at least the first term and a
staffing cap at least two. A pure composition attains the smaller term.
The lower envelope rises from zero at $.70$ to two at $.90$.

Below $.95$, preferred size rises with the human share while the staffing
cap falls. At share zero the former binds; at share one the latter binds,
since $m^*(.95)=5>2$. Their unique crossing therefore gives the upper
envelope. Substituting $\lambda=2/m$ into the preferred-size formula gives
\begin{equation}
 (1-y_*^M)m^2+(y_*^M-1.2)m-2(.95-y_*^M)=0.
 \tag{CAP}\label{eq:ces-staffing-crossing}
\end{equation}
The leading coefficient is positive and the constant is negative, so the
quadratic has exactly one positive root. At $y_*^M=.85$, that root is
$(7+\sqrt{97})/6$, with human share $12/(7+\sqrt{97})$.

At or above $.95$, no composition has preferred size above the AI-only
size, which has no staffing cap. Thus the upper envelope is
$m^*(y_*^M)=.3/(1-y_*^M)-1$. Both expressions equal five at $.95$.
These envelopes bound equilibrium sizes under \cref{thm:firm-scale};
the shaded band does not assert that every size occurs in equilibrium.
\end{proof}

\subsection{Coordination below the scale threshold}
\label{app:check-ces-coordination}
\label{app:ces-coordination}
\begin{proof}[Verification of \cref{fig:coordination-firms}]
Set $\bm y^H=(1,2,.5)$, human supply $20$, $\phi=1$, $\rho=-1$,
$r=\underline w=1$, and $P(Q)=(2-Q/10)_+$. For path progress
$\theta\in[0,1]$, we first find the best composition, then check market
fit and human supply.

\paragraph{Step 1: optimal size and composition.}
AI's first skill is two on all three paths. Write $b=y_2^M$ for its second
skill and $d=y_*^M-.5$ for its coordination advantage. Red has $b=1$ and $d=.05+.2\theta$; blue has $b=1+\theta$ and $d=.05$.
Dark green has $b=1+\theta$ and $d=0$.
At machine share $s$, the mean skills are $x=1+s$ and $y=2+(b-2)s$.
The coordination shortfall is $q=1-\overline y_*=.5-ds$.
Preferred size is $m=1/q-1\in[1,3]$, and log average product is
\[
 g(s)=\log\frac{2xy}{x+y}+1-q+q\log q.
\]
Both inputs cost one, so firms maximize $g$. Differentiation gives
\[
 g_s=\frac1x+\frac{b-2}{y}-\frac{b-1}{x+y}+d\log(1/q),
 \qquad
 g_{ss}=-\frac1{x^2}-\frac{(b-2)^2}{y^2}
          +\frac{(b-1)^2}{(x+y)^2}+\frac{d^2}{q}.
\]
On red, $g_{ss}\le-1/2+.25^2/.25<0$. On blue and dark green, $x\le2$,
$x+y\ge3$, $0\le b-1\le1$, and $q\ge.45$, so
$g_{ss}\le-1/4+1/9+.05^2/.45<0$.
Thus all three paths have a unique maximizing share, including corners.
Initially $g_s(1/2)>0$ on red and blue, so their optimal share exceeds
$1/2$. Dark green starts at share $1/2$ and size one.
On red, $g_s(0)>0$ and $g_s(1)\le-1/2+.25\log4<0$,
so the optimum stays interior. Since
\[
 g_{sd}=\log(1/q)+ds/q>0,
\]
the share rises strictly. On blue and dark green,
\[
 g_{sb}=\frac{2}{y^2}-\frac{3}{(x+y)^2}>0,
\]
because $x\ge1$, $y\le2$, and hence $y/(x+y)\le2/3$.
The share rises until it reaches one and then stays there; at $b=2$,
$g_s(1)=1/4+d\log(1/(.5-d))>0$.

\paragraph{Step 2: attain the surplus bound.}
Let $A=\exp(g(s))$ at the maximizing share. The equal-coordination mixed firm provides the lower bound
$A\ge(3/2)\exp[(1-\log2)/2]>1.5$ along all three paths.
For every composition, the skill level is at most two and the scale factor
is at most $\exp[.75+.25\log(.25)]<1.5$, so average product is below three.
Optimal output is $Q=10(2-1/A)>40/3$.
Every preferred firm produces less than $3\cdot3=9<Q$ and uses fewer than
$20$ humans. Copies of the optimal firm with activity weight $Q/(mA)$
use $(1-s)Q/A<20$ humans in aggregate. The common verification therefore
gives equilibrium at price $p=1/A$ and wage $w=1$, with surplus $Q^2/20$.
Strict composition optimality and unique preferred sizes identify the industry.

\paragraph{Step 3: separate the two size responses.}
On blue, size is $m=1/(.5-.05s)-1$, so
\[
 \frac{\de m}{\de\theta}
 =\frac{.05}{(.5-.05s)^2}\frac{\de s}{\de\theta}.
\]
Size rises only through composition and stops rising when the machine share
reaches one. On dark green, $d=0$, so mean coordination stays at $.5$
and size stays at one despite the rising machine share.
On red, differentiation gives
\[
 \frac{\de m}{\de\theta}
 =\frac{.2s+d\,\de s/\de\theta}{(.5-ds)^2}.
\]
The first term raises size at fixed composition. The second is strictly
positive throughout the interior of the path, so composition amplifies growth.
All three paths raise average product, output, and surplus. At the same time,
activity weight falls: total input $Q/A=10(2/A-1/A^2)$ falls with $A>1$,
and firm size is nondecreasing. The figure plots size and share, not firm counts.
\end{proof}

\subsection{Competence and human exit}
\label{app:check-exit}
\label{app:exit-figure}
\begin{proof}[Verification of \cref{fig:human-exit}]
Both paths set $\rho=-1$ and $r=\underline w=1$. We bound output for
every firm, then construct industries that attain the surplus bound.

\paragraph{Step 1: complementary mixing.}
Set $\bm y^H=(1,2,.9)$, $\bm y^M=(1.5,\theta,.9)$ for
$\theta\in[1,2]$, $\phi=.3$, $\mu=20$, and $P(Q)=(2-Q/10)_+$.
Every share has unique preferred size $m^*=2$. Use the optimized log
scale factor $\ell$ from \eqref{eq:ces-scale-peak}. The best average
product at human share $\lambda$ is
\[
 \overline a(\lambda;\theta)
 =e^{\ell(.9)}
 \left[\frac{2(1.5-.5\lambda)
                 (\theta+(2-\theta)\lambda)}
 {1.5+\theta+(1.5-\theta)\lambda}\right].
\]
The two productive capability profiles are not proportional anywhere on
this path. Writing $x=1.5-.5\lambda$ and
$y=\theta+(2-\theta)\lambda$, differentiation gives
\[
 \overline a_{\lambda\lambda}
 =-\frac{4e^{\ell(.9)}(\theta-3)^2}{(x+y)^3}<0.
\]
Thus mixing is strictly complementary throughout the path.

The unique maximizing share $\lambda^*$ is the interior root of
\[
 \frac{\partial}{\partial\lambda}\log\overline a
 =\left[
 \frac{-.5}{1.5-.5\lambda}
 +\frac{2-\theta}{\theta+(2-\theta)\lambda}
 -\frac{1.5-\theta}{1.5+\theta+(1.5-\theta)\lambda}
 \right]=0
\]
when $1\le\theta<1.5$, and is zero when $1.5\le\theta\le2$.
Indeed, the derivative at $\lambda=1$ is $-\theta/6<0$,
while its sign at zero is the sign of
$4.5-2.25\theta-.5\theta^2$, which vanishes at $\theta=1.5$.
Differentiating the first-order condition, with $\overline y_1$ and
$\overline y_2$ denoting the two mean productive capabilities, gives
\[
 \frac{\de\lambda^*}{\de\theta}
 =-\frac{(1-\lambda^*)\overline y_2+\overline y_1^2}
 {(2-\theta)(\overline y_1+\overline y_2)}<0
 \qquad (1\le\theta<1.5).
\]
The share thus falls and converges to zero at $1.5$. At the transition, the condition in \cref{prop:exit}(i) holds with equality.

Let $A:=\overline a(\lambda^*;\theta)$ be the largest average product.
Apply the common verification method above. The unique maximizing output is
$Q=10(2-1/A)$. Since $1<A<3$, this output exceeds $10$, while every
preferred-scale firm produces less than $6$ and uses at most two humans.
Use weight $Q/(2A)$ of the firm with share $\lambda^*$ and size two.
Total human use is $\lambda^*Q/A<20$, so human supply stays slack.
The common verification gives equilibrium at $p=1/A$, $w=1$.
Strict concavity and the unique preferred scale make the industry unique.
As $\theta$ rises, $A$ increases and $Q/A=20/A-10/A^2$ decreases because
$A>1$. Both the human share and human employment therefore fall continuously.

\paragraph{Step 2: segregating mixing.}
Set $\bm y^H=(1.1,1.1,.985)$, $\bm y^M=(\theta,\theta,.9999)$ for
$\theta\in[1,1.2]$, $\phi=.02$, $\mu=2/3$, and
$P(Q)=(2-Q/1000)_+$.
At human share $\lambda$, mean competence is
$\theta+(1.1-\theta)\lambda$ and mean coordination is
$.9999-.0149\lambda$. Thus
\[
 \frac{\partial^2}{\partial\lambda^2}\log\overline a
 =-\frac{(1.1-\theta)^2}
           {[\theta+(1.1-\theta)\lambda]^2}
   +\frac{.0149^2}{.0001+.0149\lambda}
 \ge -\frac{1}{100}+\frac{.0149^2}{.015}
 =\frac{7201}{1500000}>0.
\]
Strict log convexity implies strict convexity of $\overline a$.
Every firm with human input $h$ and machine input $a$ consequently obeys
\[
 F(h\delta_{\bm y^H}+a\delta_{\bm y^M})\le \overline a(1)h+\overline a(0)a,
 \qquad
 \overline a(1)=1.1e^{\ell(.985)},\quad
 \overline a(0)=\theta e^{\ell(.9999)}.
\]
Pure firms attain the bound at sizes $1/3$ and $199$, respectively.

The bound gives industry resource cost at least
$Q/\overline a(0)+(1-\overline a(1)/\overline a(0))T_H$, where $T_H\in[0,\mu]$ is total human employment.
Maximize the resulting surplus bound. Output is
$Q=1000(2-1/\overline a(0))$; humans supply $T_H=\mu$ if $\overline a(1)>\overline a(0)$ and $T_H=0$
if $\overline a(1)<\overline a(0)$. At equality, any $T_H\in[0,\mu]$ attains the bound.
The required machine input is $(Q-\overline a(1)T_H)/\overline a(0)>0$.
Throughout this path $1<\overline a(0)<2$ and $\overline a(1)<2$, so $Q>1000$,
whereas the machine firm produces less than $398$ and the human firm less
than $2/3$. The human firm needs $1/3<\mu$ workers. Both preferred-scale
firms can therefore be staffed and fit in the market, and their activity
weights implement every stated allocation. At prices $p=1/\overline a(0)$ and
$w=\max\{1,\overline a(1)/\overline a(0)\}$, the output bound verifies support for all firms.
The constructed industries are equilibria and attain the global surplus
bound. Strict convexity rules out mixed firms in every equilibrium.
The unique crossing solves
\[
 \theta=1.1\exp\left(\ell(.985)-\ell(.9999)\right)
 \simeq1.0796369.
\]
At this crossing, the full interval of employment fractions in
\cref{fig:human-exit}(f) consists of equilibria.
\end{proof}

\subsection{Human coordinators in AI-heavy firms}
\label{app:check-wage-sorting}
We reduce entry to the edges of the composition triangle, then check
\cref{fig:v4-wage-sorting} numerically at 99 sampled AI profiles.
Productive humans have profile $(6,6,.2)$ and supply $10$; coordinators
have profile $(1,1,.9)$ and supply $1$; AI has profile
$(\theta,\theta,.2)$ with $\theta\in[4,12]$.
The other parameters are $\rho=-1$, $\phi=3$, $r=\underline w=1$, and
$P(Q)=2(1-Q/3000)_+$.

\paragraph{Reduction to two-type edges.}
For productive-human, coordinator, and machine shares
$(s_P,s_C,s_M)$ summing to one, mean competence is
$6s_P+s_C+\theta s_M$, and mean coordination is $.2+.7s_C$.
Preferred size is $3/(.8-.7s_C)-1>0$, and optimized average product is
\[
 (6s_P+s_C+\theta s_M)e^{\ell(.2+.7s_C)},
\]
where $\ell$ is defined in \eqref{eq:ces-scale-peak} with $\phi=3$.
At fixed coordinator share, this expression is affine in the productive
human share, as is the input bill. Thus profit per unit of input is
maximized at an endpoint of each fixed-coordinator-share segment.
Checking the two-type edges therefore bounds entry by every composition;
optimizing size first also covers every scale, including the zero-size
limit.

\paragraph{Prices, allocations, and numerical bounds.}
The solver imposes zero profit on active firm types, composition
optimality for their positive shares, demand clearing, and human-supply
complementarity. A slack human supply fixes its wage at one; a wage
above one requires full employment of that type.
For each candidate state, adaptive Taylor bounds on each edge give an
upper bound on entry profit per unit of input, with a $10^{-11}$
floating-point guard. All 99 bounds are below $10^{-7}$, and the
largest equilibrium-equation residual is below $3\times10^{-14}$.
Each active firm uses less than the available stock of each human type
and produces less than market demand. These are numerical checks at the
sampled points, not an exact certificate for the interpolated segments.

At $\theta=4$, all humans work together, with wages about $1.501$ for
productive humans and $1.624$ for coordinators; AI-only firms supply the
remaining output. At $\theta=12$, productive humans are unemployed.
Coordinators remain fully employed at wage $1.005586$ in firms with
coordinator share $.0127096$ and machine share $.9872904$.

\begin{flushleft}\small
Source: \path{images/python code/make_coordinator_ai_sorting.py}.\\
Records: \path{intel-figures/coordinator_ai_sorting_checks.csv} and
\path{intel-figures/coordinator_ai_sorting_firms.csv}.
\end{flushleft}

\subsection{Separate and mixed firms in the CES example}
\label{app:ces-composition-regimes}
\begin{proof}[Verification of \cref{fig:ces-composition-regimes}]
We derive the envelopes and check prices, staffing, and market fit.
Both panels use $\rho=-1$, $\phi=1$, $r=\underline w=1$,
human supply $50$, and $P(Q)=(2-Q/100)_+$.
For mean coordination $z\in[.4,.8]$, preferred size is
$m^*(z)=1/(1-z)-1>0$. Define the optimized scale factor by
\[
 \mathcal S(z):=\frac{S(m^*(z),z)}{m^*(z)}
 =e^z(1-z)^{1-z}.
\]
Its log derivatives are $(\log\mathcal S)'(z)=-\log(1-z)$ and
$(\log\mathcal S)''(z)=1/(1-z)>0$.

\paragraph{Panel (a): separate firms.}
Humans have profile $(1,1,.8)$ and machines have profile $(1,1,.4)$.
Productive competence is one at every human share, so
\[
 \overline a(\lambda,w)
 =\frac{\mathcal S(.4+.4\lambda)}{1-\lambda+\lambda w}.
\]
Set $w=\mathcal S(.8)/\mathcal S(.4)>1$ and
$p=1/\mathcal S(.4)$. Both endpoints then equal $1/p$.
The log envelope is strictly convex because
\[
 \frac{\partial^2}{\partial\lambda^2}\log\overline a(\lambda,w)
 =\frac{.16}{.6-.4\lambda}
  +\frac{(w-1)^2}{[1+\lambda(w-1)]^2}>0.
\]
Every interior share therefore has cost-adjusted best average product
below $1/p$. Since the envelope optimizes scale, this bound supports
prices against every firm and excludes active mixed firms.

Demand gives $Q=100[2-1/\mathcal S(.4)]\simeq108.927$.
Employ all $50$ humans in human-only firms of size $4$, producing
$50\mathcal S(.8)\simeq80.651$. Supply the remaining output with
machine-only firms of size $2/3$, using total machine input
\[
 \frac{Q-50\mathcal S(.8)}{\mathcal S(.4)}\simeq25.751>0.
\]
Both active firm types break even. Human supply binds, as required by
$w\simeq1.469>1$. Each human-only firm uses $4<50$ humans and
produces $4\mathcal S(.8)\simeq6.452<Q$; each machine-only firm
produces $(2/3)\mathcal S(.4)\simeq.732<Q$.

\paragraph{Panel (b): mixed firms.}
Humans have profile $(2,1,.6)$ and machines have profile $(1,2,.6)$.
Mean coordination is $.6$ and preferred size is $3/2$ at every share.
At wage $w=1$, the harmonic-mean competence aggregator gives
\[
 \overline a(\lambda,1)
 =\mathcal S(.6)\frac{2(1+\lambda)(2-\lambda)}{3}.
\]
Its second derivative is $-4\mathcal S(.6)/3<0$, and its unique
maximum is at $\lambda=1/2$, with value $3\mathcal S(.6)/2$.
Set $p=2/[3\mathcal S(.6)]$ and
$Q=100\{2-2/[3\mathcal S(.6)]\}\simeq147.215$.
Use only firms of size $3/2$ with human share $1/2$.
Total human input is
\[
 \frac{Q}{3\mathcal S(.6)}\simeq38.854<50,
\]
so human supply is slack and $w=\underline w=1$ satisfies human scarcity.
Each firm uses $3/4<50$ humans and produces
$9\mathcal S(.6)/4\simeq2.842<Q$.
The envelope bound proves firm support, and each active firm breaks even.

In both panels, activity equals total input divided by firm size.
The stated prices clear demand and satisfy support and human scarcity,
so \cref{thm:equilibrium-efficiency} gives equilibrium.
\end{proof}

\subsection{Numerical checks and records}
\label{app:ces-check-scope}\label{app:ces-records}
The analytic proofs above cover all feasible shares and scales in their
stated cases. The table records checks for the numerical figures; their
bounds apply only at sampled parameter values and to the stated tolerances.

\begin{table}[htbp]
\centering\small
\renewcommand{\arraystretch}{1.2}
\begin{tabular}{@{}>{\raggedright\arraybackslash}p{.23\textwidth}>{\raggedright\arraybackslash}p{.42\textwidth}>{\raggedright\arraybackslash}p{.25\textwidth}@{}}
\toprule
Example & Check and scope & Record\\
\midrule
Human coordinators & Entry upper bounds below $10^{-7}$ per unit of
input at 99 points; all compositions and scales. &
\path{coordinator_ai_sorting_checks.csv}\\
Finite-market concentration & Pure-firm optimization, with additional entry
searches at selected states. &
\path{ces_support_jump_certificate.csv};
\path{wage_jump_checks.json}\\
Large changes in firms & Finite-menu optimization at 21 points; additional
entry searches find profits up to $4.7\times10^{-6}$. &
\path{ces_tipping_certificate.csv}\\
\bottomrule
\end{tabular}
\caption{Computational checks. The coordinator records are in
\protect\path{intel-figures/}; other file paths and supplementary checks
are in \protect\path{images/v5/canonical_ces/REPRODUCTION.md}.}
\end{table}

\paragraph{Finite-market concentration.}
\label{app:concentration-numerics}
The curves in \cref{fig:replication-aggregation,fig:ces-support-jump}
optimize output over pure human and machine firms. Additional entry searches
at selected states do not establish equilibrium over the full paths.
The analytic concentration result and pointwise wage limit instead follow
from \cref{prop:market-size} and the verification in
\cref{app:jump-details}, respectively.

\paragraph{Large changes in firms.}
\label{app:restored-wage-comparison}
\Cref{fig:re-amplification} uses a finite menu with
atomistic activities and the market constraint. A finer menu at the
start, wage peak, and endpoint changes either wage by less than $10^{-4}$.
The recorded entry searches do not bound profit over all feasible firms.
For the dashed curve, the calculation upgrades machines in the initial
machine-using firms, holds their inputs fixed, and adds the change in the
bid envelope to the initial wage. Human-only employers and binding staffing
limits can prevent the initial envelope from equaling that wage. The gap
between the curves is therefore a descriptive comparison, not an application
of the interior-match derivative formula throughout the path.
The snapshots show the largest human employer; the diamond denotes
atomistic production and has no firm-size meaning.

\clearpage
\section{Calculations for the example}
\label[appendix]{app:example-calc}

This appendix records the parameters, allocations, and checks behind the example in \cref{sec:example-economy,sec:displacement}. The two markets share one population of six coders, six writers, and two coordinators. Employment totals therefore count each human once. Compute costs $r=1$ per copy. The human outside option is $1/4$, except in \cref{sec:displacement}, where it is one.

\subsection{Capabilities and firm scale}
All illustrated AI profiles have equal task skills, $\bm y^M=(\theta,\theta,y_*^M)$, so AI competence equals $\theta$. The human population averages are
\[
 \bar L_H=\frac{12(1.8)+2}{14}=\frac{59}{35},
 \qquad
 \bar y_*^H=\frac{12(0.75)+2(0.97)}{14}=\frac{547}{700}.
\]
For these symmetric profiles, the regime definitions in \cref{sec:minimal-agi} reduce to
\[
\begin{array}{rll}
 \text{Minimal AGI:}&\bar L_H\le\theta\le9,&\bar y_*^H\le y_*^M<1,\\
 \text{Supercompetent:}&\theta>9,&y_*^M\le1,\\
 \text{Supercooperative:}&\theta\le9,&y_*^M>1,\\
 \text{ASI:}&\theta>9,&y_*^M>1.
\end{array}
\]
The boundary $y_*^M=1$ has a finite productivity ceiling. It must be distinguished from coordination strictly above one.

At fixed mean capabilities, write $z=\bar y_*$ for mean coordination. The scale formulas in \cref{app:ces-tools}, with $\phi=7/4$, give
\[
 m^*(z)=\frac{7}{4(1-z)}-1,
 \qquad
 G(z)=e^{7/4-(1-z)}\left(\frac{1-z}{7/4}\right)^{1-z},
 \qquad 1-\phi<z<1.
\]
Best output per worker is $L G(z)$. Every below-one coordination value used in the example lies in this range. Thus $m^*(3/4)=6$, $G(3/4)\approx2.75529$, $m^*(\bar y_*^H)\approx7.00654$, and $G(\bar y_*^H)\approx2.93511$.

At coordination one, output per worker approaches the finite, unattained ceiling $Le^{7/4}$ as firm size grows. Strictly above one, it grows without bound. In a finite market, a complete firm must fit within market output and human supplies. At specialist and coordinator shares $s_s$ and $s_c$, its productivity-maximizing feasible size is
\[
 m=\min\left\{m^*(z),\frac{12}{s_s},\frac{2}{s_c},m_{\mathrm{fit}}(Q)\right\},
 \qquad F(m_{\mathrm{fit}}(Q))=Q.
\]
Here shares remain fixed when solving for $m_{\mathrm{fit}}(Q)$; we omit a staffing bound when its share is zero and set $m^*(z)=+\infty$ for $z\ge1$.

\subsection{Markets, prices, and numerical method}
\paragraph{Positive production.}
Some example profiles fail the standing atomistic-machine viability
condition: for minimal AGI, $r/L(\delta_{\bm y^M})=35/59>0.3$.
The cost and supporting-price arguments in \cref{app:cost-duality} still
apply; that condition ensures positive surplus, not cost attainment.
Here a feasible human allocation gives positive surplus directly. Two
balanced specialist firms of size six produce $Q=60G(3/4)\approx165.32$
in the niche market. Even at outside wage one, their surplus is
$0.3Q-0.00075Q^2-12\approx17.10>0$. Each complete firm fits within total
output, and together they use the twelve specialists once. Thus the
example has positive production at either outside wage.

For large-market output, replace atomistic machines in the argument of
\cref{app:large-output} by a fixed finite AI firm with unit cost below
$0.3$. The minimal-AGI preferred firm has unit cost about $0.20211$;
the other displayed AI profiles also admit such a firm. Replicating this
recipe is feasible once market output exceeds its capacity and gives
surplus proportional to market size. Hence efficient output diverges.
This argument does not establish convergence to the reported limiting
allocations.

Firm activity weights can be fractional. A complete firm with size $m$ and activity weight $a$ contributes $am$ inputs and $aF(m)$ output. Its complete recipe must still fit within the market and human supplies. A box in \cref{fig:economy-regimes,fig:displacement-composition} shows that recipe; its multiplier gives the activity weight. These weights are not integer firm counts.

We minimize resource cost at each niche output, allowing humans to work in either market, then maximize surplus over output. We recover internal prices and wages from the cost problem and search for profitable feasible entrants. Note here that the internal price $p$ finances production and need not equal
buyers' marginal willingness to pay $P(Q)$. For one pure-AI firm
serving the whole market at an interior efficient size,
\[
 P(F(m))F'(m)=r,\qquad pF(m)=rm.
\]
When average product rises with size, $mF'(m)>F(m)$, so
$p>P(F(m))$. Transfers then reconcile these two prices.

Without AI, fixed human supply keeps deep-market output bounded and its demand price tends to $0.3$. With AI coordination below one, the limiting deep-market price is instead $p=r/[\theta G(y_*^M)]$ whenever this unit cost is below $0.3$, as it is in the displayed allocations. Deep output then satisfies
\[
 \frac{Q}{200N}\longrightarrow1-\frac{p}{0.3}.
\]
At coordination one the limiting unit cost is $r/(\theta e^{7/4})$. Strictly above one it tends to zero; a demand price tending to zero requires $Q/(200N)\to1$. We use these limiting prices in the joint allocation calculation. The numerical records do not establish convergence of finite-market equilibria to each displayed allocation.

\Needspace{5\baselineskip}
For wage calculations at positive scale, the full marginal product of an added worker with profile $\bm y$ is
\[
 D_{\bm y}F=\frac{F}{m}\left[
 1+\frac{\nabla L\cdot(\bm y_{1:2}-\bar{\bm y}_{1:2})}{L}
 +\log(1+m)(y_*-z)
 +\frac{m(z-1)}{1+m}+\frac{\phi m}{(1+m)^2}
 \right].
\]
The task-skill vectors $\bm y_{1:2}$ and $\bar{\bm y}_{1:2}$ contain the first two coordinates, and we evaluate $\nabla L$ at the firm's mean task skills. The last two terms cancel at an unconstrained best size. Equal mean task skills give $\nabla L=(1/2,1/2)$. An interior employment choice satisfies $w=pD_{\bm y}F$; a binding staffing or market constraint requires the corresponding boundary condition.

\subsection{Human-only benchmarks}
For one human type, six coders at best size produce $6(1.8)G(3/4)\approx29.76$, compared with $21.79$ from six solo coders. Twelve coders produce $59.51$ in two firms, versus $57.22$ in one merged firm. The deep-market wage is $0.3(1.8)G(3/4)\approx1.488$.

Equal coder and writer shares raise competence to five while keeping coordination at $3/4$. Two balanced six-person firms produce $2(6)(5)G(3/4)\approx165.32$, versus $158.93$ in one merged firm. The specialist wage is $0.3(5)G(3/4)\approx4.133$.

A team of three coders, three writers, and one coordinator has mean competence $31/7$ and coordination $5.47/7$. Two such seven-person teams produce $181.98$, versus $175.66$ when merged. Seven is the rounded preferred size $7.00654$: with fractional activities, the same proportions use activity $14/7.00654$ and produce the same total to two decimal places. Marginal products at size seven are $13.828$ for specialists and $8.034$ for coordinators, giving the reported wages $4.15$ and $2.41$ after rounding. These calculations explain the human benchmarks in \cref{fig:human-economy}.

\subsection{The four regime allocations}
\Cref{tab:example-regime-checks} records the current snapshots in \cref{fig:economy-regimes}. Specialist and coordinator employment cover both markets; wages for unemployed coordinators equal their outside option.

\begin{table}[htbp]
\centering\small
\setlength{\tabcolsep}{4pt}
\begin{tabular}{@{}lrrrr@{}}
\toprule
 & Minimal AGI & Supercompetent & Supercooperative & ASI\\
\midrule
AI task skills $\theta$ & $59/35$ & $12$ & $59/35$ & $12$\\
AI coordination & $547/700$ & $547/700$ & $1.60$ & $1.05$\\
Niche output & $65.259$ & $181.005$ & $180.161$ & $187.368$\\
Niche internal price & $.202112$ & $.028620$ & $.030052$ & $.019523$\\
Niche demand price & $.202112$ & $.028493$ & $.029759$ & $.018948$\\
Specialist wage & $2.90072$ & $.39428$ & $.41443$ & $.26896$\\
Coordinator wage & $1.29801$ & $.47561$ & $.25000$ & $.25000$\\
Specialists employed & $12$ & $12$ & $12$ & $12$\\
Coordinators employed & $2$ & $2$ & $.97257$ & $0$\\
\bottomrule
\end{tabular}
\caption{Joint-market calculations for the example, with outside option $1/4$.}
\label{tab:example-regime-checks}
\end{table}

\emph{Minimal AGI.} Pure-AI firms have size $7.00654$ and output per copy $4.94775$, giving unit cost $0.202112$. A human firm in the deep market employs both coordinators and $5.13592$ specialists. The remaining $6.86408$ specialists work across AI firms, with human shares per firm tending to zero. The niche uses AI-only firms with activity $1.88246$. Deep-market AI activity grows as approximately $1.88N$.

\Needspace{7\baselineskip}
\emph{Supercompetent AI.} At task skills twelve, deep AI firms retain size $7.00654$ and have output per copy $35.22127$. The two coordinators work across these firms. Their limiting wage is
\[
 \frac1{12}+(0.97-\bar y_*^H)\log(1+m^*(\bar y_*^H))\approx0.47561.
\]
The niche employs all twelve specialists in two balanced firms of six. It also uses a complete AI recipe of size $5.18036$ with activity $0.08667$. The niche internal price is $0.028620$, above the deep price $0.028392$, because the preferred deep AI recipe does not fit in the niche. Specialist pay is the niche internal price times $5G(3/4)$, or $0.39428$. Thus coordinator pay exceeds specialist pay in this snapshot.

\emph{Supercooperative AI.} At task skills $59/35$ and coordination $1.60$, the niche uses human-only firms of size $6.51455$ and mixed firms of size $9.22247$, with activities $1.91864$ and $0.07281$. Coordinators make up $0.077812$ of the human-only recipe; specialists make up $0.705170$ of the mixed recipe. Total employment is twelve specialists and $0.97257$ coordinators. Thus $1.02743$ coordinators are unemployed. Deep production has an AI-only limiting profile.

\emph{ASI.} At profile $(12,12,1.05)$, the niche uses two balanced human firms of six and a complete AI recipe of size $3.65806$ with activity $0.11769$. The human firms supply $165.317$ units; AI supplies the remaining $22.051$. All specialists work, while both coordinators are unemployed. The deep-market limiting unit cost is zero, although the finite niche's internal price is positive.

Along the competence route, preferred pure-AI size stays fixed and best output per copy is $\theta G(\bar y_*^H)$. Along the coordination route it has preferred sizes $34$ at $0.95$ and $174$ at $0.99$, with best output per copy approaching $(59/35)e^{7/4}\approx9.70062$ as coordination approaches one. \Cref{fig:routes} illustrates the wage and size changes along these routes; its intermediate wage curves have not received the same joint output and entry checks as the four snapshots.

\subsection{Within-firm and between-firm displacement}
Only \cref{sec:displacement} raises the outside-option wage to one. Both displayed paths start with AI coordination below the human coordinator's $0.97$, and coordinators remain unemployed throughout.

\emph{Within-firm displacement.} Fix AI coordination at $0.96$ and raise both task skills from $3$ to $3.4$. All employed humans work in one mixed niche firm with activity one; the deep market is AI-only. At competence three, the niche firm has size $11.00143$ and employs $4.57923$ specialists. At competence $3.15$, specialist employment is $2.39314$. It reaches zero near $3.29535$. Human wages remain at one. A pure-AI firm at its preferred size $42.75$ would produce $609.62$ units at competence three, exceeding niche output $152.80$, so market size constrains the firm's scale.

With both input costs equal to one, an interior mix equates human and AI marginal products. If $s$ is the balanced-specialist share, competence is $L=\theta+(5-\theta)s$ and coordination is $z=y_*^M+(0.75-y_*^M)s$. Differentiating output in $s$ at fixed size gives the condition
\[
 \frac{5-\theta}{\theta+(5-\theta)s}
 +(0.75-y_*^M)\log(1+m)=0.
\]
At exit, set $s=0$ and combine this condition with the pure-AI output choice $P(F(m))F'(m)=1$. These equations give the stated exit value. The displayed segment starts after the earlier replacement of human-only niche firms.

\emph{Between-firm displacement.} Hold task skills at four and raise coordination from $0.80$ to $1.04$. Balanced human firms have size six and output per worker $5G(0.75)\approx13.77645$. Near the employment threshold, pure-AI firms also operate at their preferred size. Their unit cost crosses the human unit cost at
\[
 4G(y_*^M)=5G(0.75),\qquad y_*^M\approx0.85210100.
\]
The preferred AI firm then has size $10.83240$ and produces $149.23197$, which fits within niche output $151.60823$. Numerical entry checks support separate production near this crossing. Specialist employment is twelve below the threshold and zero above it; at equality, any employment from zero to twelve has the same cost. The wage falls from $1.14155$ at coordination $0.80$ to one at exit. Initially, specialists work in mixed deep-market firms. They move into human-only firms near the threshold while total employment stays at twelve.

For \cref{fig:displacement-composition}(b), we select the pre-exit allocation with both six-person human firms in the deep market and AI-only firms in both markets. It has the same surplus as the joint-market calculation; human firms need not sort uniquely to that market. The dots show preferred sizes at the threshold, and $\times\text{many}$ denotes the number of AI firms growing without bound in the deep-market limit.

\subsection{Calculation records and scope}
\begingroup\raggedright
The regime snapshots use \path{Example/equilibrium-checks/four_regimes_refined.json}, except for the supercooperative snapshot, which uses the coordination-$1.60$ state in \path{Example/displacement-checks/refined_paths.json}. Their generating scripts refine output choices and search over entrant compositions, including boundary compositions and firms that employ coordinators. The largest recorded entry gain per input across these four snapshots is below $10^{-10}$.

The current displacement paths are stored in \path{Example/displacement-checks/within_between_paths.json}. The between-firm path has output and entry checks at 47 states; the within-firm path has four such checks, including its exit point. The path scripts compare output choices and refine continuous composition searches. These are numerical checks, not global analytic certificates over every entrant or every point on a path. The files do not prove uniqueness at the employment threshold or convergence from finite deep markets. Appendix~\ref{app:numerics} records the separate calculations for the general model's figures.

\par\endgroup

\end{document}